\documentclass[fleqn,usenatbib]{mnras}

\usepackage{newtxtext,newtxmath}

\usepackage[T1]{fontenc}
\usepackage{ae,aecompl}
\usepackage{siunitx,booktabs}

\usepackage{lastpage}
\usepackage{graphicx}	
\usepackage{amsmath}	
\usepackage{amssymb}	

\defcitealias{Kakiichi2018}{Paper I}
\defcitealias{Meyer2019}{Paper II}
\newcommand{\hone}{H~{\small I}}
\newcommand{\otwo}{O~{\small II}}
\newcommand{\othree}{O~{\small III}}
\newcommand{\cfour}{C~{\small IV}}
\newcommand{\kms}{\text{ km s}^{-1}}

\title[Galaxy-Lyman-$\alpha$ forest correlations at $z\sim 6$]{The role of galaxies and AGN in reionising the IGM - III : IGM-galaxy cross-correlations at $z\sim 6$ from 8 quasar fields with DEIMOS and MUSE}

\author[R. A. Meyer et al.]{
Romain A. Meyer$^{1}$,\thanks{E-mail: r.meyer.17@ucl.ac.uk}
Koki Kakiichi$^{1,2}$,
Sarah E. I. Bosman$^{1}$, 
Richard S. Ellis$^{1}$,\newauthor
Nicolas Laporte $^{1,3,4}$,
Brant E. Robertson $^{5}$,
Emma V. Ryan-Weber $^{6,7}$, 
Ken Mawatari $^{8,9}$ \newauthor
and Adi Zitrin $^{10}$ 
\\
$^{1}$ Department of Physics and Astronomy, University College London, Gower Street, London WC1E 6BT, UK\\
$^{2}$Department of Physics, University of California, Santa Barbara, CA 93106, USA \\
$^{3}$ Kavli Institute for Cosmology, University of Cambridge, Madingley Road, Cambridge CB3 0HA, UK \\
$^{4}$ Cavendish Laboratory, University of Cambridge, 19 JJ Thomson Avenue, Cambridge CB3 0HE, UK \\
$^{5}$ Department of Astronomy and Astrophysics, University of California, Santa Cruz, 1156 High Street, Santa Cruz, CA 95064 USA \\
$^{6}$ Centre for Astrophysics and Supercomputing, Swinburne University of Technology, Hawthorn, VIC 3122, Australia \\
$^{7}$ ARC Centre of Excellence for All Sky Astrophysics in 3 Dimensions (ASTRO 3D), Australia \\
$^{8}$ Department of Environmental Science and Technology, Faculty of Design Technology, Osaka Sangyo University, 3-1-1, Nagaito, \\ Daito, Osaka 574-8530, Japan \\
$^{9}$ Institute for Cosmic Ray Research, The University of Tokyo, 5-1-5 Kashiwanoha, Kashiwa, Chiba 277-8583, Japan \\
$^{10}$Physics Department, Ben-Gurion University of the Negev, P.O. Box 653, Be'er-Sheva 8410501, Israel \\
\vspace{-0.1cm}}
\date{Accepted 2020 March 11. Received 2020 March 11; in original form 2019 December 9}

\pubyear{2019}

\begin{document}
\label{firstpage}
\pagerange{\pageref{firstpage}--\pageref{lastpage}}
\maketitle

\begin{abstract}
We present improved results of the measurement of the correlation between galaxies and the intergalactic medium (IGM) transmission at the end of reionisation. We have gathered a sample of $13$ spectroscopically confirmed Lyman-break galaxies (LBGs) and $21$ Lyman-$\alpha$ emitters (LAEs) at angular separations $20'' \lesssim \theta \lesssim 10'$ ($\sim 0.1-4$ pMpc at $z\sim 6$) from the sightlines to $8$ background $z\gtrsim 6$ quasars. We report for the first time the detection of an excess of Lyman-$\alpha$ transmission spikes at $\sim 10-60$ cMpc from LAEs ($3.2\sigma$) and LBGs ($1.9\sigma$). We interpret the data with an improved model of the galaxy-Lyman-$\alpha$ transmission and two-point cross-correlations which includes the enhanced photoionisation due to clustered faint sources, enhanced gas densities around the central bright objects and spatial variations of the mean free path. The observed LAE(LBG)-Lyman-$\alpha$ transmission spike two-point cross-correlation function (2PCCF) constrains the luminosity-averaged escape fraction of all galaxies contributing to reionisation to $\langle f_{\rm esc} \rangle_{M_{\rm UV}<-12} = 0.14_{-0.05}^{+0.28}\,(0.23_{-0.12}^{+0.46})$. We investigate if the 2PCCF measurement can determine whether bright or faint galaxies are the dominant contributors to reionisation. Our results show that a contribution from faint galaxies ($M_{\rm UV} > -20 \, (2\sigma)$) is necessary to reproduce the observed 2PCCF and that reionisation might be driven by different sub-populations around LBGs and LAEs at $z\sim 6$. 
\end{abstract}



\section{Introduction}
Understanding cosmic reionisation is of prime importance for cosmology and galaxy evolution. The key open questions are the timing of cosmic hydrogen reionisation and the nature of the sources of ionising photons. The timing of reionisation is now constrained to the redshift range $6\lesssim z\lesssim 15$ \citep[][]{PlanckCollaboration2018}. A large set of observational probes such the Lyman-$\alpha$ forest opacity \citep[e.g.][]{Becker2001,Fan2002,Fan2006,Bosman2018,Eilers2018}, the decline of the fraction of Lyman-$\alpha$ Emitters \citep[e.g.][]{Stark2010,Stark2011,Ono2012, Schenker2012, DeBarros2017,Pentericci2018,Mason2018a}, the number of ``dark pixels" in the Lyman-$\alpha$ forests at $z\sim6$ \citep{Mesinger2010,McGreer2015} and the damping wing of quasars \citep[e.g.][]{Banados2018} seem to favour a late, rapid and patchy reionisation process down to $z\sim 5.5-6$ \citep{Greig2017a}. In comparison, the nature of the sources has remained more elusive. Galaxies are thought to provide the bulk of the ionising photon budget necessary to drive reionisation \citep[e.g.][for a review]{Robertson2015} and the contribution of active galactic nuclei (AGN) is, although still debated, thought to be relatively minor \citep[e.g.][]{Giallongo2015, Parsa2018,Kulkarni2019}.

Nonetheless, the properties of the galaxies driving reionisation and their relative contributions are less clear. Although the existence of a large population of galaxies down to $M_{\text{UV}} \lesssim -16$ is now established up to redshift of $z\sim10$ \citep[e.g.][]{Bouwens2015,Livermore2017,Oesch2018}, a fundamental issue is the challenge of measuring the escape fraction $f_{\text{esc}}$ and production efficiencies of ionising photons $\xi_{\text{ion}}$\footnote{Defined as the number of Lyman Continuum photons emitted per unit UV luminosity of the galaxy \citep[e.g.][]{Robertson2013}}, which determine their contribution to the total photoionisation budget. At high-redshift, matching the total ionisation budget to the neutral fraction evolution requires a reionisation process driven by faint galaxies ($M_{\text{UV}} \lesssim -10$) with a moderate escape fraction $f_{\text{esc}}\gtrsim 10-20\,\%$ and standard Lyman continuum (LyC) photon production efficiencies $\log \xi_{\text{ion}}/[\text{erg}^{-1} \text{Hz}] \simeq 25.2-25.5$  \citep{Ouchi2009a, Robertson2013,Robertson2015,Matthee2017,Nakajima2018}.  However the proposed picture is dependent on whether low-mass faint galaxies do contribute significantly to the reionisation of the surrounding intergalactic medium (IGM). Alternatives are possible if massive and efficiently LyC leaking galaxies and AGN dominate late reionisation \citep[e.g.][]{Naidu2020}. Direct measurements of the ionising parameters of galaxies have proven challenging and offer a fractured picture. Spectroscopy of the afterglow of gamma ray bursts indicate extremely low escape fractions $\lesssim 2\,\%$ for their hosts \citep{Tanvir2019}, whereas the peak separation of Lyman-$\alpha$ in a bright LAE (COLA1) results in an indirect measurement of $f_{\text{esc}} \sim 15-30\,\%$ \citep{Matthee2018}. Individual galaxies at high redshift have been found to have hard radiation spectra \citep[e.g.][]{Stark2017,Mainali2018} comparable to strong LAEs or high [\othree]/[\otwo] emitters  at $z\sim3$ \citep{Nakajima2016,Nakajima2018,Tang2019}. At $z\lesssim 4$ where the escape fraction is directly measurable, it is found to be highly varying in individual objects but on average is around the $\sim 5-10\,\%$ required for reionisation \citep[e.g.][]{Shapley2006,Izotov2016,Izotov2018,Vanzella2016,Vanzella2018, Steidel2018,Fletcher2019}.

In this series, we have sought to address the issue of the ensemble contribution of sub-luminous galaxies to reionisation. We have presented in \citet[][henceforth \citetalias{Kakiichi2018}]{Kakiichi2018} a new approach to uncover the contribution of faint sources by the detection and modelling of the statistical \hone\, proximity effect due to faint galaxies clustered around bright LBGs at $z\sim 6$ in cosmic volumes probed in absorption with quasar spectra. In \citet[][henceforth \citetalias{Meyer2019}]{Meyer2019}, we extended this framework to enable us to correlate metal absorbers, considered to be hosted by sub-luminous LBGs, with the IGM transmission measured in the Lyman-$\alpha$ forest of quasars. \citetalias{Kakiichi2018} was a pilot study that analysed only one quasar sightline, which raised the question of the statistical significance of the tantalising proximity effect detected. Indeed, it was shown in \citetalias{Meyer2019} that cosmic variance between sightlines is an important factor also noted independently in simulations \citep{Garaldi2019}. Though \citetalias{Meyer2019} overcame this issue by sampling \cfour\, absorbers at $4.5<z<6$ in 25 quasar sightlines and detected an excess of transmission around \cfour\, absorbers, they were a proxy for observed starlight from galaxies and raised the questions of the nature of the \cfour\, hosts. Nonetheless, both studies suggested that the faint population of galaxies at $z\sim6$ has a high ensemble-averaged escape fraction and/or ionising efficiencies required to sustain reionisation \citep[e.g.][]{Robertson2015}. In this third paper of the series, we present an improved study of the correlation between galaxies and the IGM at the end of reionisation and the resulting inference on the ionising capabilities of the sub-luminous population. We have gathered an extensive dataset of galaxies in the fields of $8$ quasars at $z>6$ through an additional observational campaign with DEIMOS/Keck as well as archival MUSE/VLT data. Moreover, we have extended the analytical framework to include the effect of gas overdensities in addition to the enhanced UV background (UVB) caused by clustered faint galaxies, spatially varying mean free path and forward modelling of peculiar velocities and observed flux uncertainties. Finally, we present a new probe of the galaxy-IGM connection by measuring and modelling the two-point cross-correlation function (2PCCF) between galaxies and transmission spikes in the Lyman-$\alpha$ forest of background quasars.

The plan of this paper is as follows. Section \ref{sec:obs} introduces our new dataset, starting with the 8 high-redshift quasar spectra used in this study. Section \ref{sec:LBG_obs} presents DEIMOS/Keck spectroscopic data of Lyman-break galaxies in three quasar fields with multi-slit spectroscopy. Section \ref{sec:MUSE_archive_data} details our dataset drawn from MUSE archival observations, and our search for Lyman-$\alpha$ Emitters in the Integral Field Unit (IFU) datacubes. In Section \ref{sec:result_compiled}, we present the galaxies detected in the field of background quasars with redshifts overlapping with the IGM probed by the Lyman-$\alpha$ forest, and the cross-correlations of galaxies with the surrounding IGM are detailed in Section \ref{sec:results_data}. Section \ref{sec:model_extension} presents an extension to the analytical framework of \citetalias{Kakiichi2018} necessary to interpret our new measurements. The final results and constraints on the ionising capabilities of galaxies at the end of reionisation are presented in Section \ref{sec:results_final}. We discuss the use of our measurement to differentiate between the relative contributions of faint and bright galaxies to reionisation and the difference between the cross-correlation statistics in Section \ref{sec:discussion} before concluding in Section \ref{sec:conclusions}.

Throughout this paper, the magnitudes are quoted in the AB system \citep{Oke1974}, we refer to proper (comoving) kiloparsecs as pkpc (ckpc) and megaparsecs as pMpc (cMpc), assuming a concordance cosmology with $H_0 = 70 \kms\, \text{Mpc}^{-1}$, $\Omega_{\rm M} = 0.3,~\Omega_\Lambda = 0.7$.
\section{Observations}
\label{sec:obs}
This series focuses on the measurement and modelling of the correlations between galaxies and the surrounding IGM at the end of reionisation. In order to achieve that goal, we have gathered different datasets of luminous galaxies acting as signposts for overdensities of less luminous sources. We have continued the approach of \citetalias{Kakiichi2018} by confirming high redshift Lyman-break galaxies (LBG) via their Lyman-$\alpha$ emission with the DEep Imaging Multi-Object Spectrograph \citep[DEIMOS,][]{Faber2003} at Keck. For convenience, we refer to those as LBGs  because of their selection technique (although formally they are all also Lyman-$\alpha$ emitters (LAEs) given they were confirmed with this line). Throughout the paper, we thus reserve the terminology LAE for galaxies detected blindly in archival IFU  data of the Multi Unit Spectroscopic Explorer \citep[MUSE,][]{Bacon2010} at the VLT. MUSE complements the early DEIMOS approach since we use a different selection method for galaxies and probe the smaller scales appropriate to the small MUSE  Field of View (FoV) ($1'$ corresponding to $\sim 360$ pkpc at $z\sim 5.8$). These complementary datasets of galaxies were gathered in the field of $z\sim 6$ quasars with existing absorption spectroscopy of the Lyman-$\alpha$ forest, which probes the IGM transmission and ultimately enables us to compute the galaxy-IGM correlations. We now proceed to describe this rich observational dataset, starting with the quasar spectra and moving then to the DEIMOS and MUSE data.

\subsection{Quasar spectroscopic observations}
\label{sec:obs_qso}

The $8$ quasar fields studied in this work were chosen to have existing moderate or high Signal-to-Noise ratio (SNR) spectroscopy of the Lyman-$\alpha$ forests and either be accessible to Keck for DEIMOS follow-up or have archival MUSE data with adequate ($\geq 2$h) exposure time. The quasar spectra used in this study were  downloaded from the ESO XShooter Archive or the Keck Observatory Archive (ESI). We use the same ESI spectrum of J1148 as in Paper I, originally observed by \citet{Eilers2017}. The quasars already presented in \citet[][J0836, J1030]{Bosman2018} were reduced using a custom pipeline based on the standard \textsc{ESORex} XShooter recipes as detailed therein. The remaining quasars (J0305, J1526, J2032, J2100, J2329) were reduced with the open-source reduction package \textsc{PypeIt} \footnote{\url{https://github.com/pypeit/PypeIt}} \citep{Prochaska2019}. The quasars have a median SNR of $\sim 16$ in the rest-frame UV. The spectra were finally normalised by a power-law $f(\lambda) = A \lambda^{b}$ fitted to the portion of the continuum redwards of Lyman-$\alpha$ relatively devoid of broad emission lines ($1270-1450$ \AA), as described in \citet{Bosman2018}, to compute the transmission in the Lyman-$\alpha$ forest. Table \ref{tab:all_detections_summary} summarises the quasar spectroscopic data information alongside the galaxy detections. 

\begin{table*}
    \centering
    \begin{tabular}{lcccccccccc}
     Quasar & $z$ & LAEs & LBGs & DEIMOS ID & MUSE ID & Total Exptime & ref  &  Spectrum & ref \\ \hline \hline
     J0305$-$3150 & 6.61 & 3 & - & - & 094.B-0893(A) & 2h30m & (2)  & XShooter & (3) \\
     J0836$+$0054 & 5.81 & - & 1  & U182 & - & 5h10m & (1)  & XShooter  & (5)  \\
     J1030$+$0524 & 6.28 & 7 & 8  & C231, U182  & 095.A-0714(A)  & 10h50m\,/\,6h40m & (1)/(4)   & XShooter & (5) \\
     J1148$+$5251 & 6.419 & - & 4 & C231, U182 & -  & 13h10m & (1,6) & ESI & (5) \\
     J1526$-$2050$^{a}$ & 6.586  & 2 & - & - & 099.A-0682(A) & 3h20m   & (7,8) & XShooter & (9) \\
     J2032$-$2114$^{b}$ & 6.24 & 3 & - & -  &  099.A-0682(A) & 5h & (7,8) & XShooter & (10)\\
     J2100$-$1715 & 6.09 & 4 & - &-  &097.A-5054(A) & 3h40m   & (7,8) & XShooter  & (11) \\
     J2329$-$0301 & 6.43 & 2 & - & - & 060.A-9321(A)  & 2h  & (7,8) & ESI & (12) \\
     \hline
    \end{tabular}
    \caption{Observational data summary of the quasar fields. Fields: 1) Quasar name (\textit{a:} Also known as P231-20.  \textit{b:} Also known as P308-21) 2) Quasar redshift 3) Number of suitable LAEs in the Lyman-$\alpha$ forest redshift range of the nearby quasar detected with MUSE 4) Number of suitable LBGs detected with DEIMOS 5) DEIMOS Keck programme ID 6) MUSE ESO programme ID 7) Total exposure time of DEIMOS/MUSE in the field 8) Reference of the original published paper on the observational programme 9) Instrument used for the quasar spectrum 10) Reference of the discovery paper of the quasar.  References: (1) This work (2) \citet{Farina2017} (3)  \citet{Venemans2013} (4) \citet{Diaz2020} (5) \citet{McGreer2015} (6) \citet{Kakiichi2018} (7) \citet{Drake2019} (8) \citet{Farina2019} (9) \citet{Mazzucchelli2017} (10) \citet{Banados2016} (11) \citet{Willott2007} (12) \citet{Bosman2018}  }
    \label{tab:all_detections_summary}
\end{table*}

\subsection{DEIMOS spectroscopy of LBGs in 3 quasar fields}\label{sec:LBG_obs}
As part of this study we have re-observed the field of quasar J1148$+$5251 explored in \citetalias{Kakiichi2018} to improve our selection of LBGs. As the slitmask design of DEIMOS does not allow small slit separations, only a selected subset of LBGs can be observed in any given mask. Accordingly, the detection of the proximity signal in \citetalias{Kakiichi2018} might be affected by the sampling of candidate LBGs in the field. We also include data for two new quasar fields: J1030$+$0524 ($z=6.3$) and J0836$+$0054 ($z=5.8$) (Table \ref{tab:DEIMOS_observing_log}).

\begin{figure}
    \centering
    \includegraphics[width=0.48\textwidth]{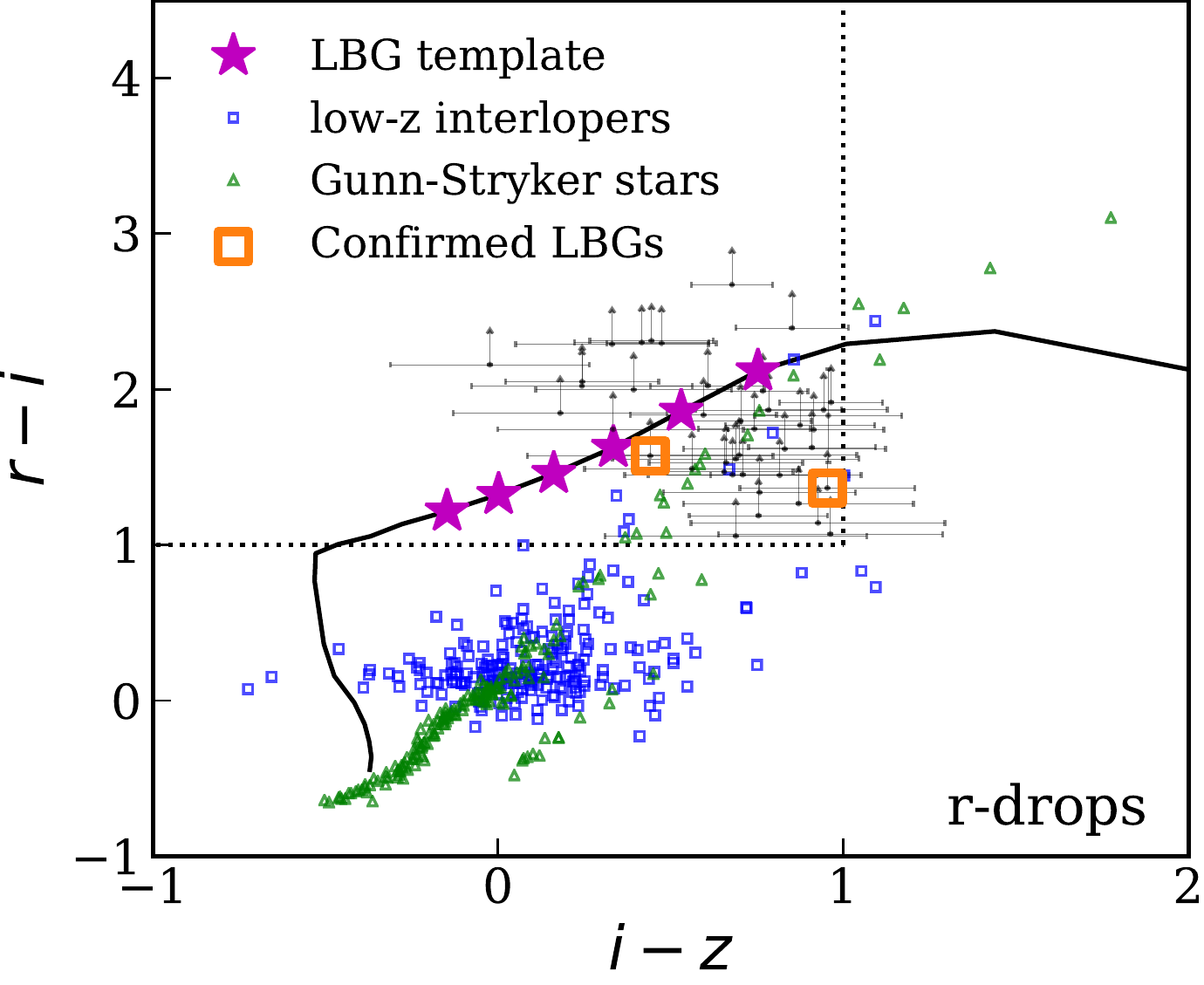}
    \includegraphics[width=0.48\textwidth]{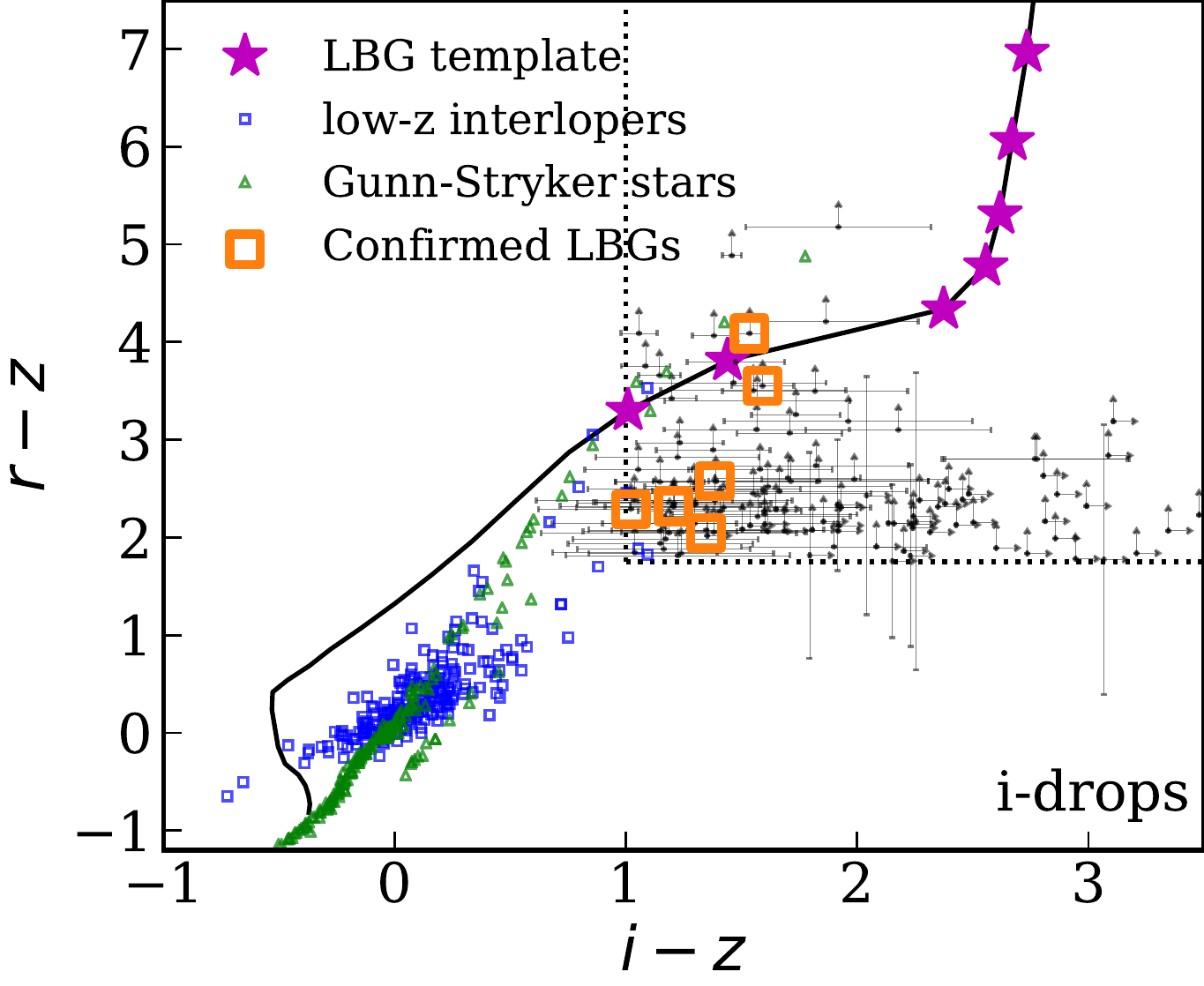}
    \caption{r-drop-out (top) and i-drop-out (bottom) selection of candidate LBGs in the fields of J0836, J1030 and J1148. A LBG template (magenta stars, black line) falls into the colour-colour cuts (dotted lines) at the redshift of interest $z\sim 5-6$. Galactic stars \citep[green triangles,][]{Gunn1983} and low-redshift interlopers \citep[blue squares, VUDS-DR1 samples from the COSMOS field,][]{LeFevre2015,Tasca2017} are however mostly rejected. The target candidates are in shown in black, and confirmed LBGs are highlighted with orange squares.}
    \label{fig:drop-out_selection}
\end{figure}

Deep ground-based photometry of the three fields was used to construct catalogs of \textit{r'} and \textit{i'} drop-outs for follow-up. The fields of J1030 and J1148 have been observed in the SDSS \textit{r'}-,\,\textit{i'}-,\,\textit{z'}-band filters with the Large Binocular Camera \citep[LBC,][]{Giallongo2008} at the Large Binocular Telescope \citep[LBT, ][]{Hill2000}, with the publicly available photometry reduced by \citet[][]{Morselli2014} \footnote{\url{http://www.oabo.inaf.it/~LBTz6/1030/lbtz6.html}}. For the field of J0836, we used \textit{r'}-, \textit{i'}-, \textit{z'}-band observations \citep[][]{Ajiki2006} taken with SuprimeCam on the 8.2m Subaru Telescope \citep{Kaifu2000, Miyazaki2002}. We chose the following colour cuts to select potential $z\sim 5-6$ LBG candidates (see Fig. \ref{fig:drop-out_selection})
\begin{equation}
    \left[r'-i' > 1.0 \right] \wedge \left[ i'-z' < 1.0 \right] \wedge \left[ z'<z'(3\sigma) \right]
\end{equation}
for $r'$-drop-outs, and
\begin{equation}
    \left[ i'-z' > 1.0 \right] \wedge \left[r'>r'(2\sigma) \vee  r'-z' > 1.75\right] \wedge \left[z'<z'(3\sigma)\right]
\end{equation}
for i-drop-outs, where $r',\,i',\,z'$ indicates the magnitude in the corresponding SDSS band, and $r'(2\sigma), z'(3\sigma)$ the limiting $2,3\sigma$ magnitude in the $r', z'$-band image respectively. All candidates were visually inspected to produce a final catalogue. In designing the DEIMOS masks, we prioritised  drop-outs based on the strength of their colour drop ($i'-z' (r'-i')>1.0,\,1.3,\,1.5$) or $r'$-band non-detection ($r'>r'(2\sigma,\,3\sigma,\,5\sigma)$) to optimise the chance of confirmation. Priority was always given to better candidates first, and then to $i'$-drop-outs over $r'$-drop-outs of the same quality. The masks contained $\sim 25-40$ slits for science targets and $5$ or $6$ square holes for alignment stars.

\begin{figure}
    \centering
    \includegraphics[width=0.45\textwidth]{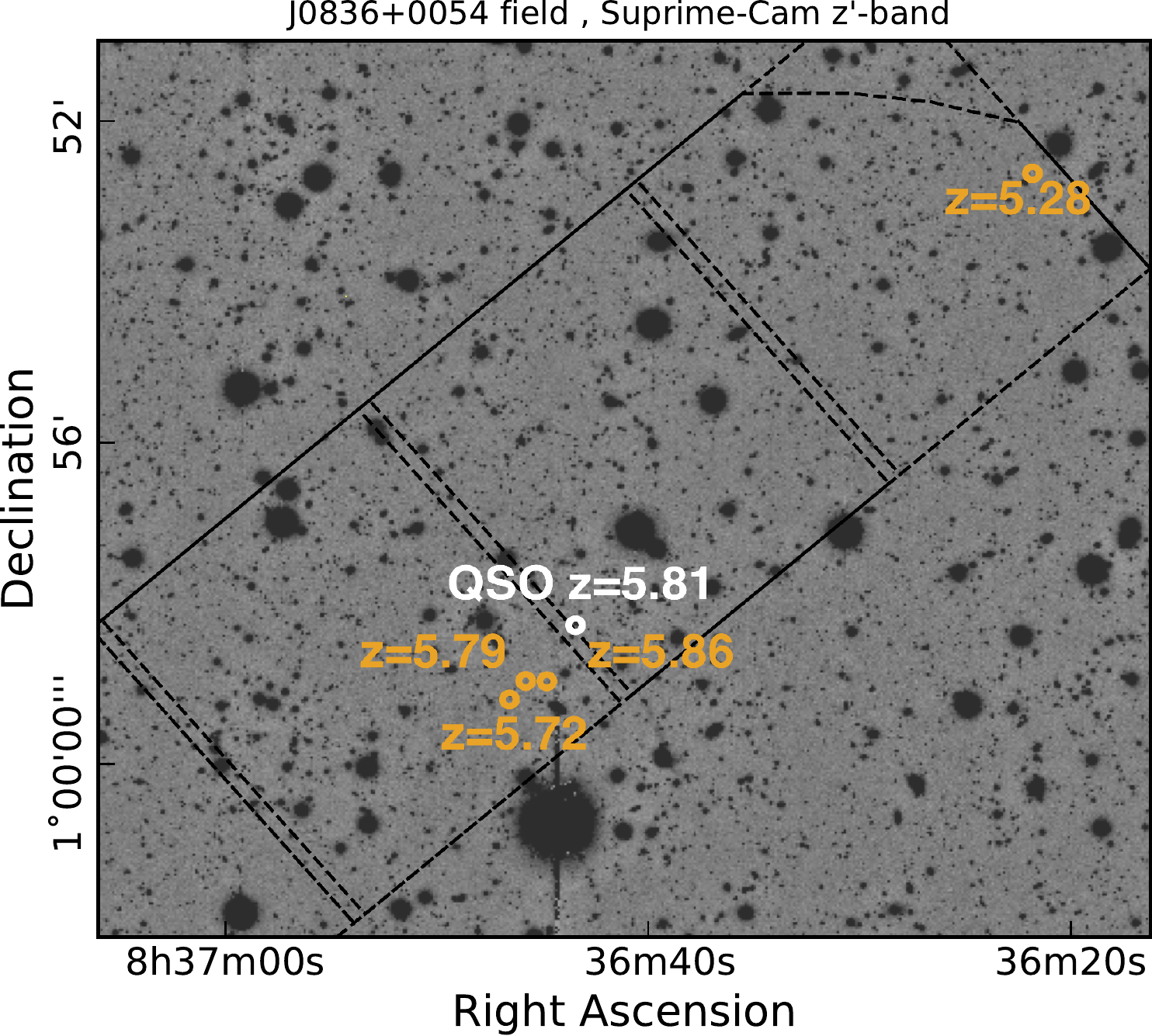}
    \includegraphics[width=0.45\textwidth]{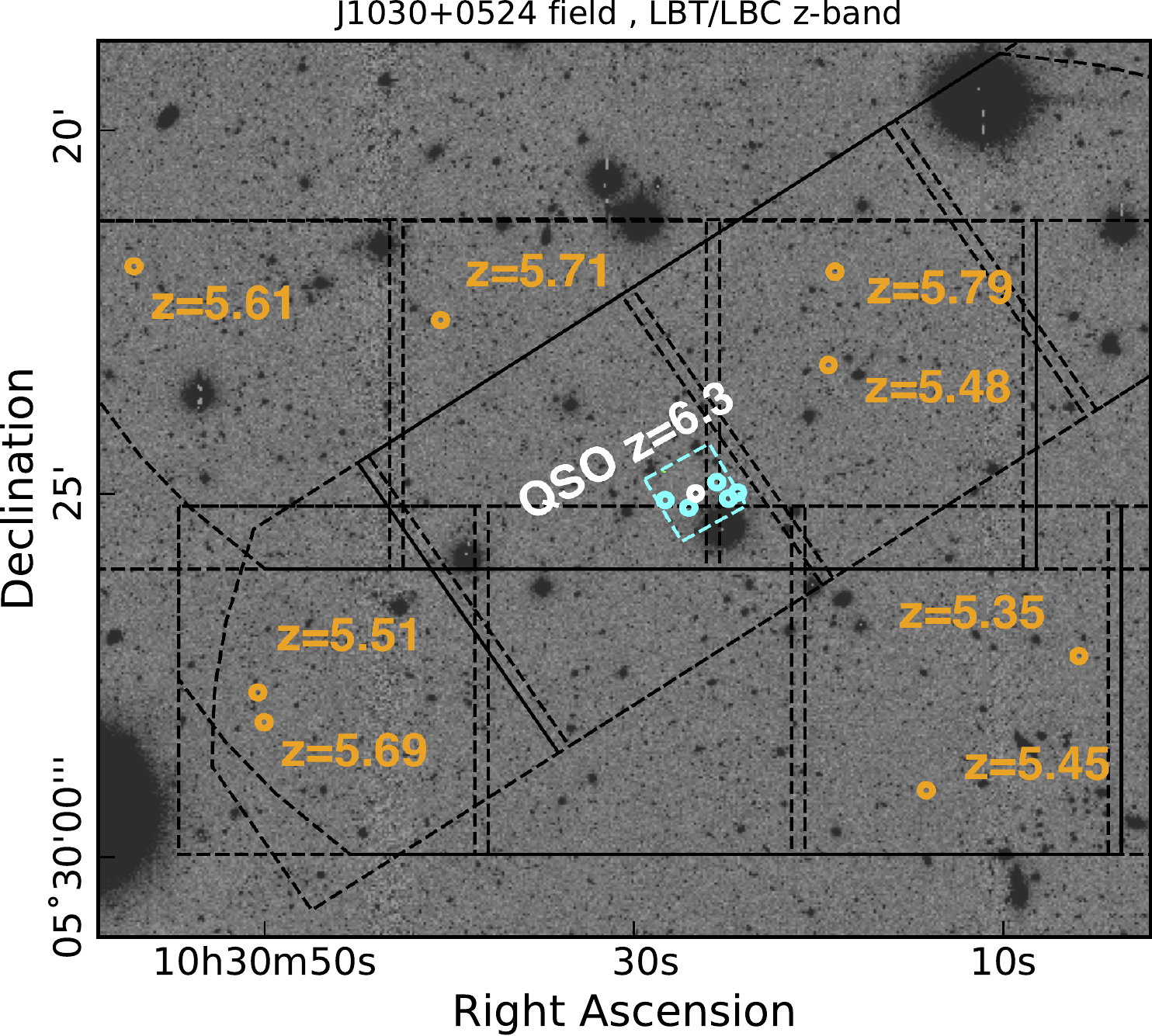}
    \includegraphics[width=0.45\textwidth]{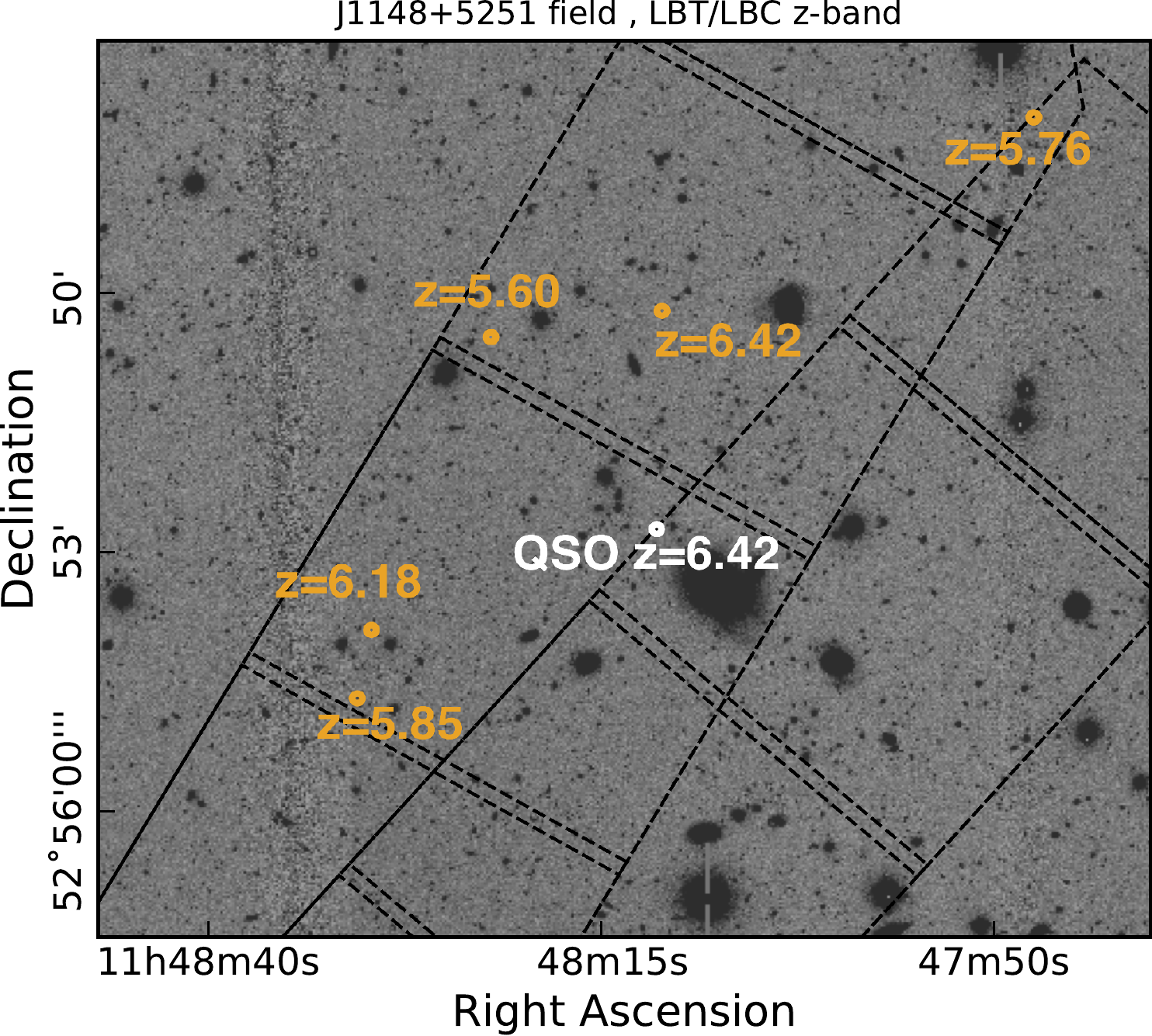}
    \caption{Spectroscopically confirmed high-redshift LBGs (orange squares) and potential LAEs (green circles) identified with DEIMOS in the quasar fields of J0836, J1030, J1148. We overlay the DEIMOS slitmask FoV in black. For J1030 (middle panel), we also add the MUSE FoV (cyan dashed square) and the LAEs detected in the MUSE datacube (cyan circles)}
    \label{fig:DEIMOS_fields_results}
\end{figure}

The candidates were observed with the DEIMOS instrument \citep{Faber2003} at the Keck II 10-m telescope during two observing runs in 2017 March 26-27 (PI: A. Zitrin) and 2018 March 07-08 (PI: B. Robertson). We confirmed 13 LBGs in the 3 fields over the course of 4 nights in good conditions with a seeing of $0.7-0.9''$ except for one night at $0.9-1.5''$, as summarised in Table \ref{tab:DEIMOS_observing_log}. The masks and the LBG detections are shown in the $3$ fields in Fig. \ref{fig:DEIMOS_fields_results}. Surprisingly we could not confirm any new LBG in the field of J1148+5251, although \citetalias{Kakiichi2018} confirmed 5(+1 AGN) in a smaller exposure time.  The other masks in J0836 and J1030, exposed for 1h30m to 5h10m, yielded 2 to 4 LBG confirmations each. 
The completeness of our search for LBGs in the relevant
redshift range is not straightforward to estimate but fortunately
not a major concern for our analysis. Whilst in principle the total number of galaxies in these fields can be estimated using the UV LF and the depth of the photometric data, we find no proportionality with the observed numbers of LBGs. The scatter from field to field detailed above is therefore mainly driven by the random selection of objects on each mask ($\lesssim 20$ 'prime' candidates) and the number of mask observed for each given field. Indeed, the number of objects confirmed per mask is roughly constant, regardless of the redshift and field. However, this does not affect our results since we are aiming to measure the average Ly-$\alpha$ transmission around the \textit{detected} bright galaxies only. As we cross-correlate them with the Ly-$\alpha$ forest and do not measure their number density, we do not need to correct for completeness (see further Section \ref{sec:results_data}). 

The data were reduced using the DEEP2 pipeline \citep{Cooper2012, Newman2013}, and the 1D spectra were extracted using optimal extraction with a $1.2''$ boxcar aperture \citep{Horne1986}. The 2D spectra were inspected visually by $5$ authors (RAM, KK, SEIB, RSE, NL) for line emitters. Candidate LBGs were retained if they were found by 3 authors or more in the 2D spectra. We show examples of a clear LBG detection and a less convincing one in Fig. \ref{fig:example_DEIMOS_LBG}. The remaining LBG detections are presented individually in Appendix \ref{appendix:DEIMOS_detections}. We also present serendipitous line emitters (without optical counterpart) which are not used in this study. 

\begin{table}
    \centering
    \begin{tabular}{lccccc}
     Quasar & $N_{\rm LBG}$  & \#$_{\text{Mask}}$ & Exptime & Seeing\\ \hline \hline
     J0836$+$0054 & 4$^{a}$  & K1 & 5h10m & $0.7''-0.9''$  \\
     J1030$+$0524 & 3 & K1 & 4h00m & $0.9''-1.5''$ \\
       & 3  & K2 & 5h20m & $0.7''-0.9''$ \\
       & 2  & K3 & 1h30m & $0.7''-0.9''$ \\
     J1148+5251 & 4$^{b}$ & K1$^{b}$ & 4h30m & $0.7''-1.5''$  \\
             & 0 & K2 & 8h40m & $0.7''-0.9''$ \\
     \hline
    \end{tabular}
    \caption{Summary of the DEIMOS observations. The masks J1030-K1, J1030-K2 and J1148-K1 were observed in 26-27 March 2017 (PI: A. Zitrin, ID: C231) and the remainder in 07-08 March 2018 (PI: B. Robertson, ID: U182). The number of confirmed LBGs is only weakly correlated to the total exposure time on the mask. \textbf{a)} 3 LBGs in J0836+0054 are in the near-zone of the quasar and hence they do not appear in Table \ref{tab:all_detections_summary} as they are not suited for our purposes. They will be studied in greater in \citet{Bosman2019}. \textbf{b)} We remove the faint AGN as well as the least convincing LBG detection (ID = 022) presented in \citetalias{Kakiichi2018}  to harmonize the LBG selection.  \label{tab:DEIMOS_observing_log} }

\end{table}

\begin{figure}
    \centering
    \includegraphics[width = 0.48\textwidth]{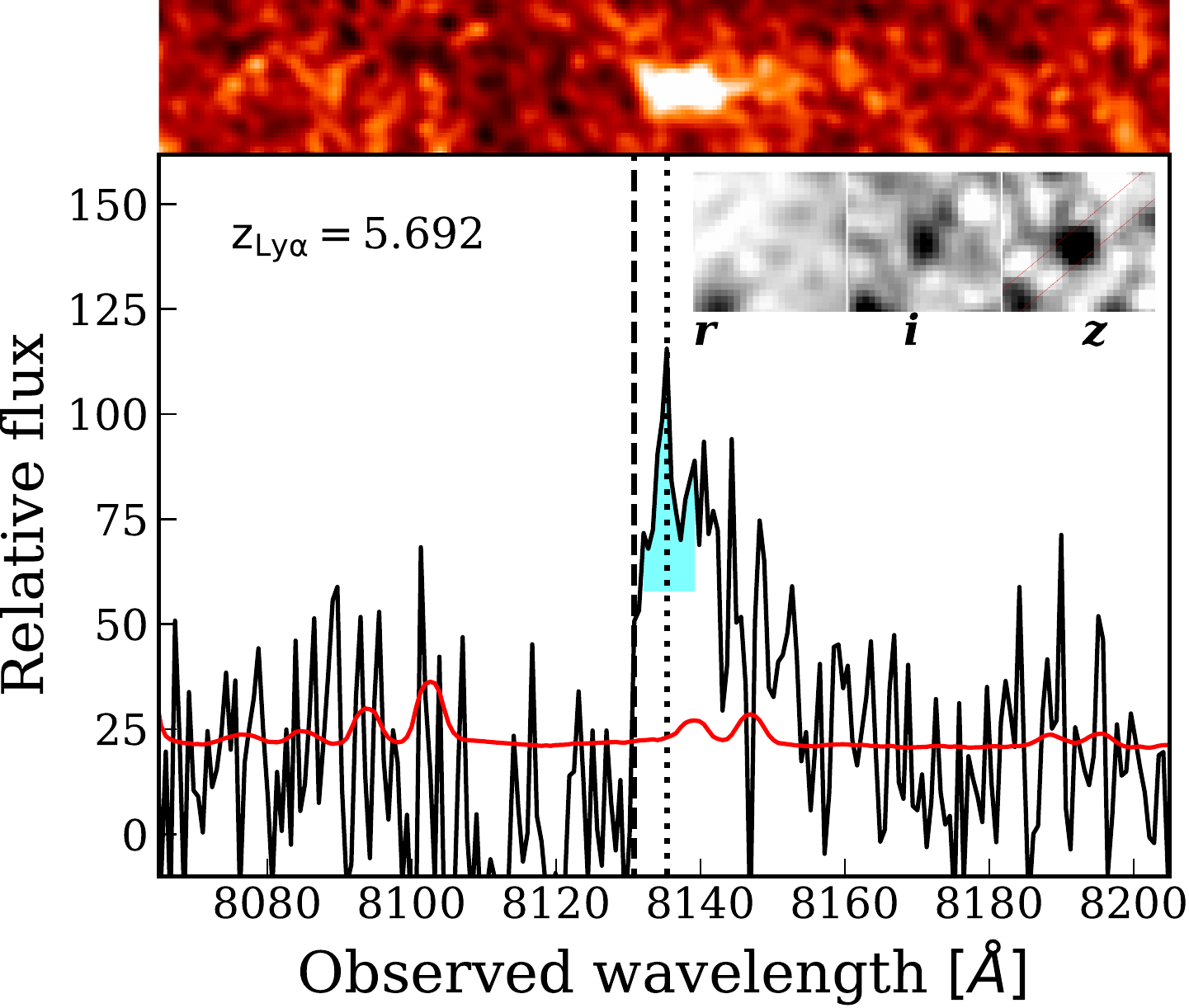} \\
    \includegraphics[width =0.48\textwidth]{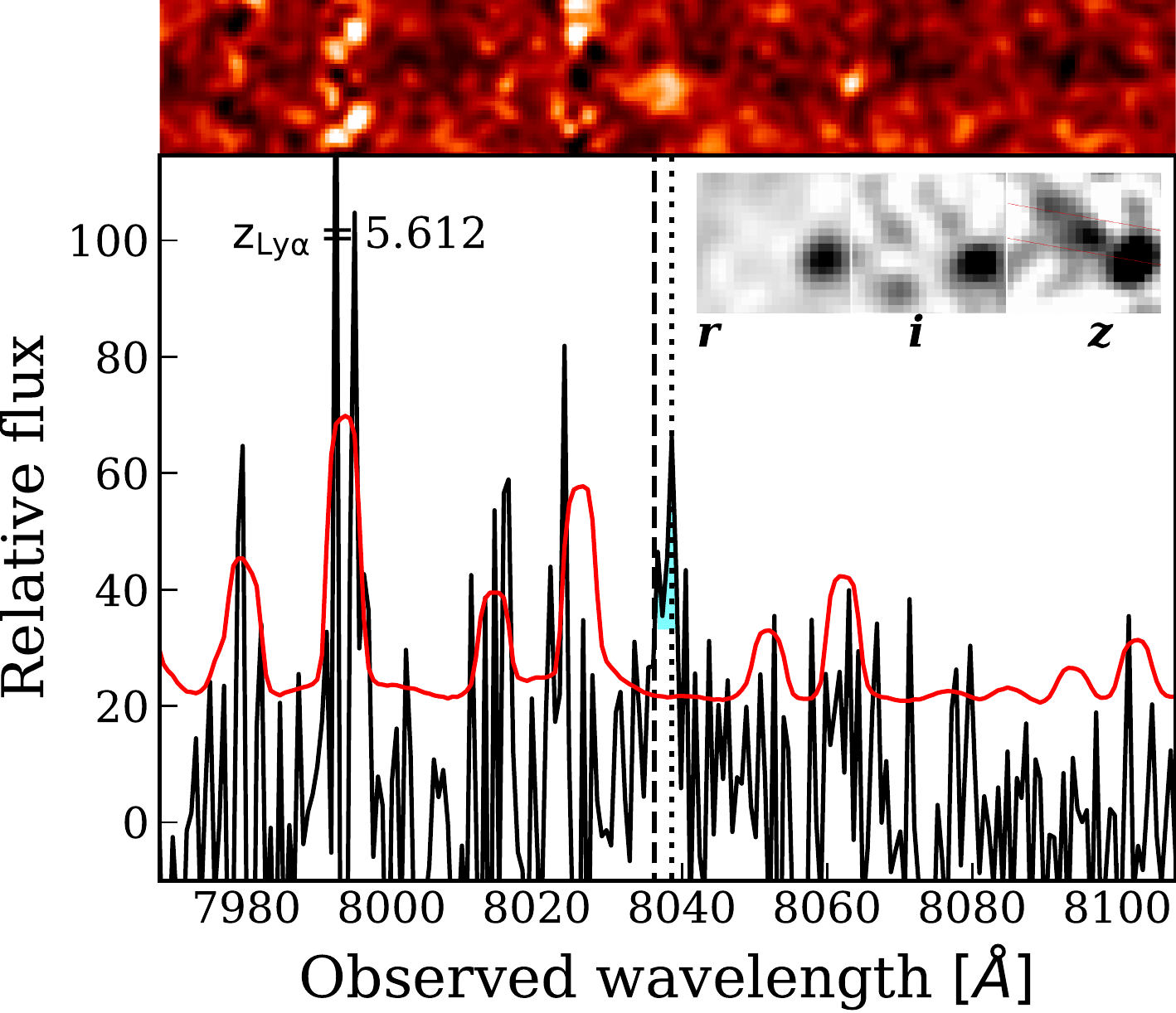}
    \caption{Examples of a clear and a marginal detection of Lyman-$\alpha$ emission for LBG-selected galaxies in the field of J1030. The top panels shows the 2D spectra from which the spectrum (black line) and noise (red) are optimally extracted using a boxcar aperture of $1.2$''. The peak of Lyman-$\alpha$ is identified with a dotted line, and the location of the systemic by a dashed line. The systemic redshift is found by applying a correction based on the FWHM of the Lyman-$\alpha$ emission (cyan, see Section \ref{sec:z_corr}). In the upper right corner is displayed the \textit{riz} photometry used for the drop-out selection. The remaining detections are presented in Appendix \ref{appendix:DEIMOS_detections}.}
    \label{fig:example_DEIMOS_LBG}
\end{figure}


\subsection{Archival MUSE quasar fields \label{sec:MUSE_archive_data} }
We exploit 6 $z\sim 6$ quasar fields with deep ($\gtrsim 2\text{h}$) archival MUSE data to search for galaxies close to the sightline. The MUSE quasar fields are listed in Table \ref{tab:all_detections_summary}.

We reduce the archival MUSE data using the MUSE v2.6 pipeline \citep{Weilbacher2012,Weilbacher2015} with the standard parameters. We further clean the skylines emission using the Zurich Atmospheric Purge (ZAP) code \citep{Soto2016}. After masking the bright sources and the edges of the data cubes, we run \textsc{MUSELET} \citep{Bacon2016} and \textsc{LSDCat} \citep{Herenz2017} to extract line emitters. We find that both algorithms are complementary due to their different search strategy. \textsc{MUSELET} reduces the IFU cube to a series of narrow band images ($6.25$\,\AA\,width) and uses \textsc{Sextractor} to identify emission lines by subtracting a median continuum constructed from the adjacent wavelength planes. Detections in several narrow bands at similar positions can be grouped together to find a redshift solution. Whilst it is a robust technique, continuum absorptions or rapid continuum variations often produce spurious detections (when the adjacent narrow bands are subtracted). \textsc{LSDCat} improves the removal of foreground continuum objects by utilizing a median filtering of the cube in the wavelength dimension. The emission lines are then detected with a 3-dimensional matched-filtering approach. \textsc{LSDCat} also allows one to mask brighter sources with custom masks. It however then fails to pick faint sources next to bright objects if the masking and/or the median filtering is too aggressive. Finally, the width of the narrow-bands in \textsc{MUSELET} and the convolution kernel sizes of the matched-filter in \textsc{LSDCat} can lead to different false positives or negatives. Therefore, we use both algorithms to generate a consolidated list of line emitter candidates which are then visually inspected to compile a robust sample of high-redshift LAEs. We check that candidates are present in the two datacubes produced with the two halves of the exposures to remove cosmic rays and other artifacts, and we remove double peaked emissions which are likely to be low-redshift [\otwo] $\lambda \lambda$ $3727$\,\AA\, doublets as it would mimic a double-peaked $z\sim5-6$ Lyman-$\alpha$ with a reasonable peak separation $\Delta v\sim 220\kms$. Double-peaked Lyman-$\alpha$ profiles at $z>5$ are exceedingly rare due to absorption by the IGM \citep{Hu2016,Songaila2018}, so we expect to lose very few high-redhift LAEs in being so conservative. Finally, we produce a white light image of the MUSE cube and check that the line detection is not caused by a poor continuum subtraction of a bright foreground object or a nearby contaminant (see Fig. B7, online material for typical false positives). We show two representative detections in Fig. \ref{fig:MUSE_detections_LAEs} and the remainder in Appendix 
B (online material).

\begin{figure*}
    \centering
    \includegraphics[width=0.98\textwidth]{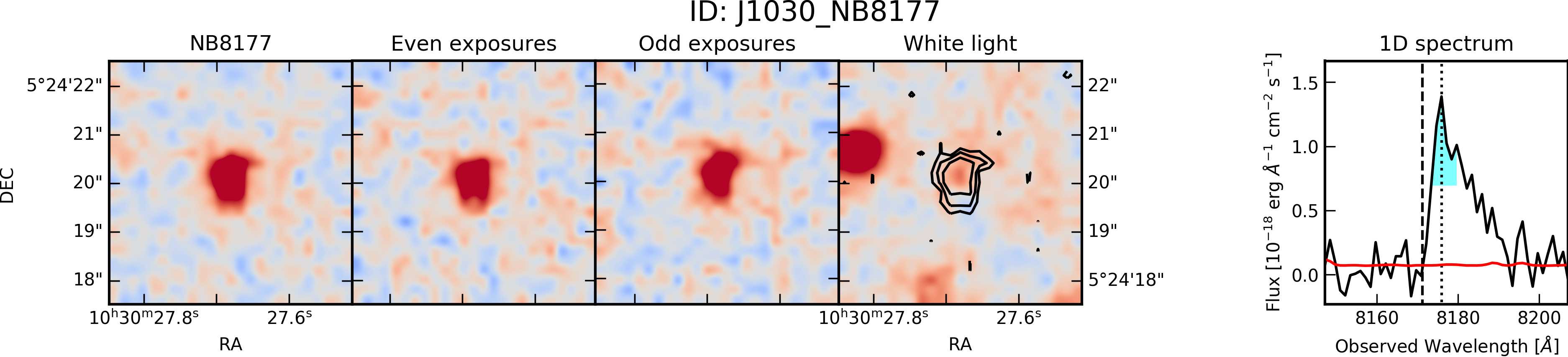} \\
    \includegraphics[width=0.98\textwidth]{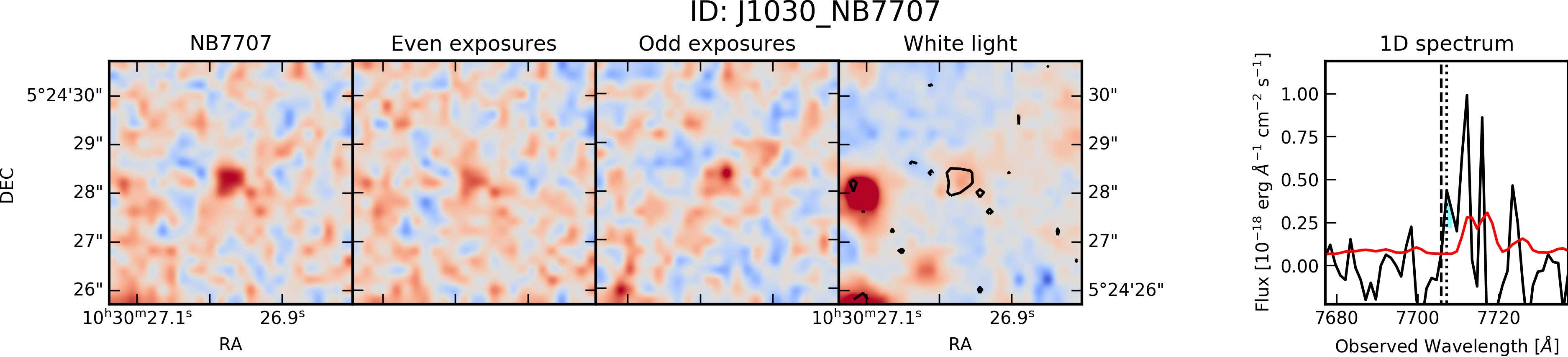}
    \caption{Two representative LAEs detected in MUSE archival data in the field of quasar J1030+0524. The first detection is a strong LAE in the field of J1030 already reported by \citet{Diaz2015}. The four first panels show, in order, the detection in the narrow band centred on the detection in the full combined cube, followed by the same location in cubes with either half of the exposures. The fourth panel shows the white light image with superimposed black significance levels ($-3\sigma, 3\sigma,5\sigma,8\sigma$) of the narrow band detection. The fifth panel shows the optimally extracted 1D spectra (black line) and the noise level (red line). The cyan shading highlights the FWHM of the line, used to correct the peak redshift (dotted vertical line) to systemic (dashed vertical line). The remainder of the LAE detections are summarised in Table \ref{tab:MUSE_LAE} and individual detections are presented in Appendix \ref{appendix:MUSE_detections}.}
    \label{fig:MUSE_detections_LAEs}
\end{figure*}

\begin{figure}
    \centering
    \includegraphics[width = 0.45\textwidth]{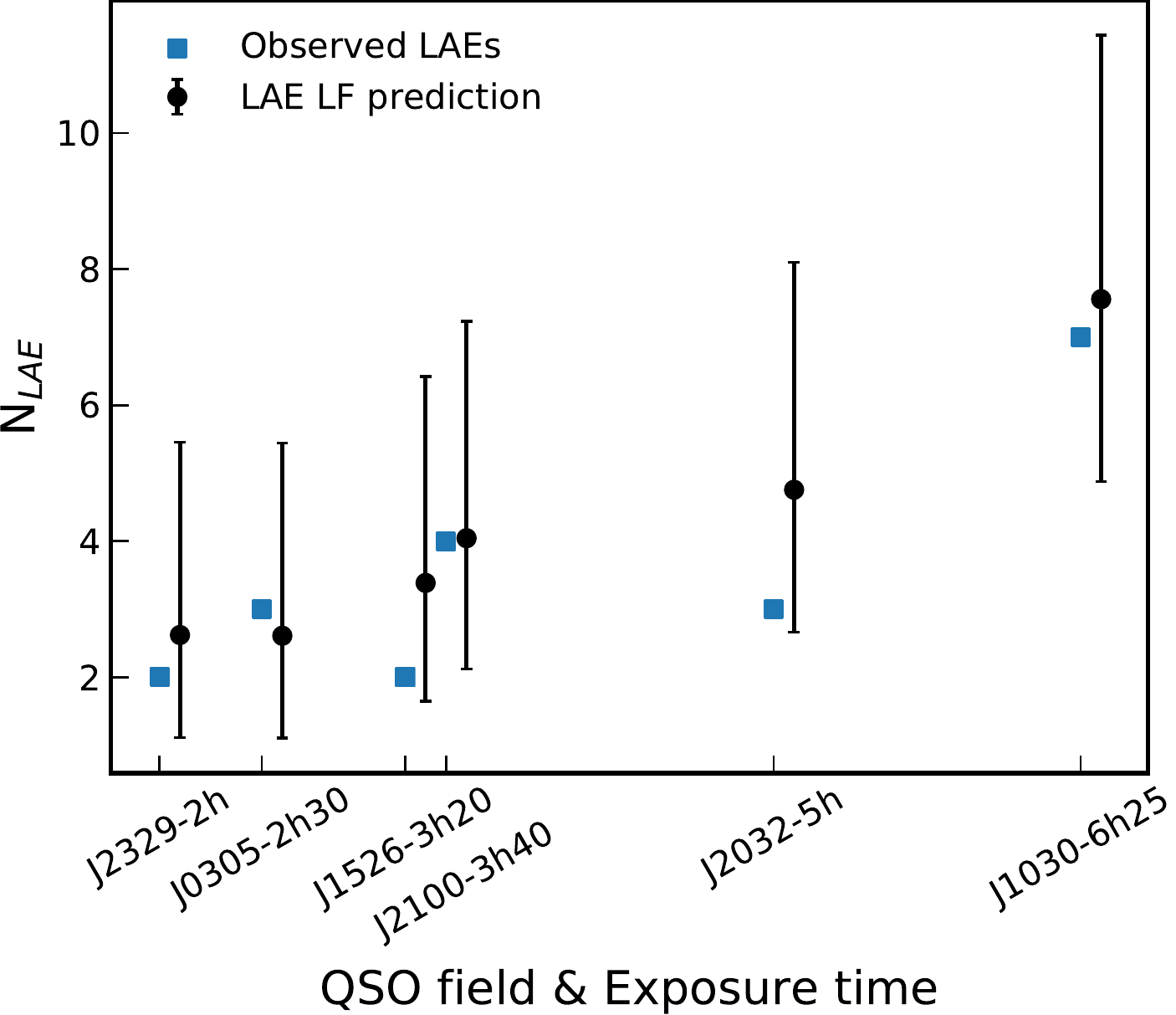}
    \caption{Predicted number of recovered LAEs in each MUSE cube (black 1-sigma Poisson ranges) compared to the number of retrieved LAEs (blue squares),  with the quasar fields sorted (x-axis) by increasing exposure time of the MUSE data. The prediction is made from the LAE LF of \citep{DeLaVieuville2019a,Herenz2019} in the redshift range of the Lyman-$\alpha$ forest of each central quasar and effective efficiencies of \textsc{LSDCat} and MUSE reduction. We apply an effective efficiency of the MUSE LAE search which is derived by comparison with the 27h GTO observation of a single field searched for LAEs at high-redshift with \textsc{LSDCat} in \citet{Bacon2015}. }
    \label{fig:comparison_LAE_LF}
\end{figure}

We checked that the number of LAEs is consistent with expectations from the LAE luminosity function (LF) integrated down to the MUSE sensitivity limit. We first compute the number of LAEs in a given comoving volume using the LAE LF from \citet{DeLaVieuville2019a,Herenz2019} which measured the LAE LF in deep MUSE datacubes with \textsc{MUSELET} and \textsc{LSDCat}, respectively, i.e. the same algorithms that we use. 
By comparing the numbers of LAEs predicted with the LAE LF on a deep $27$h field realised by the MUSE GTO team \citep{Bacon2015} to the numbers of LAEs those authors recovered with \textsc{LSDCat}, we estimate that \textsc{LSDCat} has a recovery rate of $\simeq 32\,\%$ for LAEs at $z\sim 5.5$. This is a global recovery rate for all LAEs with peak flux density greater than $f>4.8\times 10^{-19} \text{erg s}^{-1} \text{cm}^{-2}$, which is  below the sensitivity of all MUSE observations used in this study. We then predict the number of LAEs we expect to find in each of our MUSE quasar fields depending on the exposure time, effective survey area, and the redshift interval of the central quasar Lyman-$\alpha$ forest. We find good agreement between the predicted number (including \textsc{LSDCat} efficiency) and the number of retrieved LAEs (Fig. \ref{fig:comparison_LAE_LF}), indicating a successful search for LAEs and low levels of contamination.

Table \ref{tab:all_detections_summary} summarises all the LBGs and LAEs detected in our quasar fields, alongside the reference of the MUSE and DEIMOS programmes, and the quasar discovery reference study.

\subsection{Correcting the Lyman-$\alpha$-based redshifts}
\label{sec:z_corr}
The red peak of the Lyman-$\alpha$ emission line, commonly observed without its blue counterpart at high-redshift due to the increasingly neutral IGM, is often shifted from the systemic redshift of the galaxy. \citet{Steidel2010} give a typical offset of $v_{\rm red}^{\text{peak}} \sim 200\,\text{km s}^{-1}$, but the range is large and can span  $\sim 0 - 500 \kms$ at high redshift \citep[e.g.][]{Vanzella2016,Stark2017,Hashimoto2018a}. A velocity offset of $\sim 200 \kms$ translates at $z\sim 6$ to $\sim 2 \text{ cMpc}~(\sim 280 \text{ pkpc}) $, which is not negligible given the expected scale of $\sim 10 \text{ cMpc}$ for the peak of the cross-correlation signal. As the cross-correlation is computed in 3D space and then radially averaged, any offset would damp the sought-after signal.

In order to get a better estimate of the systemic redshift of the galaxy, we thus apply a correction to the Lyman-$\alpha$ redshift based on the Full-Width-Half-Maximum (FWHM) of the line. We follow the approach of  \citet{Verhamme2018} who developed an empirical calibration using the FWHM directly measured from the data without modelling
\begin{equation}
    v_{\text{red}}^{\text{peak}} = 0.9(\pm 0.14) \times \text{FWHM}(\text{Ly}\alpha) - 34(\pm 60) \kms \text{   .   }
\end{equation}

The measured FWHM values of our LAEs (LBGs) all fall in the expected range $100 \kms\lesssim \text{FWHM} \lesssim 400 \kms$, and are indicated for each LAE (LBG) in Table \ref{tab:MUSE_LAE}. Throughout this paper, we use these corrected redshifts for the purpose of computing galaxy-IGM cross-correlations.

\section{The apparent clustering of galaxies and transmission spikes from 8 quasar fields}
\label{sec:result_compiled}
\begin{figure*}
    \centering
    \includegraphics[width=0.98\textwidth]{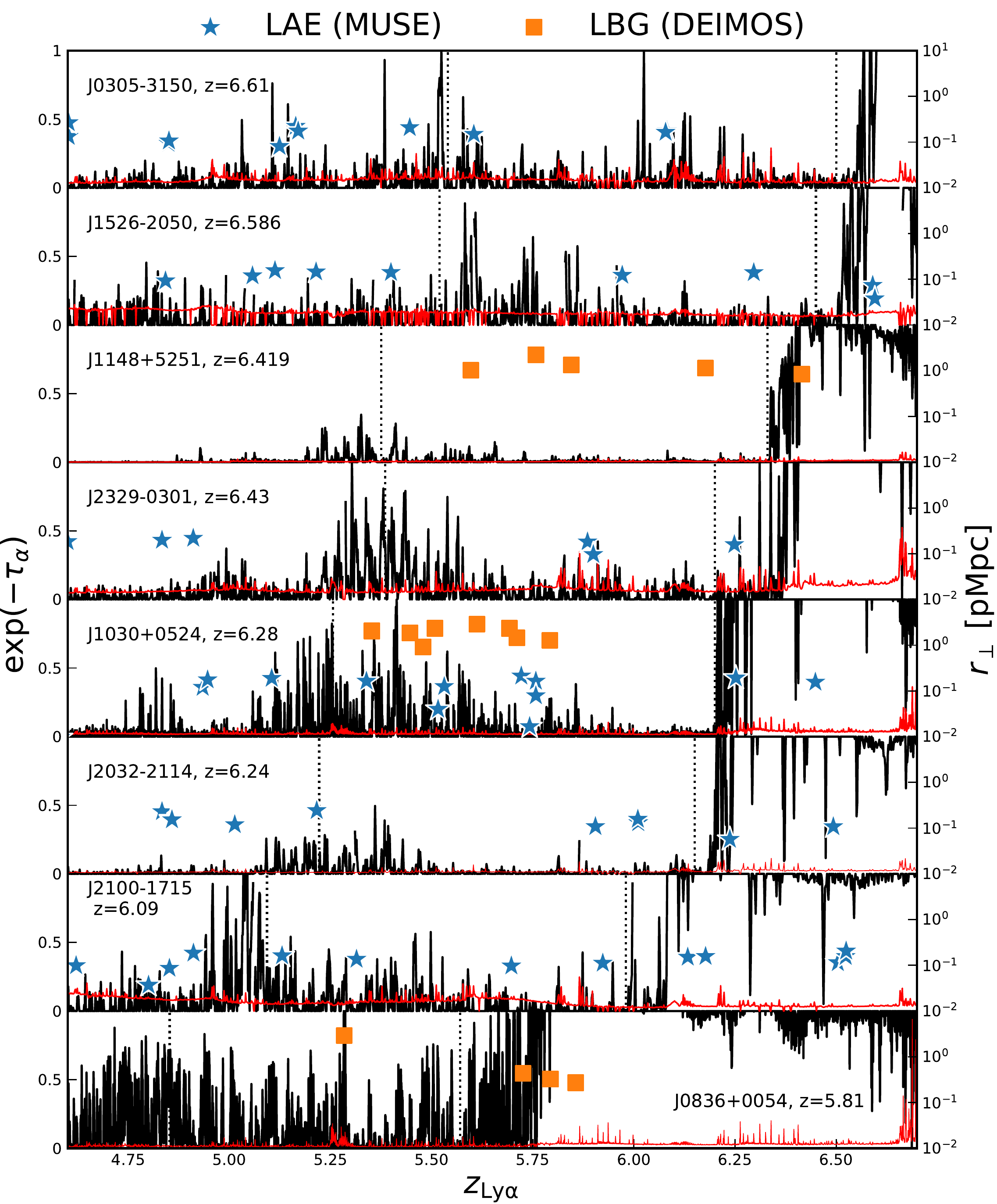}
    \caption{The Lyman-$\alpha$ forests of our high-redshift quasar sample and galaxies detected in the quasar fields. Whilst the average transmission is clearly decreasing with increasing redshift, the galaxies (LAEs in particular) seem to cluster with transmission spikes in some sightlines. The transmission in the quasar Lyman-$\alpha$ forest (black) is indicated on the left axis whilst the right axis indicates the transverse distance of foreground galaxies to the quasar sightline. LBGs  detected with DEIMOS are indicated in orange squares. MUSE LAE detections are indicated with blue stars. Only J1030 displays both LAEs and LBGs as it is the only field with DEIMOS and MUSE data. The vertical black dotted lines indicate the redshift range of the Lyman-$\alpha$ forests as defined in Section \ref{sec:results_data}.}
    \label{fig:forest_det}
\end{figure*}

Galaxies are usually thought to be responsible for reionising the Universe and driving the UVB fluctuations at $z\sim6$. Having gathered a sample of $21$ LAEs and $13$ LBGs in the redshift range of the Lyman-$\alpha$ forest of nearby quasars, we are in a position to investigate the direct impact of galaxies on the surrounding IGM. The observational result of our work is summarised by Fig. \ref{fig:forest_det}, where we overlay the detected LAEs and LBGs with the transmission features found at the same redshift in the Lyman-$\alpha$ forest of the background quasar.

The natural corollary to the large-scale underdensity of galaxies found around highly opaque sightlines \citep{Becker2018,Kashino2019} would be a close association between overdensities of transmission spikes and detected galaxies. We find that LAEs and LBGs are found close to at least one transmission spike in all quasar sightlines, but it is difficult to conclude at first sight whether they trace local spike overdensities. Moreover, this is not a reciprocal relation: some large transmission spikes are not associated with any detected galaxy. Two of our quasars, J1030 and J2032, illustrate this complicated relationship very well. The two sightlines both have a transparent region at $z\sim 5.5$, followed by a relatively opaque one at $z\sim 6$, and a similar MUSE exposure time. The transparent region in J1030 is associated with a large overdensity of LAEs and LBGs. In contrast, it is the few transmission spikes in the high-redshift opaque region in the sightline of J2032 that are associated with detected LAEs, whereas only one LAE is detected in the transparent region at $z\sim5.5$. The detection of LAEs across $5\lesssim z \lesssim 6$ in both quasar fields implies that we do not miss existing LAEs in these sightlines. 

Hence studying the correlations between galaxies and the IGM must been conducted in a more quantitative manner. In the following section, we compute the cross-correlation of the galaxies' positions with the transmission and the position of selected transmission spikes in the Lyman-$\alpha$ forest of the background quasar.

\section{The correlation of galaxies with the surrounding IGM transmission}\label{sec:results_data}

In this section, we introduce two indicators of the link between galaxies and the ionisation state of the surrounding IGM: the cross-correlation of galaxies with the transmitted flux, and the 2-point correlation function (2PCCF) of galaxies with selected transmission spikes. We present the mean transmitted flux around galaxies (the quantity used in \citetalias{Kakiichi2018}) in Appendix C (online material) given that method has been superceded by the transmitted flux cross-correlation. Although the mean transmission around galaxies is the most intuitive measure of an enhanced UVB due to faint LyC leakers, in practice this measurement is dominated by the cosmic variance between sightlines and the redshift evolution of the IGM opacity, as noted in \citetalias{Meyer2019}. For the purpose of these cross-correlation measurements, we only consider the Lyman-$\alpha$ forest between $1045$\,\AA\, (to avoid the intrinsic Lyman-$\beta$/O{~\small VI} quasar emission) and the start of the near-zone of the quasar, and consider only galaxies whose Lyman-$\alpha$ emission would fall in the same observed wavelength range\footnote{In the following, we loosely describe these galaxies as "being in the (redshift range) of the Lyman-$\alpha$ forest of the quasar".}. These boundaries are plotted in dashed black lines in Fig. \ref{fig:forest_det}.

\subsection{The cross-correlation of the IGM transmission with field galaxies}

We first compute the cross-correlation of the IGM transmission with galaxies. We measure the transmission in the Lyman-$\alpha$ forest at a comoving distance $r$ of observed galaxies $(\text{DD})$ and of random mock galaxies $(\text{DR})$. The distance $r=( r_\perp + r_\parallel )^{1/2}$ is computed from the angular diameter distance $r_\perp$ of the galaxy to the quasar sightline and the comoving distance parallel to the quasar line-of-sight $r_\parallel$. For each quasar field, we compute the probability of detecting a LAE at a given redshift in the quasar Lyman-$\alpha$ forest considering the LAE LF \citep{DeLaVieuville2019a,Herenz2019} and the depth of the MUSE data. The redshifts of random galaxies are sampled from this probability distribution and the angular distances from the quasar sightlines are chosen uniformly up to $1'$ to mimic the MUSE FoV. For LBGs we sample the UV LF \citep{Bouwens2015, Finkelstein2015, Bowler2015,Ono2018} at the $2\sigma$ depth of the photometry of our 3 fields with an appropriate k-correction ($2.5(\alpha-1)\log_{10}(1+z),$ with $\alpha=2$) and the angular separation from the quasar is drawn uniformly in the $4'\times16'$ DEIMOS FoV. As noted in Section 2.2, the number of LBGs detected in each field depends primarily on the number of slitmasks observed in each field, rather than the depth of the photometric data used for selection. If we sampled each field down to the $3\sigma$ detection limit in the z band, we would thus predict similar numbers of predicted LBGs per field. However, a random sample created in this way would have a mean number of transmission spikes around the LBGs larger than is actually observed around our spectroscopically confirmed galaxies. Indeed, those observations which targeted higher redshift fields with reduced IGM transmission (e.g. J1030,J1148) have greater spectroscopic coverage (due to the use of more than one DEIMOS mask) than for the lower-redshift field of J0836. We therefore construct random LBGs by  sampling the UV LF to the limiting depth (\textit{z’}) of each field matching the observed redshift distribution, but oversampling by the number of spectroscopic confirmations in each field. This procedure still reproduces the decline of the number of galaxies with redshift in each individual field.

The cross-correlation is then estimated using the standard estimator
\begin{equation}
    \xi^{\exp(-\tau_\alpha)}_{\text{Gal}-{\text{Ly}\alpha}}(r) = \frac{DD(r)}{DR(r)} -1 \text{   .  }
\end{equation}

Normalising the transmitted flux in this way removes the bias introduced by the evolving IGM opacity, and allows us to average sightlines without being biased by the most transparent ones (see Appendix C, online material). 

\begin{figure}
     \centering
    \includegraphics[width=0.48\textwidth]{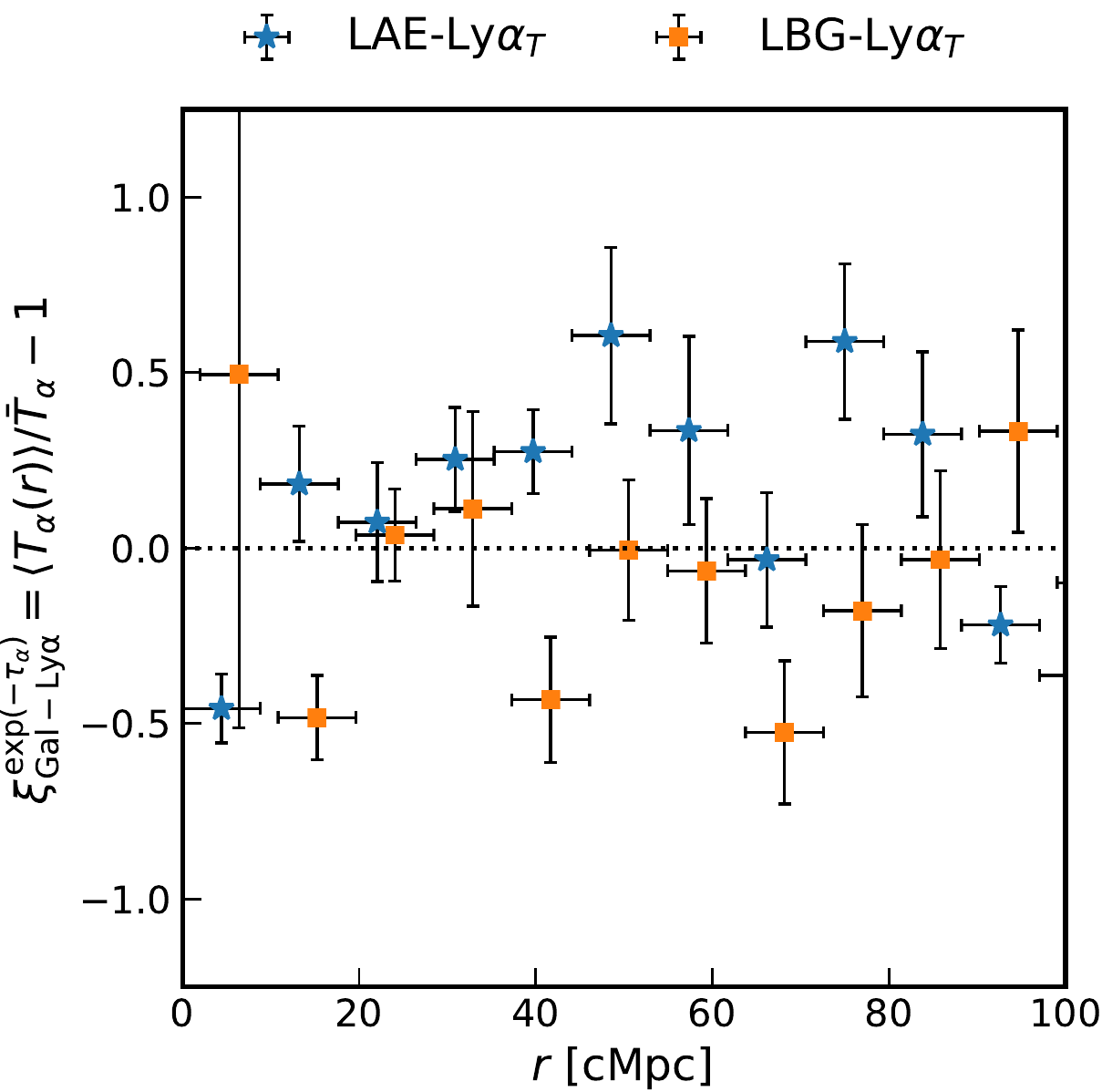}
     \caption{Cross-correlation of the position of LAEs (blue stars) and LBGs (orange squares) with the IGM transmission in the Lyman-$\alpha$ forest of the background quasar. The errorbars are computed by bootstrap resampling of the sample of detected galaxies. The significant deficit of transmission in the first bin of the LAE transmission cross-correlation is consistent with that measured around high-redshift \cfour\, absorbers \citepalias{Meyer2019}.}
     \label{fig:XCorr_LAE_LBG}
\end{figure}

We present the cross-correlation independently for LAEs and LBGs in Fig. \ref{fig:XCorr_LAE_LBG}. We do not find any evidence for an excess transmission, unlike that seen around lower-redshift \cfour\, absorbers in \citetalias{Meyer2019}. The signal is still dominated by the small number of objects and sightlines as the large uncertainties show. The errors are estimated by bootstrapping the sample of detected galaxies, and thus they might be even underestimated given the small sample of sightlines and the large cosmic variance seen between Lyman-$\alpha$ forest at that redshift \citep{Bosman2018}. The issue is potentially more acute for LBGs as the selection is not complete down to a given luminosity as 1) only a fraction of drop-out candidates could be observed per field due to the instrument and time constraints 2) only a fraction of LBGs have a bright Lyman-$\alpha$ line \citep[e.g.][]{Stark2010,Ono2012,DeBarros2017,Mason2019}. Finally, we did not remove completely opaque parts of the Lyman-$\alpha$ where the flux is below the noise level unlike in \citetalias{Meyer2019}, as it would greatly reduce our sample. The measured fluxes are therefore dominated by noise in some sections of the Lyman-$\alpha$ forest, which decreases the signal.
Nonetheless, we find that the absorption on small scales $\lesssim 1\,\text{pMpc}$ around LAEs is similar to that seen for \cfour\, absorbers in \citetalias{Meyer2019}. The direct association or not of \cfour\, with LAEs is outside the scope of this paper and is studied in \citet{Diaz2020}

\subsection{The 2-point correlation of galaxies with selected transmission spikes}
At $z\gtrsim 5.5$, the opacity of the IGM has increased sufficiently that the Lyman-$\alpha$ forest resembles more a ``savannah" than a forest: a barren landscape of opaque troughs occasionally interrupted by transmission spikes. At these redshifts the effective opacity can only be constrained with an upper limit ($\tau_{\rm eff} \gtrsim 3-4$), and the average opacity in large sections of the Lyman-$\alpha$ forest falls below this limit. It is thus not surprising that the previous transmission cross-correlation fails to capture the link between galaxies and the reionising IGM. Indeed, the normalisation term $DR(r)$ is often ill-defined at $z\gtrsim 6$ since the average flux measured is below or at the level of the noise of the spectrograph. To circumvent this issue we examine the extrema of the opacity distribution rather than its mean by utilizing the \textit{2-Point Cross-Correlation Function (2PCCF) of galaxies with selected transmission spikes} in the Lyman-$\alpha$ forest of quasars. We expect the observed Lyman-$\alpha$ transmission to be more sensitive to fluctuations of the extrema of the distribution, making the 2PCCF theoretically more suited to capturing small perturbations due to clustered faint contributors to reionisation.

\begin{figure*}
    \centering
    \includegraphics[width=1\textwidth]{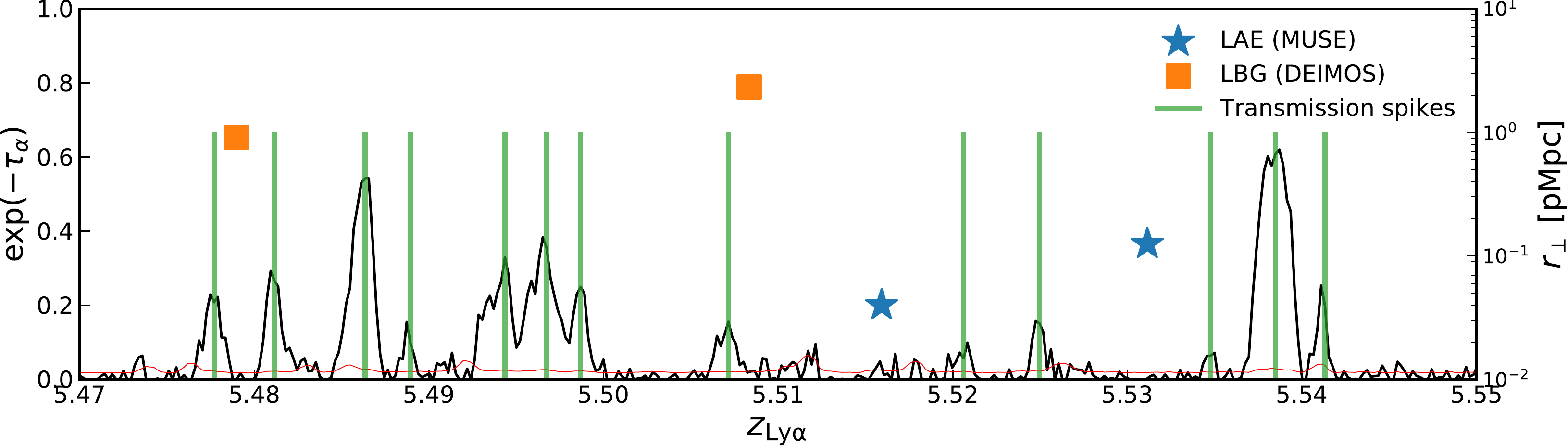}
    \caption{Zoom on the Lyman-$\alpha$ forest of J1030 (Fig. \ref{fig:forest_det}). LAEs and LBGs are indicated with blue stars and orange squares, respectively, and the location of transmission spikes identified with a Gaussian-matched filter with green vertical lines. }
    \label{fig:zoom_spikes}
\end{figure*}

We thus measure the 2-point cross-correlation function (2PCCF) between galaxies and selected transmission spikes in the Lyman-$\alpha$ forest. We identify transmission spikes with a Gaussian matched-filtering technique \citep[e.g.][]{Hewett1985, Bolton2004}. We use Gaussian kernels with $\sigma = [10,14,20] \kms$ to pick individual small spikes and compute the SNR for each kernel at each pixel. We keep the best SNR at each pixel, and we select local peaks at SNR $>3$, with $T>0.02$ (corresponding to $\tau_\alpha \lesssim 4$) as the positions of our transmission spikes. The transmission threshold ($T>0.02$) was chosen to balance recovery of the small transmission spikes in J1148 and J2032 whilst limiting spurious detections in sightlines with worse SNR (J1526, J2100), and the SNR threshold as a compromise between purity of the spike sample and large enough numbers to have reasonable bootstrap error estimates. We present an example of the successful recovery of transmission spikes in the high-redshift Lyman-$\alpha$ forest in Fig. \ref{fig:zoom_spikes}. 

We then estimate the 2PCCF with the estimator
\begin{equation}
    \xi_{\text{Gal}-\text{Ly}\alpha}^{\text{2PCCF}} = \frac{D_G D_\curlywedge (r) }{R_GD_\curlywedge (r)} -1 \text{  ,  }
\end{equation} 
where $D_GD_\curlywedge$ is the number of transmission spikes-galaxy pairs normalised by the number of pixels in each radial bin and $R_G D_\curlywedge$ is the normalised number of transmission spikes - random galaxies pairs, and $r$ the 3D comoving distance. As for the transmission cross-correlation, the redshift of random galaxies are sampled from the LAE or UV LF for LAEs and LBGs, respectively, the angular separation drawn from adequate uniform distributions, and the errors are computed by boostrapping the sample of detected galaxies.

\begin{figure}
    \centering
    \includegraphics[width=0.48\textwidth]{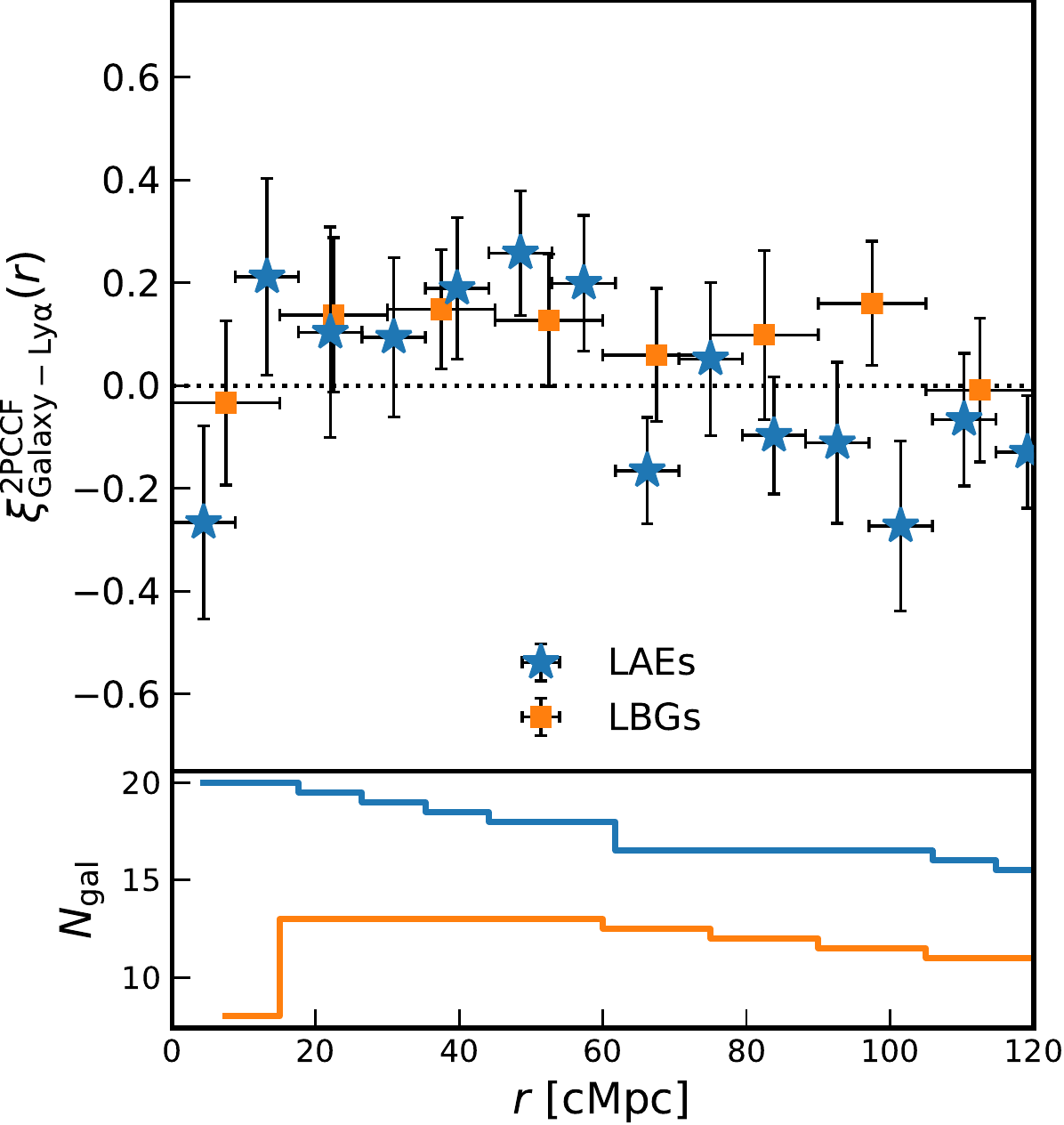}
    \caption{\textbf{Upper panel:} The two-point cross-correlation function of LBGs (orange squares) and LAEs (blue stars) with selected Lyman-$\alpha$ transmission spikes at $z\sim 6$. The errorbars are bootstrap errors on the number of detected galaxies. We find a significant excess of transmission spikes on scales $10-60$ cMpc around LAEs ($3.2 \sigma$). The excess of transmission spikes around LBGs is significant at $1.9 \sigma$, but might extend to larger scales. We point out however that the LBG selection is less complete than the LAEs due to the DEIMOS mask design and that one of the three quasar fields (J0836) has only one detected LBG. On the smaller scales ($\lesssim 1$ pMpc or $\lesssim 7$ cMpc), a deficit of transmission spikes is possibly present. The scales of the excess and the deficit are in good agreement with the measurements of \citetalias{Kakiichi2018} and \citetalias{Meyer2019}. \textbf{Lower panel:} Number of galaxies contributing to the 2PCCF in each radial bin. Note that due to the redshift distribution of galaxies and the limits of the Lyman-$\alpha$ forests, at larger distances some galaxies can only be correlated with transmission spikes at lower or higher redshift. In that case, we count these as contributing $N=0.5$ instead of $N=1$ to the total number of galaxies.}
    \label{fig:2PCCF_LAE}
\end{figure}

We show in Fig. \ref{fig:2PCCF_LAE} the 2PCCFs for both LAEs and LBGs. We detect a positive signal at $3.2 \sigma$ as an excess of transmission spikes on $r\sim 10-60$ cMpc scales around LAEs and a deficit of transmission spikes at $r \lesssim 10$ cMpc. We also find an excess ($1.9 \sigma$) of transmission spikes on large scales around LBGs. The significance of the LAE(LBG) 2PCCF excess is decreased by $-1.5\sigma(-0.2\sigma)$ if the redshift correction is not applied (Section \ref{sec:z_corr}). We compare in Fig. D1 (online) the 2PCCFs with and without correction, with the excesses being reduced in the latter case. The absence of correlation (or even an anti-correlation) on the smaller scales stems both from increased absorption by dense gas around the central LAE (which we model in Section 
\ref{sec:model_pdf_gas}) and the redshift error which dampens the signal ($\sim 200\kms$ corresponding to $\sim 1.8$ cMpc at $z\sim 6$). The reduced significance of the LBG 2PCCF could stem from an inappropriate normalisation due to the difficulty of creating randomly samples of LBGs. As detailed previously, we have conservatively decided to scale the number of random galaxies to that of the observed ones. However if some of the fields \textit{are} indeed slightly overdense, we would be overestimating the normalisation in the cross-correlation and thus decrease the significance of the signal.
The difference in the strength of the signal between the transmission cross-correlation and the 2PCCF can be attributed to the uncertainty in the mean flux at high-redshift. We defer the discussion of this difference to Section \ref{sec:discussion} where we examine the impact of noise on our measurements of the flux transmission and transmission spikes cross-correlation with galaxies. 

\section{Modelling the galaxy-Lyman-$\alpha$ transmission and 2-point cross-correlations }\label{sec:model_extension}

In order to interpret the observed galaxy-Lyman-$\alpha$ forest cross-correlations, we use a radiative transfer model based on the halo occupation distribution (HOD) framework introduced in \citetalias{Kakiichi2018}. Here we summarise the key ingredients and extensions used in this paper.

\citetalias{Kakiichi2018} derived the average \hone\, photoionisation rate at a distance $r$ from a galaxy due to the clustered faint population
\begin{align}
&\langle\Gamma_{\rm HI}^{\mbox{\tiny CL}}(r | M_h, z)\rangle=\frac{\bar{\Gamma}_{\rm HI}}{\lambda_{\rm mfp}(z)}\int \frac{e^{-|\boldsymbol{r}-\boldsymbol{r}'|/\lambda_{\rm mfp}(z)}}{4\pi|\boldsymbol{r}-\boldsymbol{r}'|^2}\left[1+\langle\xi_g(|\boldsymbol{r}'|)\rangle_L\right]{\rm d}^3r',  \nonumber \\
&~~~~~~~~~~=\bar{\Gamma}_{\rm HI}\left[1+\int_0^\infty \frac{k^2{\rm d}k}{2\pi^2} R(k\lambda_{\rm mfp}(z))\langle P_g(k |M_h,z)\rangle_L\frac{\sin kr}{kr}\right],\label{eq:GammaCL}
\end{align}
where $\lambda_{\rm mfp}(z) = 6.0 \left(\frac{1+z}{7} \right)^{-5.4}$ \citep{Worseck2014} is the mean free path of ionising photons and $R(k\lambda_{\rm mfp})=\arctan(k\lambda_{\rm mfp})/(k\lambda_{\rm mfp})$ is the Fourier transform of the radiative transfer kernel $e^{-r/\lambda_{\rm mfp}}/(4\pi r^2\lambda_{\rm mfp})$. The luminosity-weighted galaxy power spectrum is
\begin{equation}
\langle P_g(k|M_h,z)\rangle_L=\frac{\int^\infty_{L_{\rm UV}^{\rm min}} L_{\rm UV}\Phi(L_{\rm UV}|z)P_g(k,L_{\rm UV}|M_h,z){\rm d}L_{\rm UV}}{\int^\infty_{L_{\rm UV}^{\rm min}} L_{\rm UV}\Phi(L_{\rm UV}|z){\rm d}L_{\rm UV}},\label{eq:14}
\end{equation}
where  $P_g(k,L_{\rm UV}|M_h,z)$ is the Fourier transform of the correlation function between bright tracers (i.e. detected LBGs and LAEs) with host-halo mass $>M_h$  and galaxies with luminosity $L_{\rm UV}$.  We assume only central galaxies will be detected as LBGs or LAEs and therefore populate each halo with a HOD using a step function, $\langle N|M_h\rangle=1$ for halo mass $>M_h$ and zero otherwise. Fainter galaxies are populated using the conditional luminosity function pre-constrained by simultaneously fitting the $z\sim6$ UV luminosity function \citep{Bouwens2015, Finkelstein2015, Bowler2015,Ono2018} and the galaxy auto-correlation function \citep{Harikane2016} as in \citetalias{Kakiichi2018}.

\subsection{From the cross-correlation of galaxies with transmitted flux to the 2PCCF}

As in \citetalias{Kakiichi2018} and \citetalias{Meyer2019}, the enhanced UVB can be used to compute the mean Lyman-$\alpha$ forest transmission at a distance $r$ of galaxy,
\begin{align}
&\langle \exp(-\tau_\alpha)(r|M_h,z)\rangle =\nonumber \\  &~~~~~~~\int \exp\left[-\tau_\alpha\left(\Delta_b,\langle\Gamma_{\rm HI}^{\mbox{\tiny CL}}(r|M_h,z)\rangle\right)\right] \times P_V(\Delta_b|r,M_h) \text{d}\Delta_b \text{   ,  } \label{eq:transmitted_flux}
\end{align}
where $P_V(\Delta_b| r, M_h)$ is the volume-averaged PDF of the baryon overdensities $\Delta_b$ at a distance $r$ from our galaxy tracer with a halo of mass $M_h$ at redshift $z$, and $\langle\Gamma_{\rm HI}^{\mbox{\tiny CL}}(r)\rangle$ is the clustering-enhanced photoionisation rate modelled previously. The optical depth $\tau_\alpha$ is derived using the fluctuating Gunn-Peterson approximation \citep[see, e.g.][for a review]{Becker2015},
\begin{equation}
\tau_\alpha\simeq 11 \Delta_b^{2-0.72(\gamma-1)}  \left( \frac{\Gamma_{\rm HI}}{10^{-12} \text{ s}^{-1}}\right)^{-1}  \left( \frac{T_0}{10^4 \text{ K}}\right)^{-0.72} \left( \frac{1+z}{7}\right)^{9/2} \text{   ,  } \label{eq:opt_depth}
\end{equation}
where $\Delta_b$ is the baryon overdensity and $T_0$ is the temperature of the IGM at mean density. We include thermal fluctuations of the IGM using the standard power-law scaling relation \citep{Hui1997,McQuinn2016},
\begin{equation}
T(\Delta_b) = T_0 \Delta_b^{\gamma-1},
\end{equation}
assuming the fiducial values $T_0=10^{4} \text{ K}$ and $\gamma =1.3$.

We now expand this framework to predict a new statistic: the probability of seeing a transmission spike in the Lyman-$\alpha$ forest. Given a transmission threshold over which a detection is considered secure, we can derive an equivalent optical depth threshold. We fix the transmission threshold at $\exp(-\tau_\alpha) \gtrsim 0.02$, corresponding to $\tau_\alpha^{\rm th}\simeq4$, to match our measurement of the 2PCCF. By substituting the predicted clustering-enhanced photoionisation rate and the threshold optical depth in Eq. \ref{eq:opt_depth}, we obtain the maximum baryon underdensity $\Delta_b^{\text{\tiny max}}$ required to produce a detected transmission spike in the Lyman-$\alpha$ forest at a distance $r$ of a tracer galaxy,
\begin{align}
& \Delta_b \leq \Delta_b^{\text{\tiny max}}(\Gamma_{\rm HI}) \simeq \nonumber \\
&~~~~0.57 \left(\frac{\tau^{\rm th}_\alpha}{4} \right)^{0.56}  \left( \frac{\Gamma_{\rm HI}}{10^{-12} \text{ s}^{-1}}\right)^{0.56} \left( \frac{T_0}{10^4 \text{ K}}\right)^{0.4}  \left( \frac{1+z}{7}\right)^{-2.52}   \text{  .}
\end{align}
Thus the occurrence probability of Lyman-$\alpha$ transmission spike at a location with \hone\, photoionisation rate $\Gamma_{\rm HI}$ is given by the probability to reach such an underdensity:
\begin{equation}
P[<\Delta_b^{\text{\tiny max}}(\Gamma_{\rm HI})\, | r,M_h] = \int_0^{\Delta_b^{\text{\tiny max}}(\Gamma_{\rm HI})} P_V(\Delta_b| r, M_h) \text{d}\Delta_b \text{   .  }
\end{equation}
The cross-correlation between galaxies and the Lyman-$\alpha$ transmission spikes can therefore be modelled as the excess occurrence probability, $P[<\Delta_b^{\text{\tiny max}} (\langle \Gamma_{\text{HI}}^{\text{\tiny CL}}(r) \rangle) | r,M_h]$, of transmission spikes around an object with host halo mass $M_h$ and an enhanced photoionisation rate $\langle \Gamma_{\rm HI}^{\text{\tiny CL}}\rangle$ relative to one at mean photoionisation rate $\bar{\Gamma}_{\rm HI}$ and average density fluctuations, i.e. $P[<\Delta_b^{\text{\tiny max}}(\bar{\Gamma}_{\rm HI}) | r\rightarrow\infty,M_h ]$.
It is then straightforward to deduce the cross-correlation between galaxies and the transmitted Ly$\alpha$ spikes as
\begin{equation}
\xi^{\rm 2PCCF}_{\rm Gal-Ly\alpha}(r) = \frac{ P[<\Delta_b^{\text{\tiny max}} (\langle \Gamma_{\text{HI}}^{\text{\tiny CL}}(r) \rangle) | r,M_h] }{ P[<\Delta_b^{\text{\tiny max}}(\bar{\Gamma}_{\rm HI}) | r\rightarrow\infty,M_h ]} - 1    \text{   ,  }
\label{eq:2PCCF_final}
\end{equation}
The advantage of such a statistic over the transmission cross-correlation is that given the large number of pixels in high-resolution spectra of high-redshift quasars, a very low probability of transmission can still be measured with acceptable significance, whereas often only an upper limit on the mean flux can be measured at $z\gtrsim 6$.

\subsection{Extending our UVB model with varying mean free path and gas overdensities}
\label{sec:model_pdf_gas}
We now proceed to extend the model of UVB enhancement due to galaxy clustering by adding a varying mean free path and taking into account the gas overdensities associated with LAEs and LBGs on scales of several cMpc.

We first consider the effect of gas overdensity using the relevant probability distribution function. We derive the conditional PDF of overdensities around suitable haloes $P_V(\Delta_b|r, M_h)$ from the IllustrisTNG simulations \citep{Nelson2018}. We utilise the TNG100-2 simulation for host halo masses $10^{10.5}\rm\,M_\odot < M_h<10^{11.7}\rm\,M_\odot$ whereas for larger host halo masses ($M_h>10^{11.7}\rm\,M_\odot$) we use TNG300-3 in order to get higher number of such halos at the cost of larger gas and dark matter particle masses. We present in Appendix D (online material) the extracted conditional PDFs for a range of halo masses and radii.

Following \citet{Miralda-Escude2000,Pawlik2009} we then fit each conditional PDF with a parameterisation of the form
\begin{align}
   &  P_V(\Delta_b|r,M_h) \text{d}\Delta_b  = \nonumber \\
   & ~~~ A(r,M_h) \exp{\left[ - \frac{(\Delta_b^{-2/3} - C_0(r,M_h))^2}{2(2\delta_0(r,M_h) / 3)^2}   \right]} \Delta_b^{-\beta(r,M_h)} \text{d}\Delta_b  \text{    ,   }
    \label{eq:analytical_form_PDF}
\end{align}
with the parameter $A(r,M_h)$ being determined by requiring that the integral of the PDF is unity ($\int P_V (\Delta_b|r,M_h)\text{d}\Delta_b$ = 1).
The fitted values of $A,\,C_0,\,\delta_0$ and $\beta$ are listed in Table D1 (online material) for a choice of the relevant range of $(r,M_h)$ whilst the rest are given in Appendix D (online material). We show in Fig. \ref{fig:fit_pdf_illustrisTNG} the good agreement of the fits with the simulated PDF in a snapshot at $z=5.85$ and  a chosen central halo mass $M_h \sim 10^{11.2\pm 0.1}\,\rm M_{\odot}$ corresponding to the one derived from the clustering of LAEs \citep{Ouchi2018}.

\citetalias[][]{Kakiichi2018} considered a constant mean free path for simplicity. Introducing a full self-consistency of the mean free path down to ckpc scales in Eq. \ref{eq:GammaCL} is the realm of numerical simulations if a real distribution of gas and discrete sources is to be considered (and not the average distribution we use here). However we can approximate variations of the mean free path to first order. Following \citet{Miralda-Escude2000, McQuinn2011, Davies2016, Chardin2017}, the mean free path of ionising photons is dependent on the photoionisation rate and the mean density of hydrogen,
\begin{equation}
\lambda_{\text{mfp}}(r) = \lambda_0 \left( \frac{\langle\Gamma_{\rm HI}^{\mbox{\tiny CL}}(r)\rangle}{\bar{\Gamma}_{\rm HI}} \right)^{\beta_{\text{mfp}}}\left[\int \Delta_b P_V(\Delta_b|r,M_h){\rm d}\Delta_b\right]^{-\gamma_{\text{mfp}}}  \text{    ,   }
\end{equation}
where $\beta_{\text{mfp}}$ and $\gamma_{\text{mfp}}$ reflects a simple parameterisation of the mean free path dependence on the local UVB and gas overdensity. In this work, we chose to use the values $\beta_{\text{mfp}} = 2/3$, $\gamma_{\text{mfp}} =-1$ for our fiducial model following previous works cited above.

\begin{figure}
    \centering
    \includegraphics[width = \columnwidth]{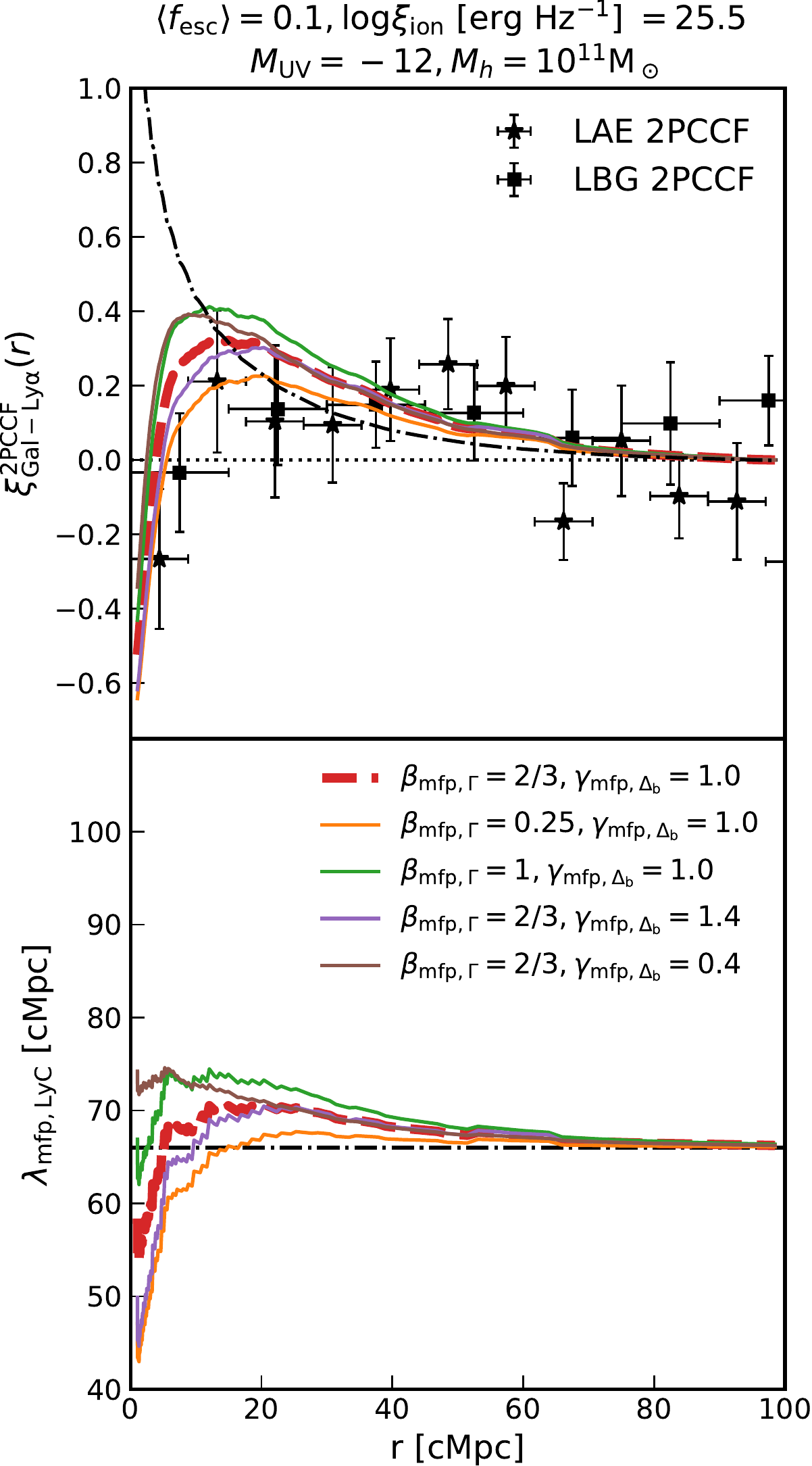}
   \caption{The impact of a spatially varying mean free path on the modelled  2PCCF of galaxies with transmission spikes in the Lyman-$\alpha$ forest. Variation of either the mean free path power-law dependence on the photoionisation rate ($\beta_{\rm mfp}$) or the gas overdensity ($\gamma_{\rm mfp}$) do not affect significantly the predicted 2PCCF. The models are generated with the fiducial parameters, $\langle f_{\text{esc}}\rangle = 0.1$, $M_h = 10^{11}\rm\,M_{\odot}$, $\log \xi_{\text{ion}}/[\text{erg}^{-1} \text{Hz}] =25.5$, and $\gamma_{\rm mfp} = 1.3$. The black dashed-dotted line in the upper panel show a model with a fixed mean free path. }
    \label{fig:examples_models_mfp}
\end{figure}

Given the mutual dependence between $\langle\Gamma_{\rm HI}^{\mbox{\tiny CL}}(r)\rangle$ and $\lambda_{\text{mfp}} (r)$, we iterate the computation until $\langle\Gamma_{\rm HI}^{\mbox{\tiny CL}}(r)\rangle$ is converged at the $1\,\%$ level at every distance $r$. As expected, a varying mean free path does not affect the photoionisation rate on large scales but decreases it by a factor 2-3 on scales $\lesssim 1 \text{ cMpc}$. We show the impact on the predicted 2PCCF in Fig. \ref{fig:examples_models_mfp}. We find that any reasonable choice of $(\beta_{\text{mfp}},\gamma_{\text{mfp}})$ modifies the 2PCCF only by a factor $<2$ on scales $r<10$ cMpc.

\begin{figure}
    \includegraphics[width=0.48\textwidth]{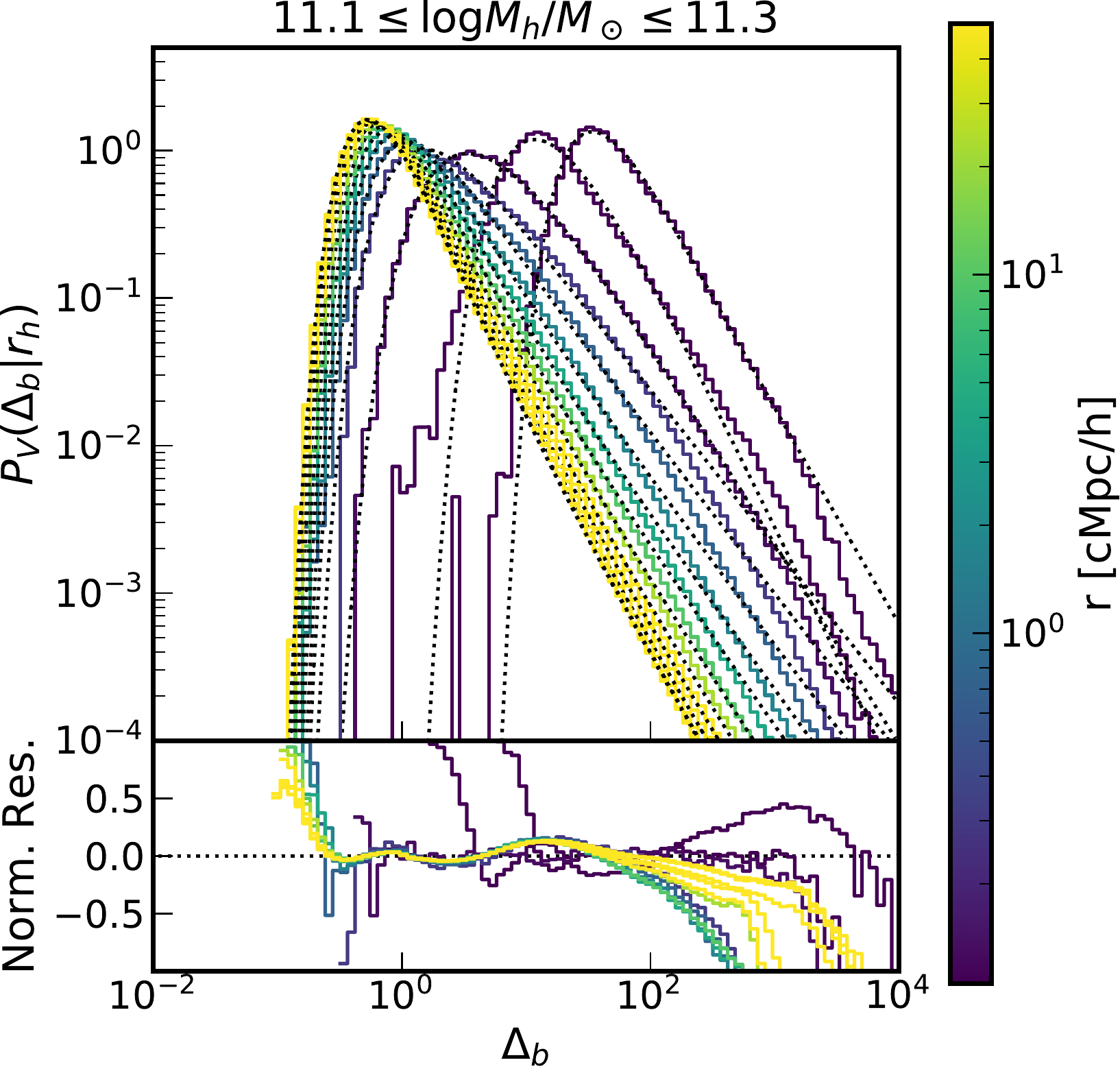}
    \caption{  \textbf{Upper panel:} A subsample of the conditional PDFs from $r=10^{-1}$ cMpc/h to $r=10^{1.5}$ cMpc/h in increments of $0.045$ dex. The fits with Eq. \ref{eq:analytical_form_PDF} are overlaid (dotted black) on top of the PDF extracted  from the IllustrisTNG 100-2 simulation box in a snapshot at $z\sim 5.85$ (coloured histograms). \textbf{Lower panel:} Residuals of the PDF fit, coloured by distance from the centre of the halo, showing good agreement in the validity limits of the prescribed analytical form between $10^{-1} \leq \Delta_b \leq 10^{2}$.}
    \label{fig:fit_pdf_illustrisTNG}
\end{figure}

\subsection{The observed 2PCCF}
\begin{figure}
    \centering
    \includegraphics[width = 0.48
    \textwidth]{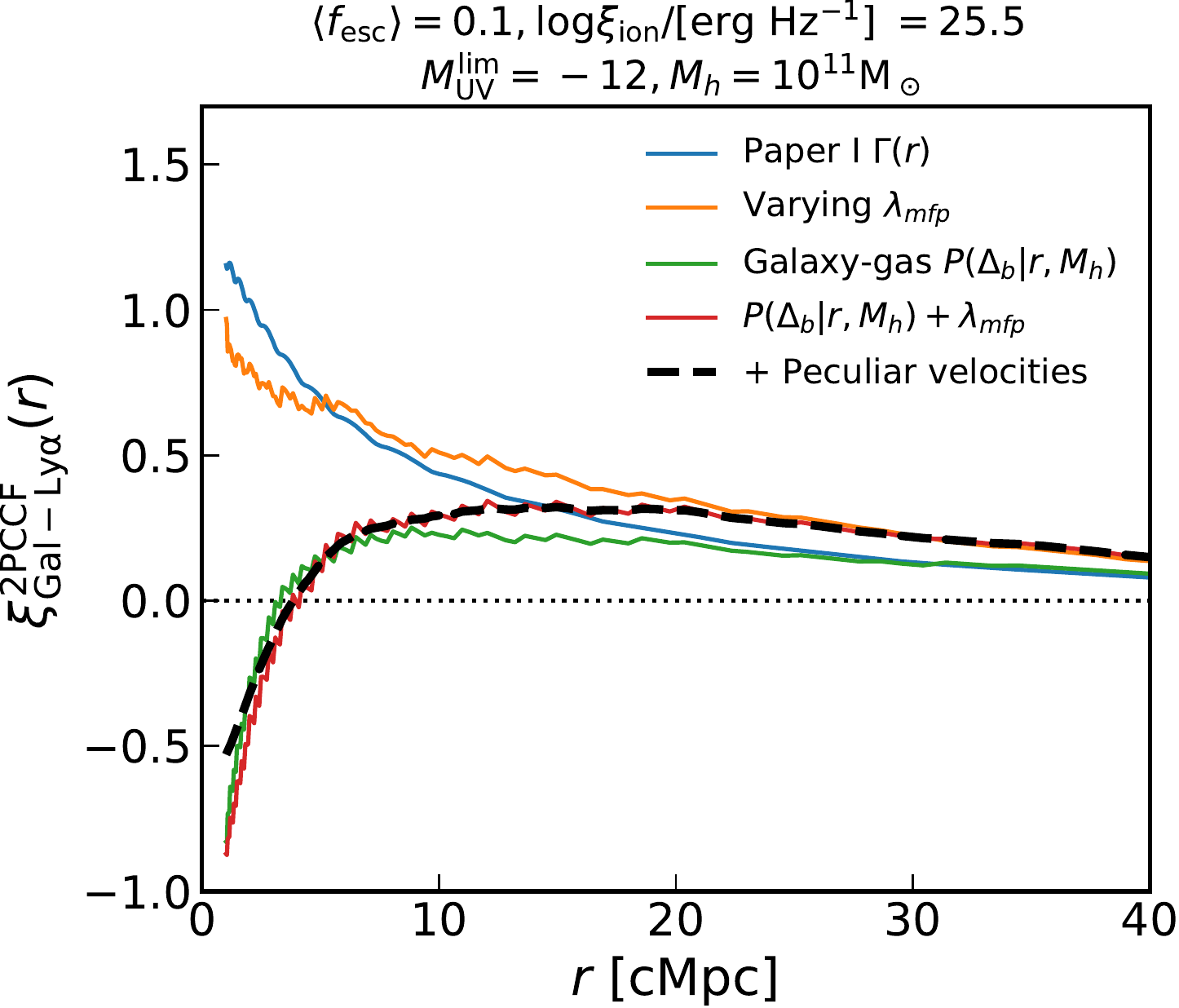}
    \caption{Successive versions of our model showcasing the increasing levels of refinement detailed throughout Section \ref{sec:model_extension}. The resulting predicted 2PCCF is mainly sensitive to the original UVB computed as in \citetalias{Kakiichi2018}, but is dampened on small scales by the addition of a realistic gas overdensity PDF. The implementation of a variable mean free path enhances the signal on large scales.}
    \label{fig:model_ingredients}
\end{figure}

We have so far only considered the cross-correlation in real space. However, the observed two-point correlation is distorted by peculiar velocities and infall velocities. We consider here only the impact of random velocities and redshift errors. Following \citet{Hawkins2003, Bielby2016}, the real-space 2D correlation $\xi'(\sigma, \pi)$ is convolved with a distribution of peculiar velocities along the line of sight direction ($\pi$),
\begin{equation}
    \xi(\sigma, \pi) = \int_{-\infty}^{+\infty} \xi'(\sigma, \pi - v/H(z))f(v){\rm d}v \text{    ,   }
\end{equation}
with an Gaussian kernel for the velocity distributions $f(v)=(2\pi\sigma_v^2)^{-1}\exp\left(-\frac{v^2}{2\sigma_v^2}\right) $. We use $\sigma_v=200 \kms$, which is the observed scatter in the difference between Lyman-$\alpha$ and systemic redshifts at $z\sim 2-3$ \citep{Steidel2010}, encapsulating both redshift errors and the random velocities of galaxies. We finally take the monopole of the 2D correlation function,
\begin{equation}
    \xi_0(s) = \frac{1}{2} \int_{-1}^{-1} \xi(\sigma,\pi) P_0(\mu) \rm d\mu  \text{    ,   }
\end{equation}
where $s=\sqrt{\sigma^2+\pi^2}$, $\mu =\pi/s$,  and $P_0(\mu)=1$ is the zeroth order Legendre polynomial. As shown in Fig. \ref{fig:model_ingredients}, the peculiar velocities reduce slightly the signal on small scales.

We show in Fig. \ref{fig:model_ingredients} various realisations of our model of the 2PCCF. We present here the impact of the modelling improvements that we described previously. The addition of gas overdensities decreases the correlation on the smallest scales ($r\lesssim 20$\,cMpc). The varying mean free path has little impact on the final shape of the predicted two-point correlation function, but boosts it slightly at $r>20$ cMpc. Finally, the redshift errors and random velocities have a negligible impact on scales larger than few cMpc.

\begin{figure*}
    \centering
    \includegraphics[width = \textwidth]{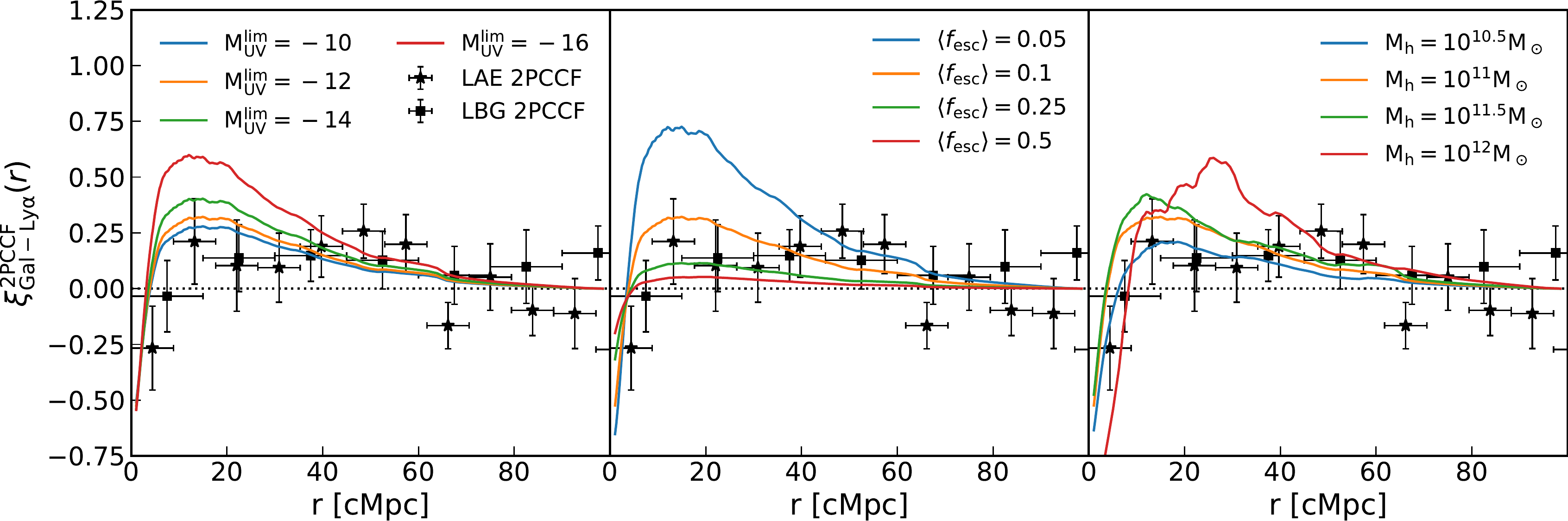}
    \caption{Examples of our model of the 2PCCF given a range of parameters (limiting luminosity of contributors to reionisation  $M_{\rm UV}^{\text{lim}}$, escape fraction of LyC photons $f_{\text{esc}}$ and host halo mass of the detected LAE/LBG $M_h$). In each panel, one parameter is varying while the others are kept fixed at the fiducial values of $M_{\rm UV}^{\text{lim}} = -12$, $\langle f_{\text{esc}} \rangle= 0.1$, $M_h = 10^{11} \rm M_{\odot}$. All models plotted here assume a redshift of $z=5.85$ and ionising efficiency $\log \xi_{\text{ion}}/[\text{erg}^{-1} \text{Hz}] =25.5  $, and a mean free path dependence on the mean overdensity with exponent $\gamma_{\rm mfp} = -1$.  }
    \label{fig:examples_models}
\end{figure*}

We conclude this modelling section by comparing the 2PCCF to the data for various fiducial parameters of the limiting luminosity of contributors to reionisation  $M_{\rm UV}^{\text{lim}}$, escape fraction of the LyC photons $\langle f_{\text{esc}}\rangle$ and host halo mass $M_h$ of the detected bright galaxy in Fig. \ref{fig:examples_models}. We adopt a fiducial $\log \xi_{\text{ion}}/[\text{erg}^{-1} \text{Hz}] =25.5 $, $\beta_{\rm mfp} = 2/3$, $\gamma_{\rm mfp} = -1$. Increasing the minimum UV luminosity increases the correlation as more sources contribute to the local photoionisation rate. We find that the host halo mass of the tracer galaxy is correlated positively with the 2PCCF signal strength, as they cluster more strongly with other galaxies. Finally, the escape fraction has a non-trivial effect on the cross-correlation: because it affects both the local and the overall UVB, an increase in the escape fraction decreases the total 2PCCF. Indeed, excesses of ionised gas close to clustered galaxies become harder to detect as the Universe becomes fully ionised and transmission spikes are ubiquitous. We defer to Appendix E (online material) for a full mathematical derivation of the role of $f_{\rm esc}$ in our modelled 2PCCF.


\section{Constraints on the ionising capabilities of $z\sim 6$ contributors clustered around LAEs }\label{sec:results_final}

Our model of the statistical proximity effect of galaxies based on their correlation with Lyman-$\alpha$ transmission spikes can be applied at different redshifts, across absorbed and transparent sightlines, and to galaxy populations with different halo masses. We have detected a signal in the 2PCCF of high-redshift LAEs and LBGs with Lyman-$\alpha$ transmission spikes, which we will now proceed to fit.

The median redshift of our LAE (LBG) sample is $\langle z \rangle = 5.82 (5.597)$. We therefore use the gas overdensity PDF from the Illustris TGN100-2 (TNG300-3 for LBGs) at $z= 5.85$ (the closest snapshot in redshift to the larger LAE sample), and we fix the redshift at the same value for the computation of the CLF and our 2PCCF model in general for consistency. We use the fiducial values of $\beta_{\rm mfp} = 2/3, \gamma_{\rm mfp} = -1$ for the mean free path of ionising photons, and a temperature density relation $T\propto \Delta_b^{\gamma -1}$ with $\gamma=1.3$.

Our model constrains the number of ionising photons emitted around galaxies, and thus the \textit{luminosity-weighted-average contribution}\footnote{Which we shorten to \textit{luminosity-averaged} for convenience.} of sources to reionisation
\begin{equation}
    \langle f_{\text{esc}} \xi_{\text{ion}} \rangle_L = \frac{\int^{\infty}_{M_{\text{UV}}^{\text{lim}}} f_{\text{esc}}(L_{\text{UV}}) \xi_{\text{ion}}(L_{\text{UV}})L_{\text{UV}} \Phi(L_{\text{UV}}) \rm dL_{\text{UV}}}{\int^{\infty}_{M_{\text{UV}}^{\text{lim}}} L_{\text{UV}} \Phi(L_{\text{UV}}) \rm dL_{\text{UV}}}  \text{    ,   }
\end{equation}
which for simplicity we have recast with a fixed $\log \xi_{\text{ion}} / [\rm erg^{-1} Hz] = 25.5$, such that our main results will the \textit{luminosity-averaged escape fraction}. We emphasize that the limiting luminosity of contributing sources simply marks the truncation of the UV LF. A Gaussian prior on the turnover of the UV LF at $M_{\text{UV}}^{\text{lim}} \sim -12 \pm 1$ encompasses the scatter between different studies  \citep[e.g.][]{Livermore2017,Bouwens2015, Atek2018} and the recent constraint via the extragalactic background light measurement \citep{Fermi-LATCollaboration2018}.

We fit the LAE 2PCCF signal with the \textsc{emcee} Monte Carlo sampler \citep{Foreman-Mackey2013} using a flat prior in the range $0\leq \langle f_{\text{esc}} \rangle \leq 1$, a Gaussian prior over $M_{\text{UV}}^{\text{lim}} \simeq -12$ with $\sigma_{M_{\rm {UV}}} = 1$, and another Gaussian prior for the host halo masses based on the angular clustering measurements of LAEs \citep{Ouchi2018}. We use the values of $\log M_h^{\text{LAE}}/[\rm M_\odot] =  11.1^{+0.2}_{-0.4}$  derived at $z=5.7$ for all our LAE detections at $5.5<z<6.2$. We marginalise over LAE host mass and minimum luminosity priors get our final constraint from the LAE-spike 2PCCF
\begin{equation}
    \langle f_{\text{esc}} \rangle_{M_{\rm UV}\lesssim-12} = 0.14_{-0.05}^{+0.28} \,\,\,\,\,\,\,\,  \,\,\,\,\,\,\,\,\, (\log \xi_{\text{ion}}/[\text{erg}^{-1} \text{Hz}] =25.5 ) \text{ , }
\end{equation}
where the errors represent a $1\sigma$ credible interval. The LBG-spike 2PCCF, where the host halo mass prior of LBGs at $z\sim 6$ ($M_h^{\text{LBG}}/[\rm M_\odot] = 12.02^{+0.02}_{-0.01}$) is based on the clustering measurement with Hyper Suprime-Cam (HSC) at the Subaru telescope  \citep{Harikane2018}, gives the following constraint
\begin{equation}
    \langle f_{\text{esc}} \rangle_{M_{\rm UV}\lesssim-12} = 0.23_{-0.12}^{+0.46}  \,\,\,\,\,\,\,\,  \,\,\,\,\,\,\,\,\, (\log \xi_{\text{ion}}/[\text{erg}^{-1} \text{Hz}] =25.5 ) \text{ . }
\end{equation}

\begin{figure}
    \centering
    \includegraphics[width = 0.48\textwidth]{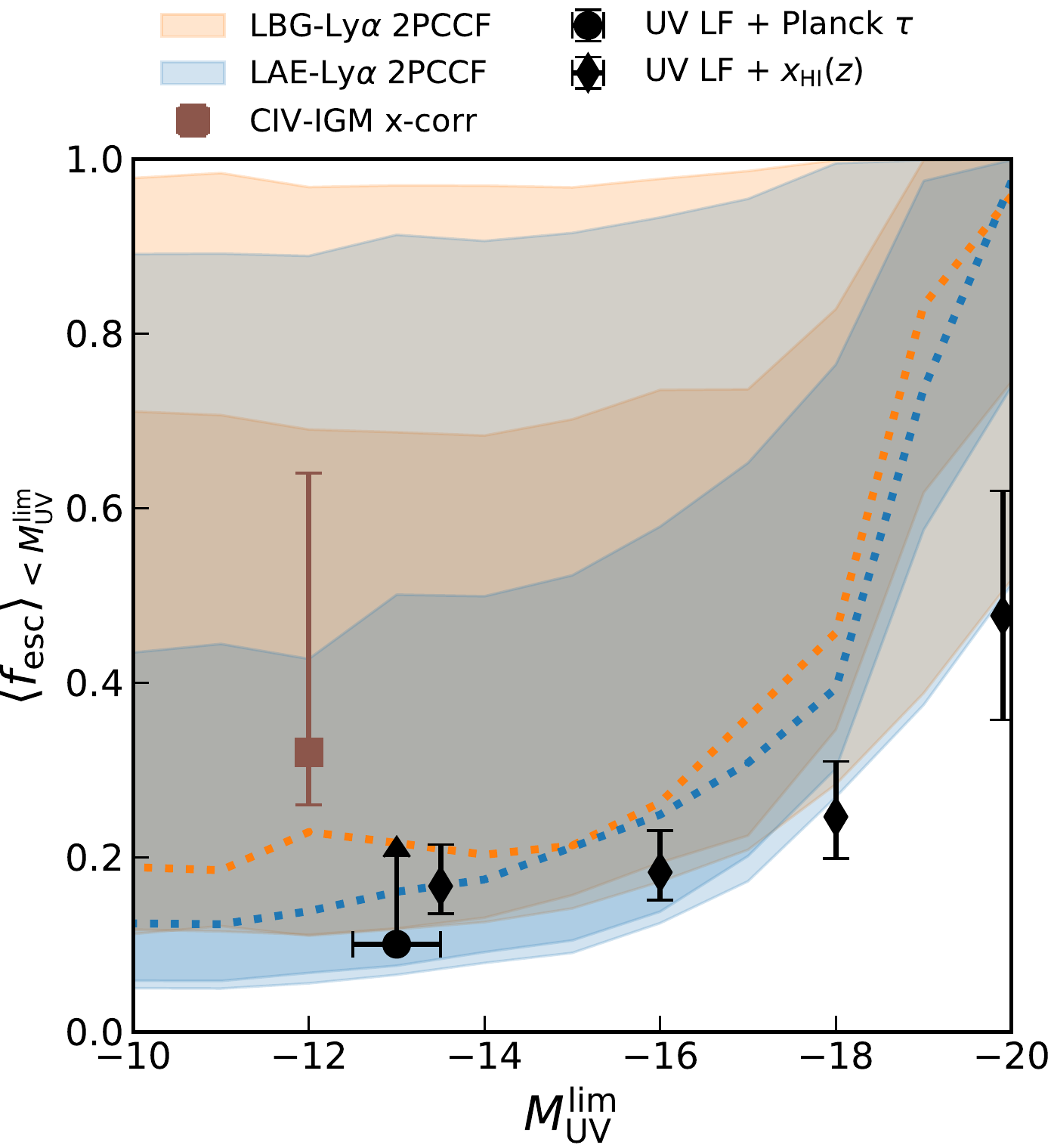}
    \caption{Luminosity-averaged escape fraction of galaxies contributing to reionisation as a function of their minimum UV luminosity. The two measurements of the 2PCCF from this work (LAE/LBG $1\sigma,2\sigma$ blue/orange contours) are in agreement with the \cfour-IGM transmission cross-correlation  from \citetalias[][(brown square)]{Meyer2019}. The irregular shape of the LBG 2PCCF posterior is due to the gas overdensities PDF which is poorly constrained by Illustris TNG due to the large mass and limited volume, and thus varies quickly from one mass bin to another, making the convergence of the MCMC chain difficult. The LAE-IGM 2PCCF is in better agreement with the average escape fraction derived from the UV LF and the Planck optical depth measurement \citep[][black circle]{Robertson2015} or the neutral fraction history when the minimum UV luminosity of contributors is small \citep[][black diamonds]{Naidu2020}. The escape fractions are (re-)derived assuming $\log \xi_{\text{ion}}/[\text{erg}^{-1} \text{Hz}] =25.5 $.}
    \label{fig:fesc_Muv_plot}
\end{figure}

These average constraints on the entire luminosity range can be of course rearranged to test any given functional form of the escape fraction and/or the ionising efficiencies, and accommodate other fiducial values of $\xi_{\text{ion}}$. For example, we present in Fig. \ref{fig:fesc_Muv_plot} the average escape fraction of galaxies as a function of the minimum UV luminosity of contributors between $-20 <M_{\rm UV}^{\rm min} < -10$. Our results are in good agreement with literature estimates derived from neutral fraction histories \citep[e.g.][]{Robertson2015,Naidu2020}, especially for models invoking a substantive contribution of faint galaxies to reionisation. Both LAE-IGM and LBG-IGM 2PCCF constraints are in agreement with the \cfour-IGM transmission cross-correlation of \citetalias{Meyer2019}. Although the three measurements' maximum likelihood value differ, the uncertainties are still too large to conclude yet on any significant tension between the escape fraction of the galaxies traced by LAEs, LBGs and \cfour\, absorbers.

\section{Discussion}\label{sec:discussion}
\subsection{Relative contribution of sub-luminous sources}
As the cross-correlation slope is sensitive to the minimum UV luminosity of contributing sources (Fig. \ref{fig:examples_models}), it is theoretically possible to measure simultaneously the luminosity-averaged escape fraction of reionisation sources and their minimum or maximum luminosity. We now proceed to extend our analysis to test whether we can infer the relative contribution of bright and faint sources to reionisation. We examine two simple cases: a model where all galaxies fainter than a certain UV luminosity solely contribute to reionisation and, conversely, a model where such faint galaxies do not contribute at all. To do so, we treat the minimum/maximum UV luminosity as a parameter and fit the model with a flat prior on this quantity. We then fit these two models to the LAE/LBG-IGM 2PCCF. 

\begin{figure}
    \centering
    \includegraphics[width=0.45\textwidth]{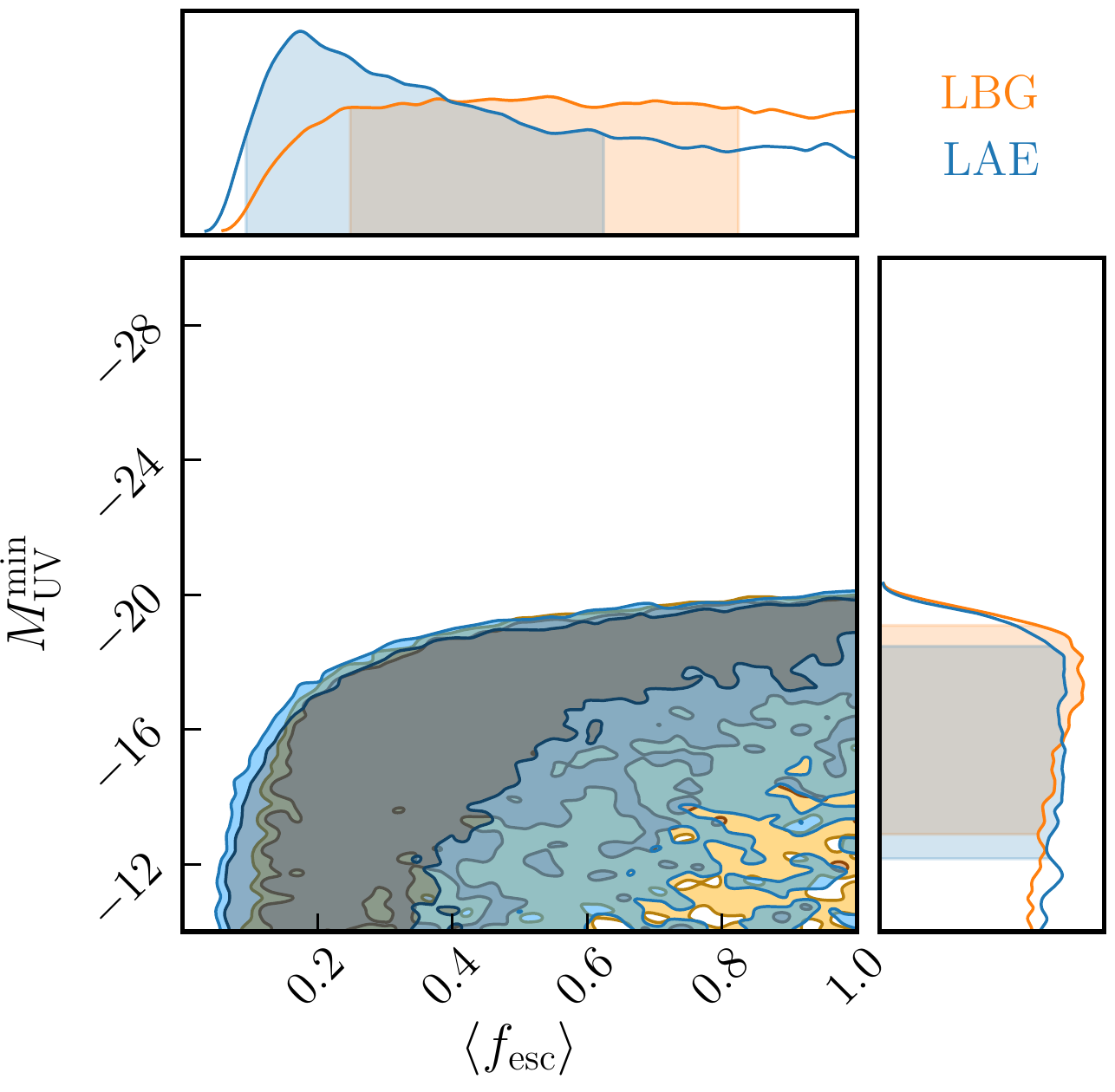}
    \caption{Posterior distributions of our fiducial model parameters with reionisation dominated by more luminous galaxies fitted to the LAE/LBG-transmission spikes 2PCCF. The 2PCCF constrain the minimum UV luminosity of contributors to be at least $M_{\rm UV} \lesssim -20$. Our fiducial models have the following fiducial parameters $\gamma=1.3,\beta_{\text{mfp}}=2/3, \gamma_{\text{mfp}} = -1$ and $\log \xi_{\text{ion}}/[\text{erg}^{-1} \text{Hz}] =25.5 $.    } \label{fig:posterior_bright}
\end{figure}

We present the posterior distribution of our parameters in Fig. \ref{fig:posterior_bright} for the model where bright galaxies dominate, and the inferred constraints in Table \ref{tab:results_fesc_summary}. The LAE/LBG 2PCCF were fitted with the parameters described above except for a flat prior on the minimum UV luminosity of contributing sources, $-10<M_{\rm UV}^{\rm min} < -30$. In both cases, the minimum UV luminosity of the contributing sources is $M_{\rm UV}^{\rm min}<-20.0 \,(2\sigma)$. In practice however, a model where \textit{only} galaxies brighter than $M_{\rm UV}=-20.0$ is implausible because it would require an overhelmingly high luminosity-averaged escape fraction of $\approx 1$, contradicting existing $z\sim 6$ measurements \citep{Matthee2018} and marking a stark departure from measurements at lower-redshift \citep[e.g.][]{Vanzella2016,Izotov2016,Izotov2018,Tanvir2019,Fletcher2019}. It thus more likely that, \textit{if only the brightest objects contribute}, they include at least relatively faint galaxies down to $M_{\rm UV} \sim -18(-16)$.

\begin{figure}
    \centering
    \includegraphics[width=0.45\textwidth]{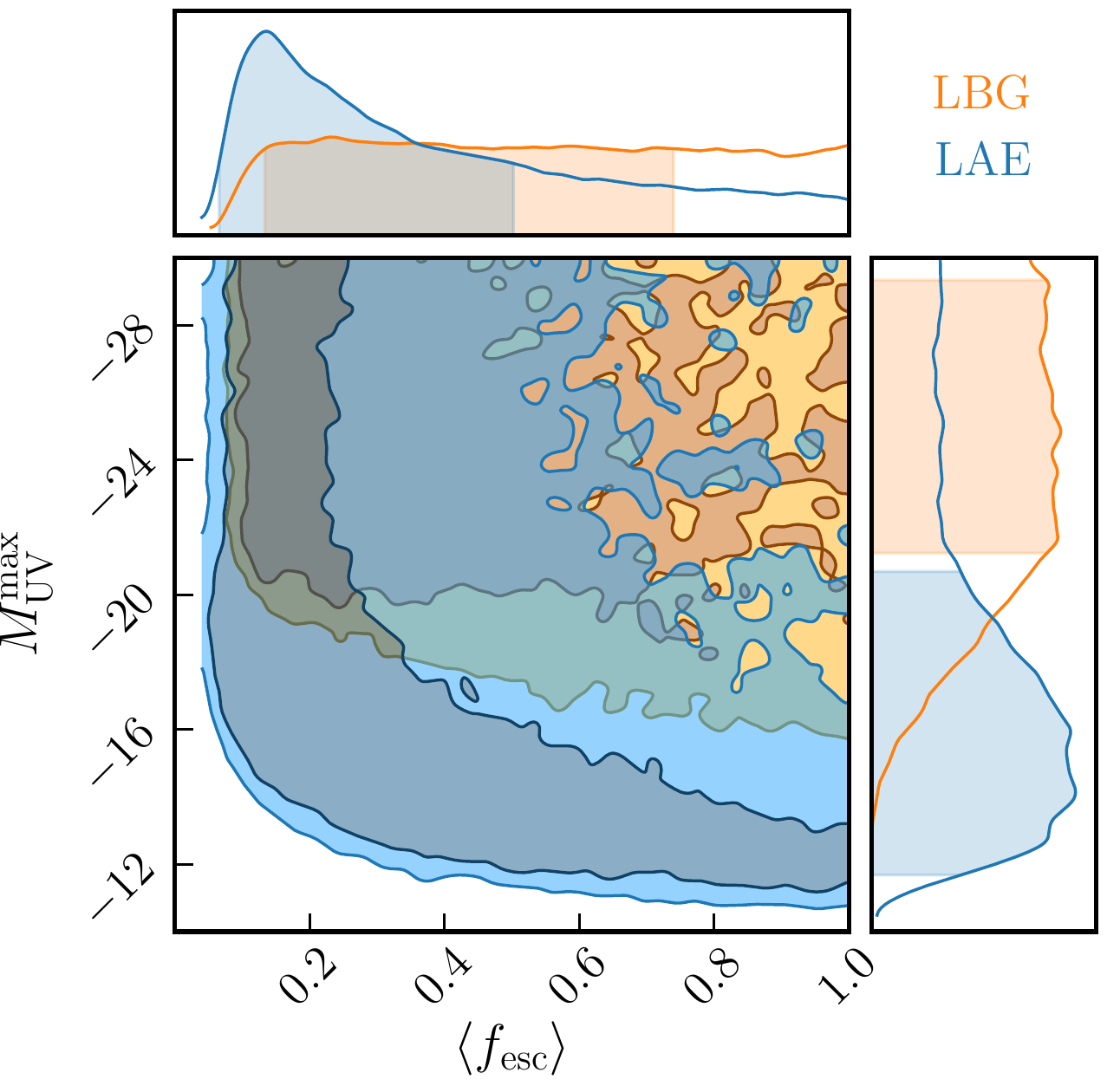}
    \caption{Posterior distributions of our fiducial model parameters where faint galaxies dominate, fitted to the LAE/LBG-transmission spikes 2PCCF. The LAE and LBG constraints are in tension, with the LAE 2PCCF favouring a model where only low luminosity galaxies contribute ($-10 \gtrsim M_{\rm UV} \gtrsim -17$) and the LBG 2PCCF requiring the contribution of more luminous objects up to at least $M_{\rm UV} \lesssim-21$. Our fiducial model has the following IGM parameters: $\gamma=1.3,\beta_{\text{mfp}}=2/3, \gamma_{\text{mfp}} = -1$ and $\log \xi_{\text{ion}}/[\text{erg}^{-1} \text{Hz}] =25.5 $. 
    }  \label{fig:posterior_faint}
\end{figure}

We now examine our model where faint galaxies dominate. The LAE/LBG 2PCCF were fitted with the parameters described in Section \ref{sec:results_final} except for a flat prior on the maximum UV luminosity of contributing sources $-10<M_{\rm UV}^{\rm max} <-30$, and the minimum UV luminosity of LyC contributing sources was fixed at $M_{\rm UV}^{\rm min}=-10$. We present the posterior distribution of our parameters in Fig. \ref{fig:posterior_faint}, and the inferred constraints in Table \ref{tab:results_fesc_summary}. The posteriors for the LAE and LBG 2PCCF are strikingly different: whereas the LAE signal is well fitted by a model where faint galaxies ($-17 \lesssim M_{UV} \lesssim -10$) drive reionisation, the LBG 2PCCF constrains the maximum luminosity of contributors to be at least $<-18.4 \,(2\sigma)$. In other words, the 2PCCF signal around more luminous tracers (LBGs) of galaxies is consistent with a contribution of brighter objects, whereas faint tracers (LAEs) favour an ionising environment dominated by faint sources. Because clustering is already included in our model, this is not simply a consequence of LAEs likely sitting in smaller overdensities than LBGs, therefore tracing less massive and fainter objects. This result rather indicates that bright objects ($M_{\rm} \lesssim -20$) traced by LBGs have increased ionising efficiencies. One natural explanation is that they would create early ionised bubbles which would in turn enhance the confirmation rate \textit{with a Lyman-$\alpha$ emission line detection} of such LBG candidates. This results is in agreement with the study by \citet{Mason2018} which found that the boosted transmission around bright ($M_{\rm UV} < -22$) objects cannot only be explained by their biased environment, and thus they must have increased ionising efficiencies.  

As a conclusion, it is interesting to note that the 2PCCFs can be fitted with two mutually exclusive populations of galaxies:  the sources or reionisation can either be galaxies fainter or brighter than $M_{\rm UV}\sim-18$ . These two results show how the 2PCCF is able to constrain the parameters of a \textit{given specific model of escape fraction dependence on luminosity}. However, identifying which model is correct is difficult with the current data as the likelihood ratio of the two best (LAE 2PCCF) fits favours only very marginally ($1.3\sigma$) the faint-dominated scenario. Future measurements of galaxy-Lyman-$\alpha$ forest cross-correlations are required to distinguish between the two scenarios, as well the possible differences in the sub-populations traced by LAEs, LBGs and other potential overdensity tracers. 

\begin{table}
    \centering
    \renewcommand{\arraystretch}{1.2}
    \begin{tabular}{c| c c c}
         Tracer & $\langle f_{\rm esc} \rangle $ & $M_{\rm UV}^{\rm min}$ & $M_{\rm  UV}^{\rm max}$ \\ \hline \hline 
         & $0.14_{-0.05}^{+0.28}$ & $-12.1^{+1.1}_{-0.9}$ & $-30$ \\
         LAE & $0.18_{-0.09}^{+0.44}$ & $>-19.0 \, (2\sigma)$ & $-30$ \\
         & $0.14_{-0.07}^{+0.36}$ &  $-10$ &$-14.1_{-6.6}^{+2.4}$  \\\hline
         & $0.23_{-0.12}^{+0.46}$ & $-12.0^{+0.9}_{-1.0}$ & $-30$ \\
         LBG  & $>0.17 \,(2\sigma)$ & $>-19.1 \,(2\sigma)$ & $-30$\\ 
         & $>0.14 \,(2\sigma)$ & $-10$ & $<-17.3 \,(2\sigma)$ \\
    \end{tabular}
    \caption{Summary of our constraints on the luminosity-weighted average escape fraction of galaxies at $z\sim 6$. For each galaxy overdensity tracer (LAEs or LBGs), we fit the galaxy-Lyman-$\alpha$ transmission spike 2PCCF for three different scenarios: a turnover of the luminosity function at $M_{\rm UV}=-12$ obtained by imposing a Gaussian prior (Section \ref{sec:results_final}), reionisation dominated by luminous galaxies (Fig. \ref{fig:posterior_bright}), and the reverse scenario where only low luminosity galaxies contribute (Fig. \ref{fig:posterior_faint}). Our models assume the following IGM parameters: $\gamma=1.3,\beta_{\text{mfp}}=2/3, \gamma_{\text{mfp}} = -1$ and $\log \xi_{\text{ion}}/[\text{erg}^{-1} \text{Hz}] =25.5 $.
    }
    \label{tab:results_fesc_summary}
\end{table}

\subsection{Impact of noise on the detection of transmission spikes and the non-detection of a transmission cross-correlation}

We now investigate whether or not we can explain the apparently contradictory absence of a transmission cross-correlation but the detection of the transmission spike two-point correlation (Fig. \ref{fig:XCorr_LAE_LBG} and \ref{fig:2PCCF_LAE}).

In order to do so, we use our improved model of the galaxy-IGM cross-correlation including the varying mean free path and the gas overdensities. We sample $P_V(\Delta_b | r,M_h)$ to generate $1000$ values of transmission $\exp(-\tau_\alpha)$ for each distance $r$ from the tracer LAE. We then sample the distribution of errors $\sigma$ as measured in the Lyman-$\alpha$ forest pixels used in the cross-correlation measurement (i.e. after masking). We then add a flux error drawn from the normal distribution $\Delta_T \sim \mathrm{N}(0,\sigma)$ to every computed flux value to mimic the effect of noise. Finally, we bin the data to match the measurement binning using the same number of mock Lyman-$\alpha$ forest ``pixel" points as the ones measured in the real quasar spectra. The transmission cross-correlation is computed as the mean flux value in each bin, whereas the 2PCCF is the fraction of transmission values above $T>0.02$ (the same threshold used for the previous measurement).

\begin{figure*}
    \centering
    \includegraphics[width = \textwidth]{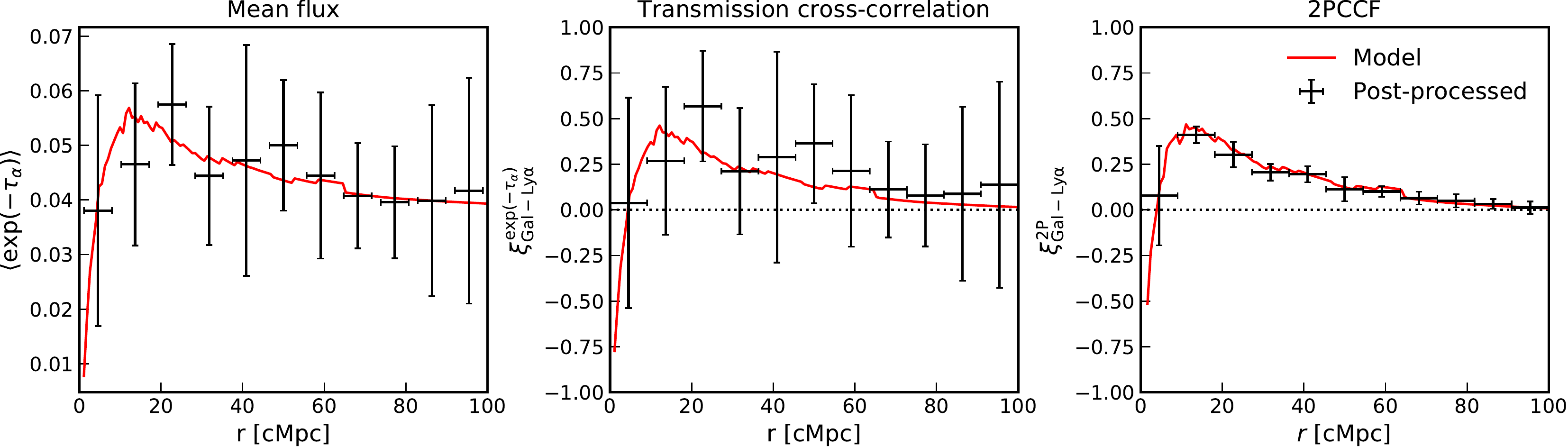}
    \caption{Post-processing of different statistics probing the Lyman-$\alpha$ forest - galaxy correlation from our improved model. The addition of noise directly drawn from the spectrsocopic data of quasars used in this study highlights the extreme difficulty of measuring the mean flux in the Lyman-$\alpha$ forest around galaxies \citepalias[left][]{Kakiichi2018} or the transmission cross-correlation with galaxies \citepalias[centre][]{Meyer2019}. The 2PCCF is however robust to such perturbations, hence its use in this study.}
    \label{fig:model_postprocessing}
\end{figure*}

The resulting mock observations are shown in Fig. \ref{fig:model_postprocessing} alongside the original model without errors. Clearly, the mean flux or the transmission cross-correlation are difficult to measure with any certainty. This also explains why an increase in the mean transmission or a transmission cross-correlation is much harder to detect than the 2PCCF at $z\sim 6$, as we found in Section \ref{sec:results_data}. The addition of noise is crucial because the noise level is comparable to the mean transmission ($T = 0.01-0.1$). It is thus no surprise that an increase in the average flux is difficult to measure. The 2PCCF however is shown to be rather unaffected by the addition of noise as the spikes we consider are at high enough SNR and transmission. Indeed, because the distribution of transmission pixels is log-normal \citep{Bosman2018}, there will be more pixels with intrinsic transmission below any given threshold ($T>0.02$) than above. As the observational error is drawn from a normal distribution, there will be more pixels observed to have a higher transmission than the given threshold but with lower intrinsic transmission than the reverse, increasing the number of spurious spike detections.
In practice, however, this only slightly decreases the 2PCCF and therefore the addition of noise is neglected in our modelling. We conclude that the 2PCCF is less biased by fluctuations of the mean opacity in different sightlines and should be less affected by continuum normalisation uncertainties.

\begin{figure}
    \centering
    \includegraphics[width=0.48\textwidth]{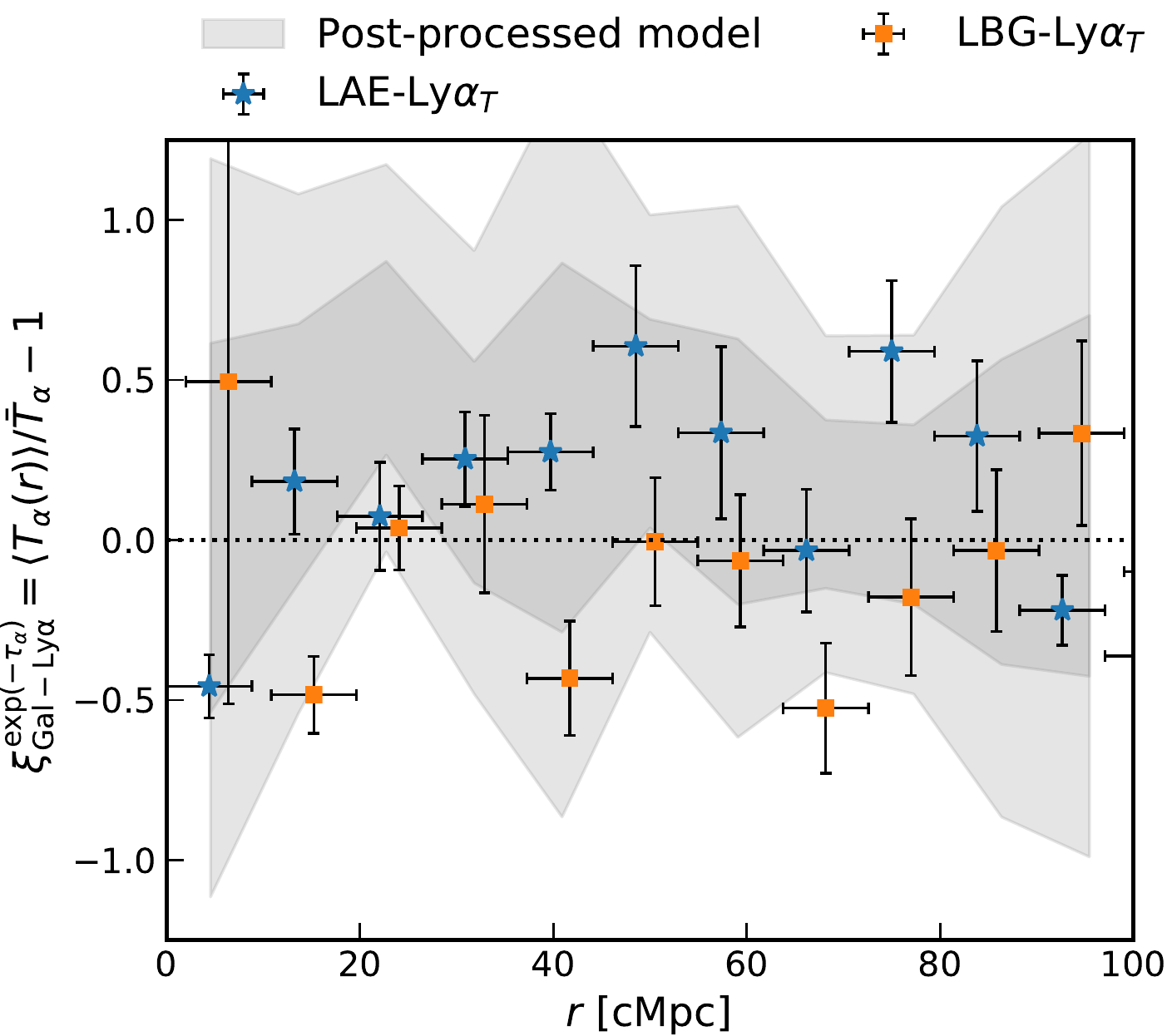}
    \caption{Comparison of the observed LAE/LBG-Ly$\alpha$ transmission cross-correlation (blue stars, orange square) with our post-processed (noisy) model (grey shaded areas, $1,2\sigma$) generated using the best-fit parameters for the LAE-Ly$\alpha$ 2PCCF. The observation are in agreement with the predicted uncertainties stemming from the fact that the SNR of the Lyman-$\alpha$ at $z\sim6$ is extremely low on average due to the IGM opacity, and that our quasars were primarily chosen to have MUSE coverage rather than deep spectroscopy.} 
    \label{fig:xcorr_withnoisemodel}
\end{figure}

We now conclude by examining whether the observed transmission cross-correlation (Fig. \ref{fig:XCorr_LAE_LBG}) is consistent with the predicted uncertainty on the modelled signal generated using the best-fit physical parameters of the 2PCCF LAE-transmission spike detection. We find that our LAE transmission cross-correlation measurement is in agreement with the predicted $1\sigma$ uncertainty range of the model. (Fig. \ref{fig:xcorr_withnoisemodel}). There appears to be a slight tension between the (LAE) post-processed model and the LBG measurement, but it is not very significant. The potential tension is more likely due to the smaller number of objects (and quasar sightlines) for the LBG transmission measurement which would lead us to underestimate the errors on the measurement. This is expected as the bootstrap uncertainties are primarily limited by cosmic variance and small sample size, and this measurement might be accurate with a larger sample of quasar sightlines and foreground objects (e.g.  \citetalias{Meyer2019}). 

\section{Summary}\label{sec:conclusions}

We have assembled a new sample of galaxies in the field of $8$ high-redshift quasars in order to examine various correlations between galaxies and the fluctuations in the Lyman-$\alpha$ forest at the end of reionisation. We have extended the approach pioneered in \citetalias{Kakiichi2018} and  \citetalias{Meyer2019} to model the galaxy-Lyman-$\alpha$ flux transmission and the two-point correlation with transmission spikes. We report the following key findings:

\begin{itemize}
    \item  We find a $3.2(1.9)\sigma$-significant excess in the 2PCCF of transmission spikes with LAEs(LBGs) at  on scales of $\sim 10-60$ cMpc. Our model of the LAE(LBG) 2PCCF constrains sources with $M_{\text{UV}}<-12$ to contribute to reionisation with a luminosity-averaged escape fraction $\langle f_{\text{esc}} \rangle_{M_{\text{UV}}<-12}= 0.14_{-0.05}^{+0.28} (0.23_{-0.12}^{+0.46})$ assuming a fixed $\log \xi_{\text{ion}}/[\text{erg}^{-1} \text{Hz}] =25.5 $. 
    \item We present a new model of the two-point cross-correlation function (2PCCF) of detected Lyman-$\alpha$ transmission spikes with LAEs which includes a consistent treatment of gas overdensities around detected LAEs and their peculiar velocities. We find that a spatially varying mean free path does not impact the 2PCCF significantly. We demonstrate that this model is more robust than the transmission cross-correlation at high-redshift.
    \item We show how parametric models of the escape fraction dependence on the galaxy luminosity can be constrained by the LAE-IGM 2PCCF. We find that the LAE 2PCCF is consistent with a local UVB enhanced either by \textit{faint} galaxies with $M_{\rm UV}^{\rm max}= -14.1_{-6.6}^{+2.4}$ or \textit{brighter} than $M_{\rm UV}<-19.0 \, (2\sigma)$. The LBG 2PCCF favours brighter objects with at least $M_{\rm UV}< -19 \,(2\sigma)$ contributing to reionisation. Differentiating between these hypotheses will however require a larger dataset of galaxies in quasar fields.
    \item We find no evidence for a correlation between the transmission in the Lyman-$\alpha$ forest and LAEs/LBGs at $z\sim 6$. We show how this absence of signal is consistent with scatter and noise of our quasar sightlines. Nonetheless, the deficit of transmission on scales up to $\sim 10$ cMpc is seen in the Lyman-$\alpha$ forest around LAEs as previously reported around C{~\small IV} absorbers \citepalias{Meyer2019}.
\end{itemize}

\section*{Acknowledgements}

The authors thank the anonymous referee for useful comments that improved the manuscript. RAM, SEIB, KK, RSE, NL acknowledge funding from the European Research Council (ERC) under the European Union's Horizon 2020 research and innovation programme (grant agreement No 669253). NL also acknowledges support from the Kavli Foundation. B.E.R. acknowledges NASA program HST-GO-14747, contract NNG16PJ25C, and grant 17-ATP17-0034, and NSF award 1828315. ERW acknowledges support from the Australian Research Council Centre of Excellence for All Sky Astrophysics in 3 Dimensions (ASTRO 3D), through project number CE170100013. RAM thanks A. Font-Ribeira and F. Davies for useful discussions.
\\
Based on observations made with ESO Telescopes at the La Silla Paranal Observatory
under programmes ID 060.A-9321(A), 094.B-0893(A), 095.A-0714(A), 097.A-5054(A), 099.A-0682(A), 0103.A-0140(A).
This research has made use of the Keck Observatory Archive (KOA), which is operated by the W. M. Keck Observatory and the NASA Exoplanet Science Institute (NExScI), under contract with the National Aeronautics and Space Administration. \\
Some of the
data used in this work was taken with the W.M. Keck Observatory on Maunakea, Hawaii, which is operated as a scientific partnership among the California Institute of Technology, the University of California and the National Aeronautics and Space Administration. This Observatory  was  made  possible  by  the generous financial support of the W. M. Keck Foundation. The authors wish to recognise and acknowledge the very significant cultural role and reverence that the summit of Maunakea has always had within the indigenous Hawaiian community. We are most fortunate to have the opportunity to conduct observations from this mountain.
\\
 The analysis pipeline used to reduce the DEIMOS data was developed at UC Berkeley with support from NSF grant AST-0071048. 
 \\
The authors acknowledge the use of the UCL Myriad High Performance Computing Facility (Myriad@UCL), and associated support services, in the completion of this work.
\\
The authors acknowledge the use of the following community-developed packages \textsc{Numpy} \citep{VanderWalt2011}, \textsc{Scipy} \citep{Virtanen2020}, \textsc{Astropy}  \citet{TheAstropyCollaboration2013,TheAstropyCollaboration2018}, \textsc{Matplotlib} \citep{Hunter2007}, \textsc{emcee} \citep{Foreman-Mackey2013}, \textsc{ChainConsumer} \citep{Hinton2016} .

\bibliographystyle{mnras}
\bibliography{phd} 



\appendix

\section{Summary of all detections and plots from DEIMOS}
\label{appendix:DEIMOS_detections}
We present the DEIMOS spectroscopic confirmation of new LBGs in the field of J0836 (Fig. A1, online material) and J1030 (Fig. A2, online) used in this work for the cross-correlations. We leave the presentation and analysis of the 3 objects detected in the near-zone of J0836 to \citet{Bosman2019}. Table \ref{tab:DEIMOS_LBG} lists the LBG detections with their coordinates, redshift, Lyman-$\alpha$ FWHM and corrected redshift. 
\begin{table*}
    \centering
    \begin{tabular}{cccccccccc}
Quasar & RA & DEC & $z_{peak}$ & FWHM & $z_{corr}$ & $\text{r (mag)}$ & $\text{i (mag)}$ & $\text{z (mag)}$ & $M_{UV}$ \\ \hline 
$\text{J0836}$ & 129.09106 & 1.00954  & 5.283 & 95.812 &5.281 &  $>27.62$ & 26.35 & 25.33 &  -21.16 \\
$\text{J1030}$ & 157.71105 & 5.36851 & 5.508 & 92.518 & 5.507 &  $>27.5$ & 25.54 & 23.95 &  -22.61\\
& 157.58161 & 5.46687 &   5.791 & 66.852 & 5.790 & $>27.50$ & 24.95 & 23.41 & -23.23 \\
& 157.58308 &  5.44516 & 5.481 & 69.325 & 5.480 &  $>27.50$ & 26.51 & 25.12 & -21.43  \\
& 157.67004 & 5.45504 & 5.712 & 176.514 & 5.709 &  $>27.50$ & 26.13 & 25.18 & -21.44 \\
& 157.73887 & 5.46775 & 5.612 & 137.348 & 5.610 &  $>27.50$ & $>26.80$ & 25.45 & -21.13\\
& 157.70962& 5.36157 & 5.692 & 223.848 & 5.688 &  $>27.50$ & 26.39 & 25.19 & -21.42 \\
& 157.52691 & 5.37737 & 5.352 & 118.119 &5.351 &  $>27.50$ & 26.38 & 25.18 & -21.33 \\
& 157.56116  & 5.34611 & 5.446 & 186.830 & 5.443 &  $>27.50$ & 25.93 & 25.49 &  -21.05\\
    \end{tabular}
    \caption{Summary of the detected LBGs in the DEIMOS fields}
    \label{tab:DEIMOS_LBG}
\end{table*}

\begin{figure*}
    \centering
    \includegraphics[width = 0.45\textwidth]{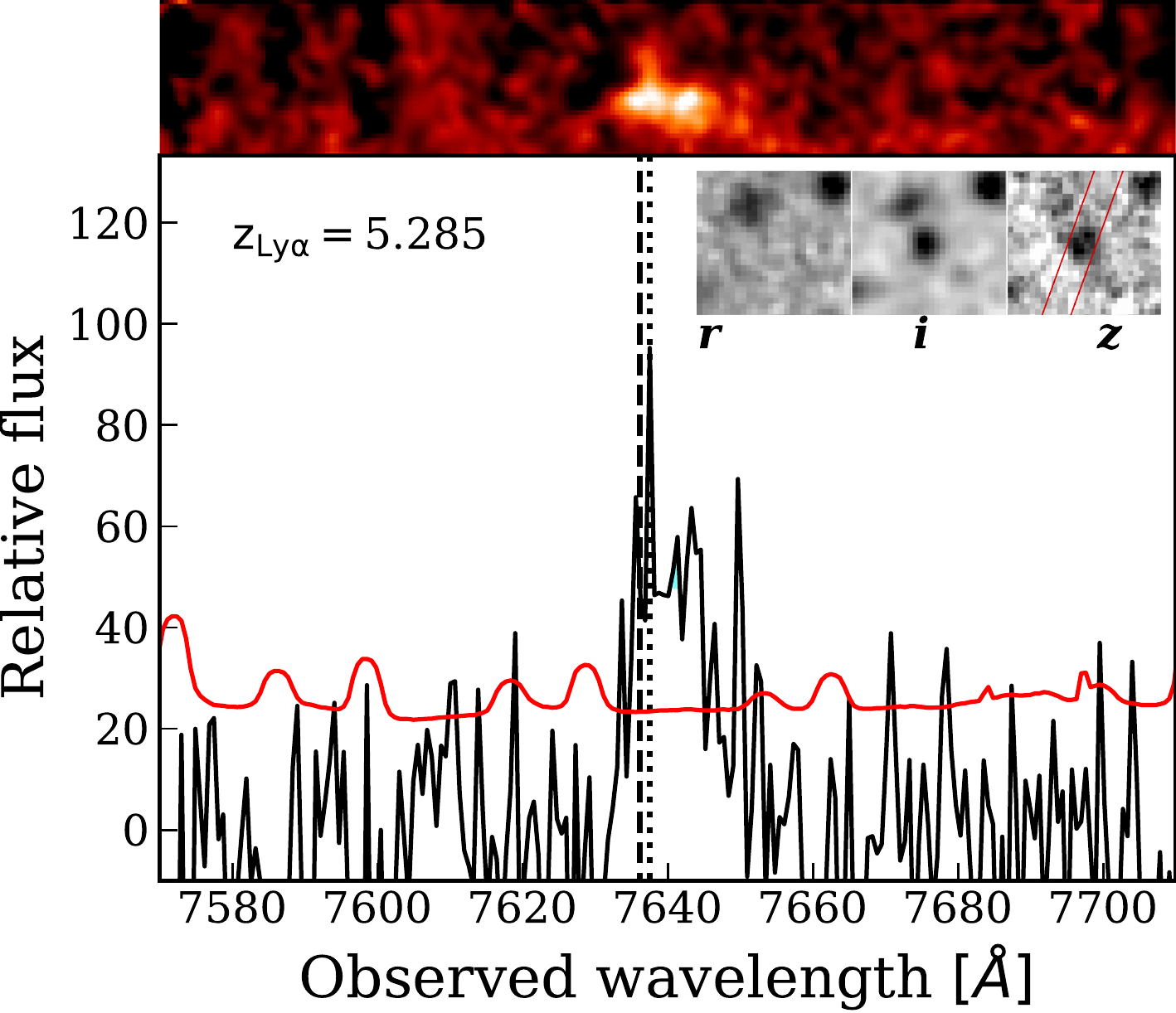} 
    \caption{Confirmed LBG observed with DEIMOS in the field of J0836 (see also Fig. \ref{fig:example_DEIMOS_LBG}). The top panels show the 2D spectra from which the 1D spectrum (black line) and noise (red) are optimally extracted using a boxcar aperture of $1.2''$. In the upper right corner is displayed the \textit{riz} image used for the drop-out selection. }
    \label{fig:J0836_DEIMOS}
\end{figure*}

\begin{figure*}
    \centering
        \includegraphics[width = 0.45\textwidth]{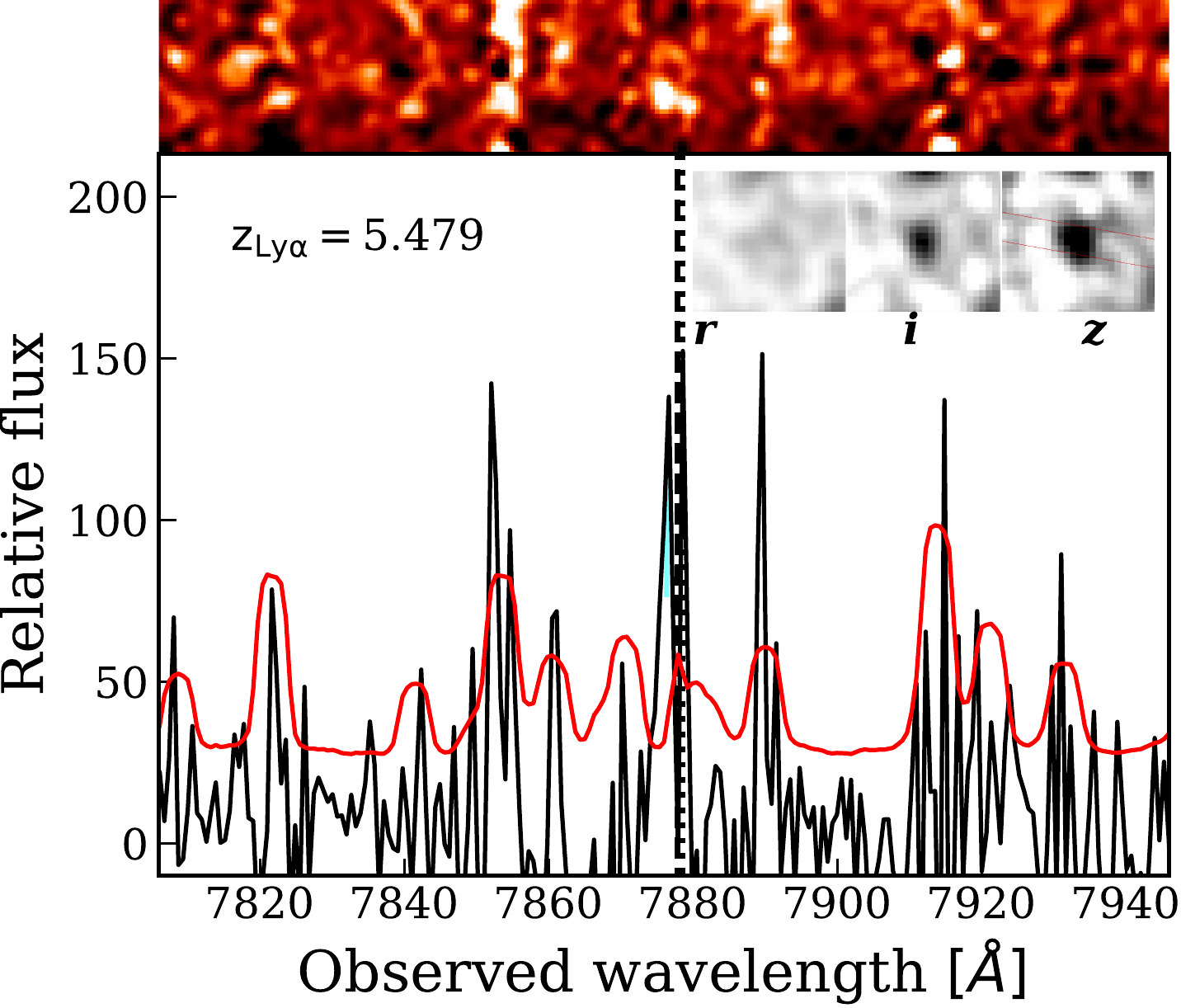} \hfill
        \includegraphics[width = 0.45\textwidth]{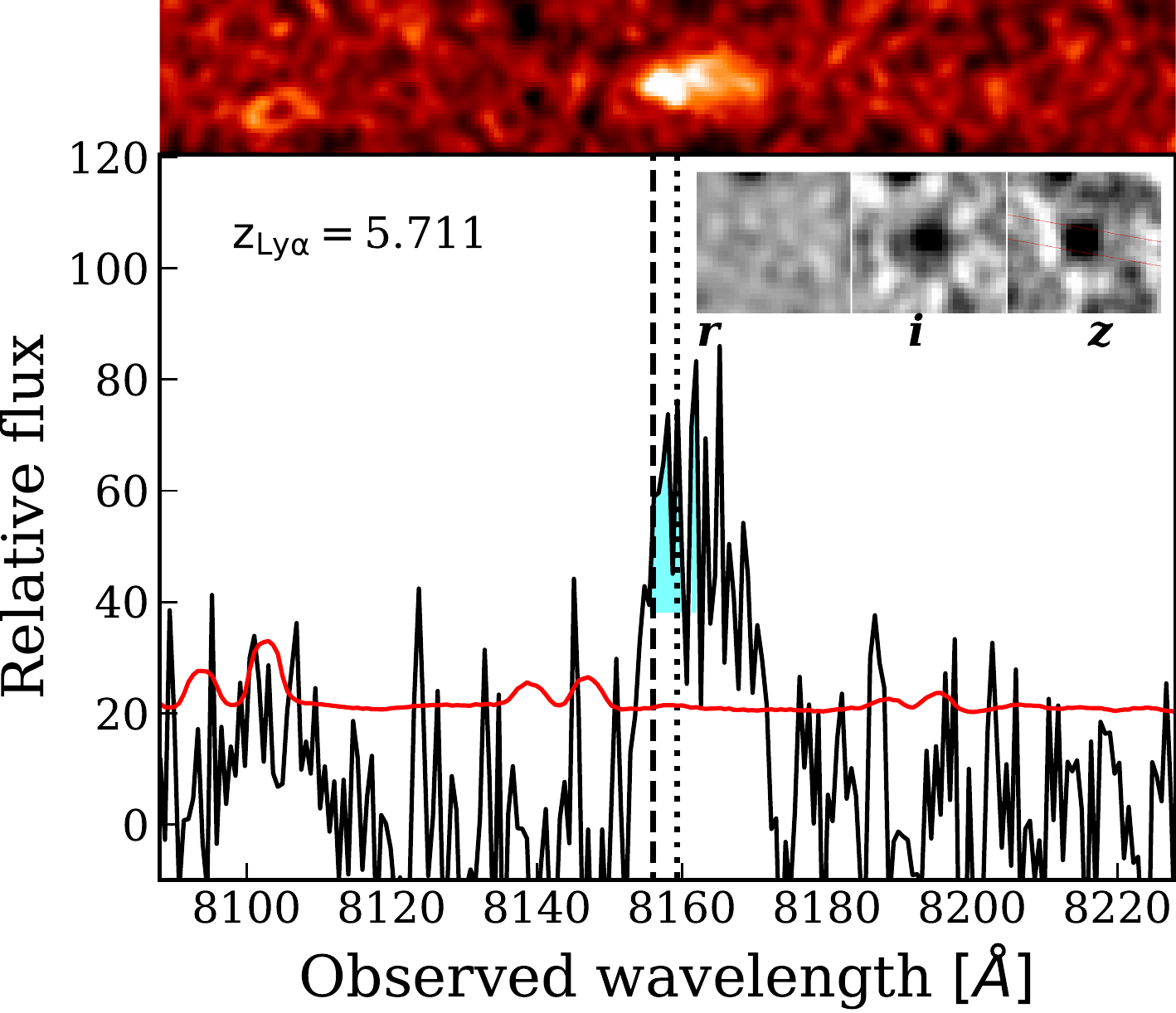}  \\ \vspace{0.5cm}
        \includegraphics[width =0.45\textwidth]{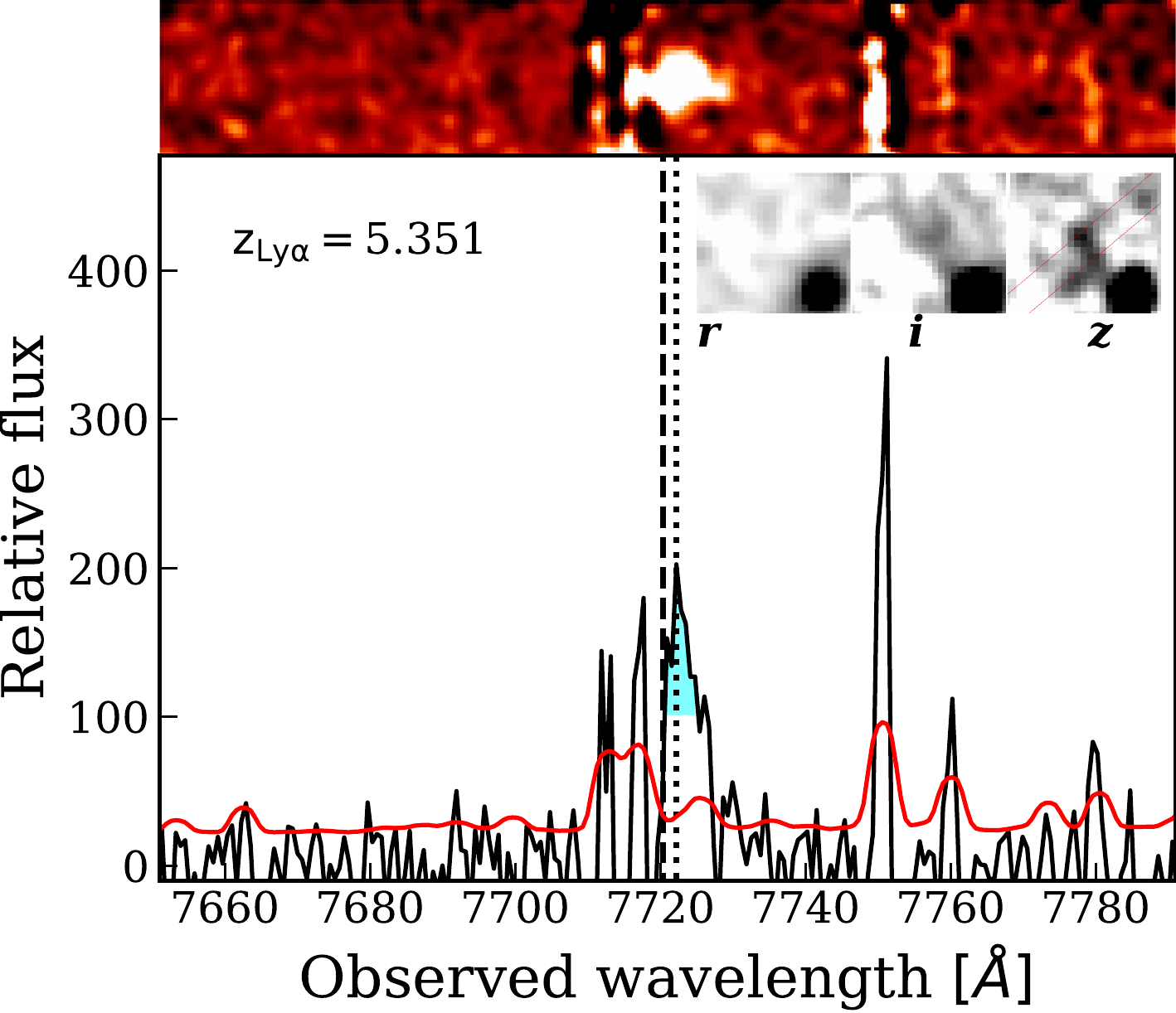} \hfill
        \includegraphics[width = 0.45\textwidth]{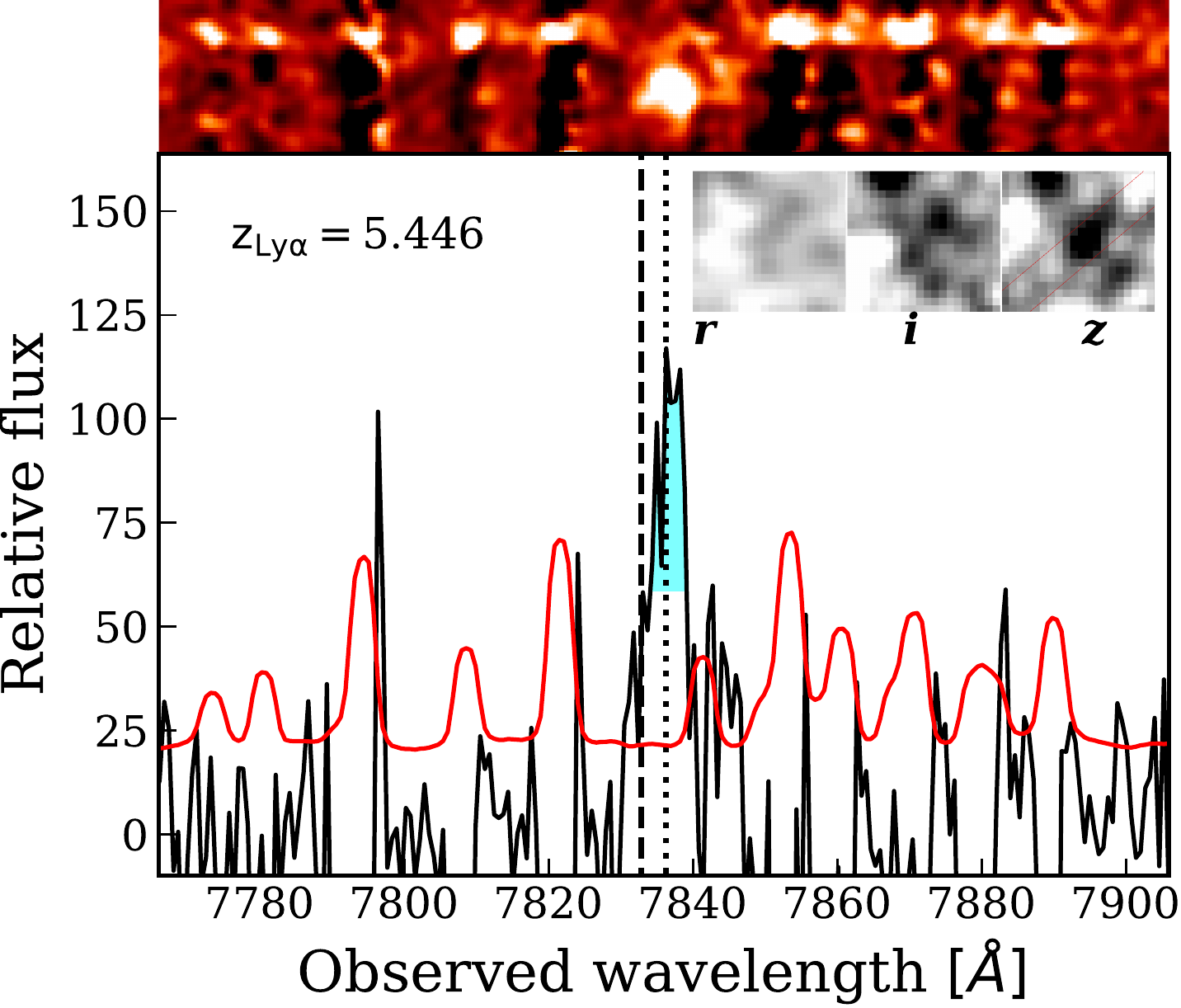}
        \includegraphics[width = 0.45\textwidth]{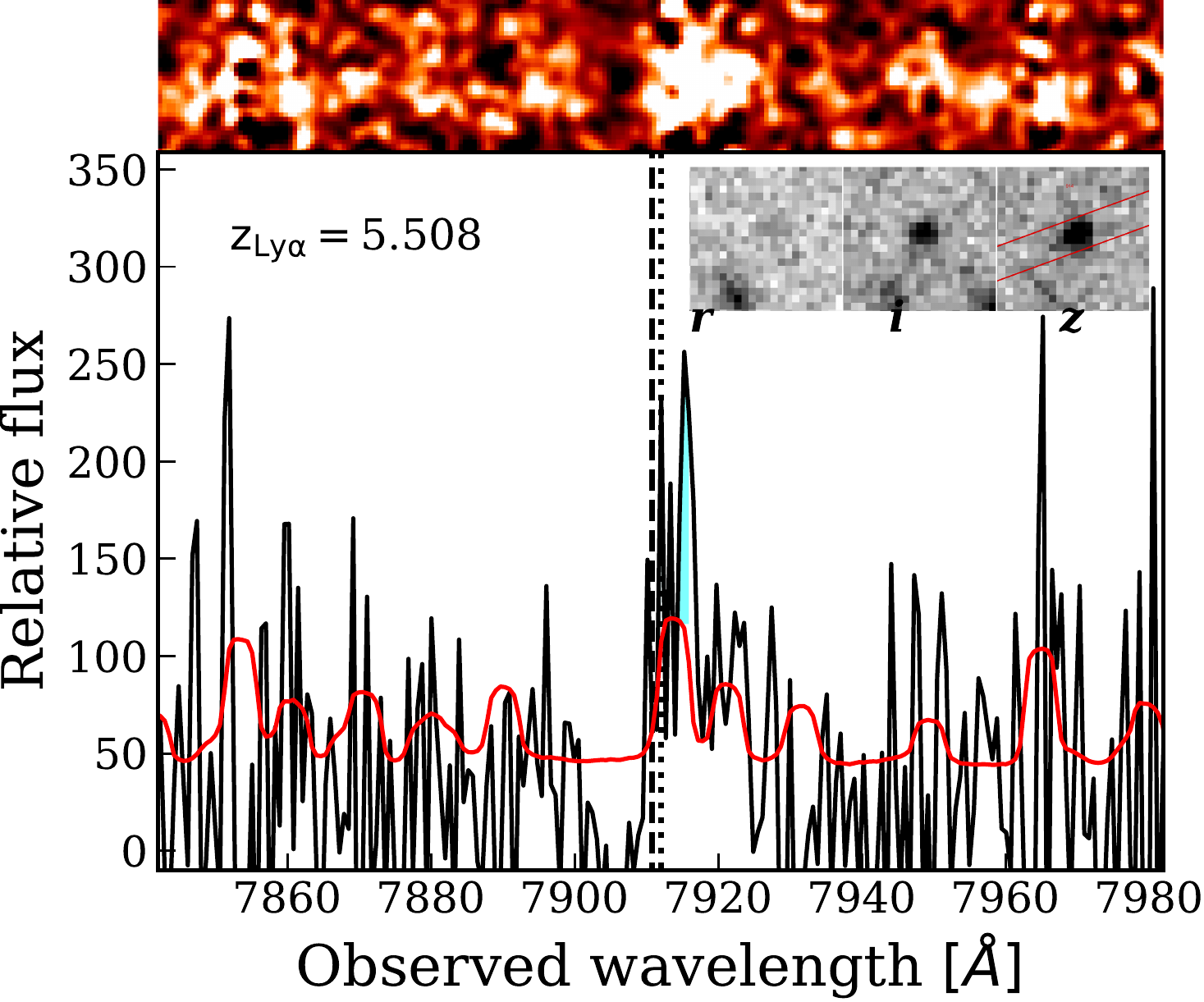} \hfill
        \includegraphics[width = 0.45\textwidth]{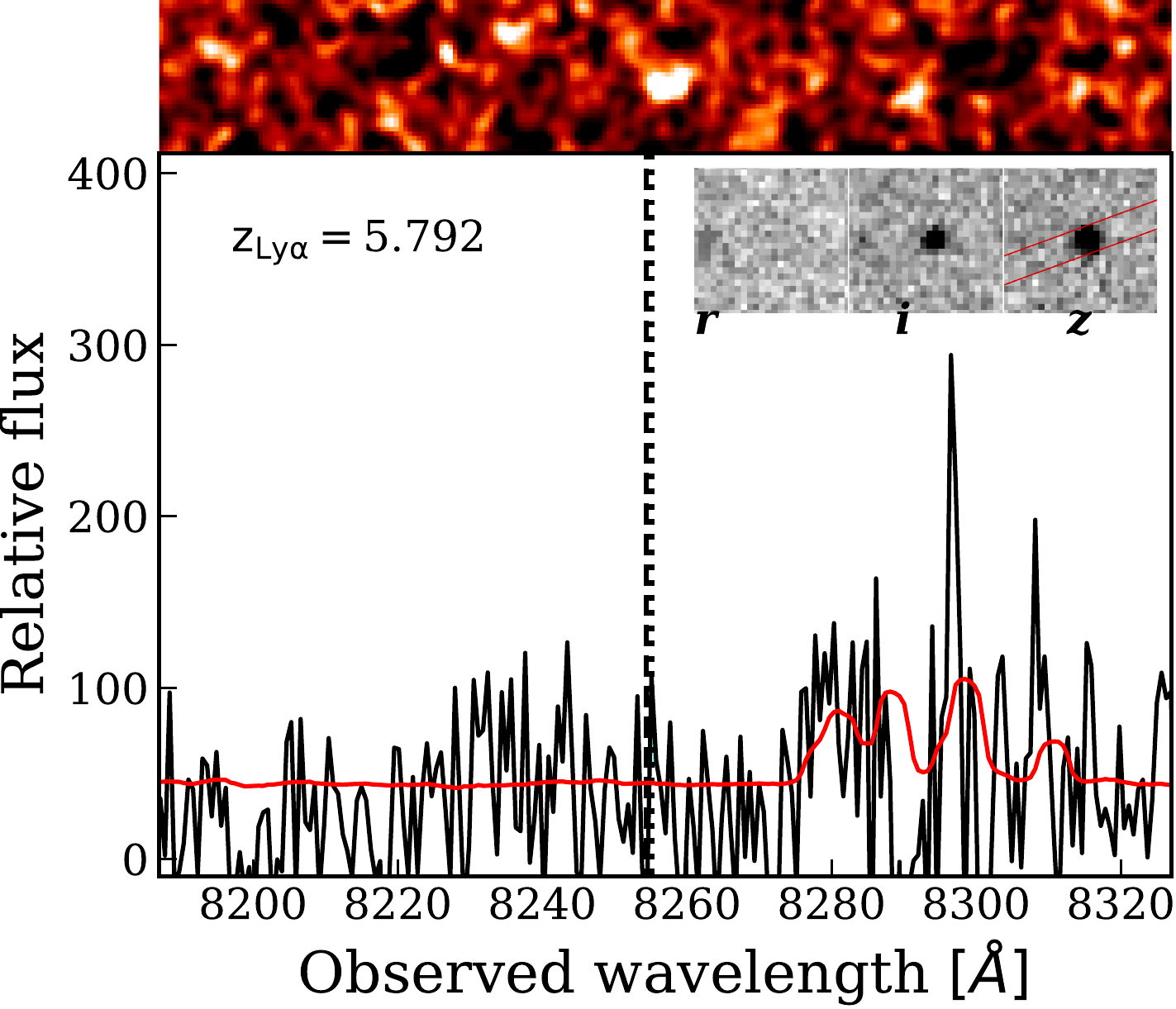}  \\ \vspace{0.5cm}
    
    \caption{Confirmed LBGs observed with DEIMOS in the field of J1030. The top panels shows the 2D spectra from which the 1D spectrum (black line) and noise (red) are optimally extracted using a boxcar aperture of $1.2''$ . In the upper right corner is displayed the \textit{riz} image used for the drop-out selection.}
    \label{fig:J1030_DEIMOS}
\end{figure*}

\section{Individual LAE detections with MUSE in the Lyman-$\alpha$ forest of our quasars}

We present a summary of all detected LAEs in the redshift range of the Lyman-$\alpha$ forest of the nearby quasar in Table \ref{tab:MUSE_LAE}. We adopt an identification scheme where each LAE is named 'JXXXX\_NBYYYY', where XXXX denotes the hours and minutes of the RA coordinates of the central quasar and YYYY the rounded wavelength of the narrowband (NB) frame in which \textsc{MUSELET} or \textsc{LSDCat} found the highest signal of the detection, which is often very close to the wavelength of the emission peak. Individual plots similar to Fig. \ref{fig:MUSE_detections_LAEs} for each LAE can be found online in Fig. B1 (J0305, online), B2 (J1030, online), B3 (J1526, online), B4 (J2032, online), B5 (J2100, online), B6 (J2329, online). Finally, we provide an example of common misdetections that are removed by visual inspection in Fig. B7 (online )such as low-redshift \otwo or continuum emitters, bright nearby foreground objects or defects or cosmic rays impacting only one of the exposures.

\label{appendix:MUSE_detections}
\begin{table*}
    \centering
    \begin{tabular}{lcccccc}
ID & RA & DEC & $\lambda_{Lya}$ & $z_{Lya}$ & FWHM [km/s] & $ z_{corr}$  \\  \hline
J0305\_NB8032 & 46.32776 & -31.84569 & 8034.7&5.607 & 186.7 & 5.604\\ 
J0305\_NB8609 & 46.31154 & -31.85152 & 8609.7&6.081 & 174.2 & 6.078\\ 
J0305\_NB8612 & 46.31095 & -31.85202 & 8612.2 &6.084 & 174.2 & 6.081\\ 
 \hline 
J1030\_NB7707 & 157.61238 & 5.40784 & 7707.2&5.340 & 97.3 & 5.339\\ 
J1030\_NB7927 & 157.61109 & 5.41578 & 7927.2&5.520 & 236.5 & 5.516\\ 
J1030\_NB7942 & 157.61054 & 5.40995 & 7942.2&5.533 & 141.6 & 5.531\\ 
J1030\_NB8177 & 157.61534 & 5.40556 & 8177.2&5.725 & 229.3 & 5.722\\ 
J1030\_NB8202 & 157.61366 & 5.41512 & 8202.2&5.746 & 228.6 & 5.742\\ 
J1030\_NB8220a & 157.62069 & 5.41484 & 8220.9&5.761 & 228.1 & 5.758\\ 
J1030\_NB8220b & 157.61321 & 5.41900 & 8220.9&5.761 & 228.1 & 5.758\\ 
 \hline 
J1526\_NB8476 & 231.66377 & -20.83180 & 8475.9&5.972 & 88.5 & 5.971\\ 
J1526\_NB8874 & 231.65771 & -20.82652 & 8874.7&6.299 & 169.0 & 6.296\\ 
 \hline 
J2032\_NB8396 & 308.04785 & -21.23293 & 8396.4&5.907 & 134.0 & 5.905\\ 
J2032\_NB8524 & 308.04240 & -21.22620 & 8523.9&6.012 & 132.0 & 6.010\\ 
J2032\_NB8525 & 308.03598 & -21.23630 & 8525.2&6.013 & 132.0 & 6.011\\ 
 \hline  
J2100\_NB7454 & 315.23399 & -17.26017 & 7454.8&5.132 & 150.9 & 5.130\\ 
J2100\_NB7678 & 315.23219 & -17.26062 & 7678.6&5.316 & 146.5 & 5.314\\ 
J2100\_NB8146 & 315.22375 & -17.25901 & 8146.1&5.701 & 230.2 & 5.697\\ 
J2100\_NB8419 & 315.22404 & -17.26045 & 8419.8&5.925 & 133.6 & 5.923\\ 
 \hline 
J2329\_NB8372 & 352.28913 & -3.04041 & 8372.7&5.887 & 134.4 & 5.885\\ 
J2329\_NB8390 & 352.28769 & -3.03636 & 8390.2&5.902 & 134.1 & 5.900\\ 
 \hline  
    \end{tabular}
    \caption{Summary of the detected LAEs in the MUSE fields (and in the suitable redshift range for the cross-correlation). The last column gives the corrected redshift using the method of \citet{Verhamme2018}, as described in Section \ref{sec:z_corr}. }
    \label{tab:MUSE_LAE}
\end{table*}

\begin{figure*}
    \centering
    \includegraphics[width=\textwidth]{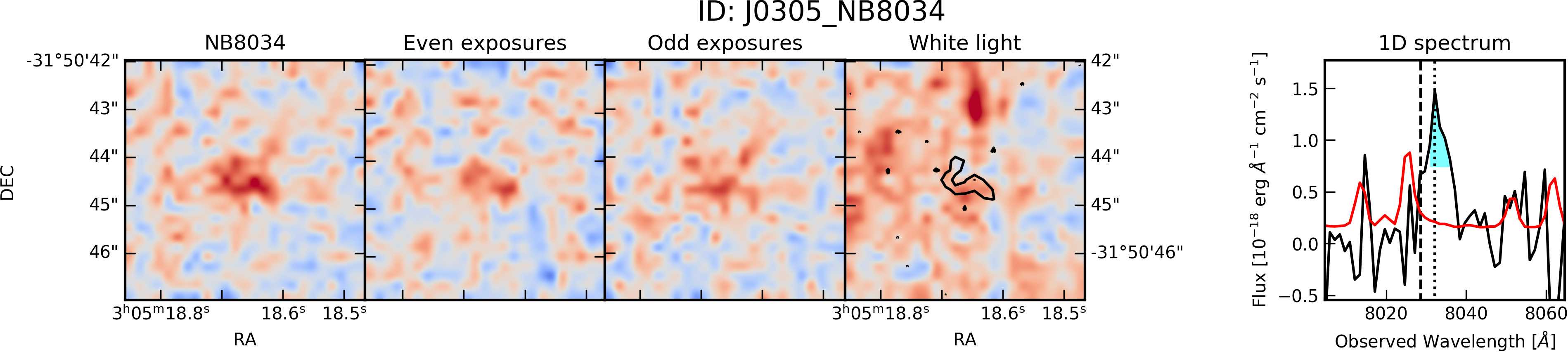} \\    \includegraphics[width=\textwidth]{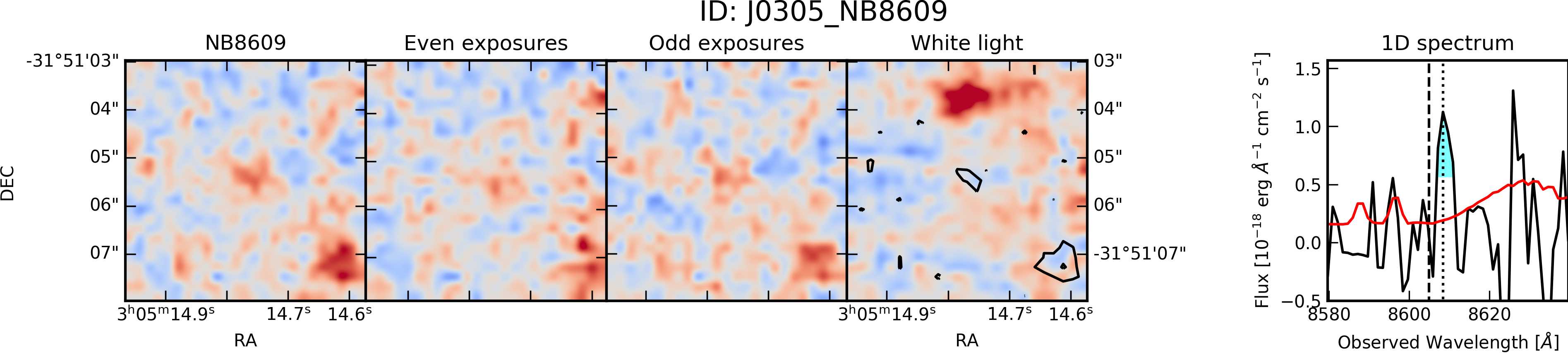} \\
    \includegraphics[width=\textwidth]{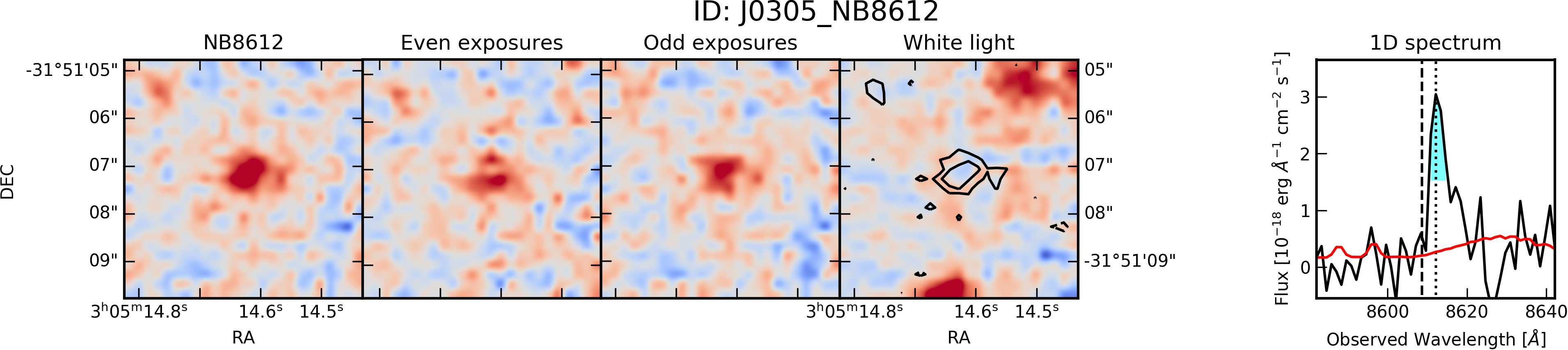} 
    \caption{LAEs detected in the field of J0305 LAEs used in this work}
    \label{fig:J0305_MUSE_detections_LAEs}
\end{figure*}

\begin{figure*}
    \centering
    \includegraphics[width=\textwidth]{Figures/ID_J1030_NB7707.jpeg} \\
    \includegraphics[width=\textwidth]{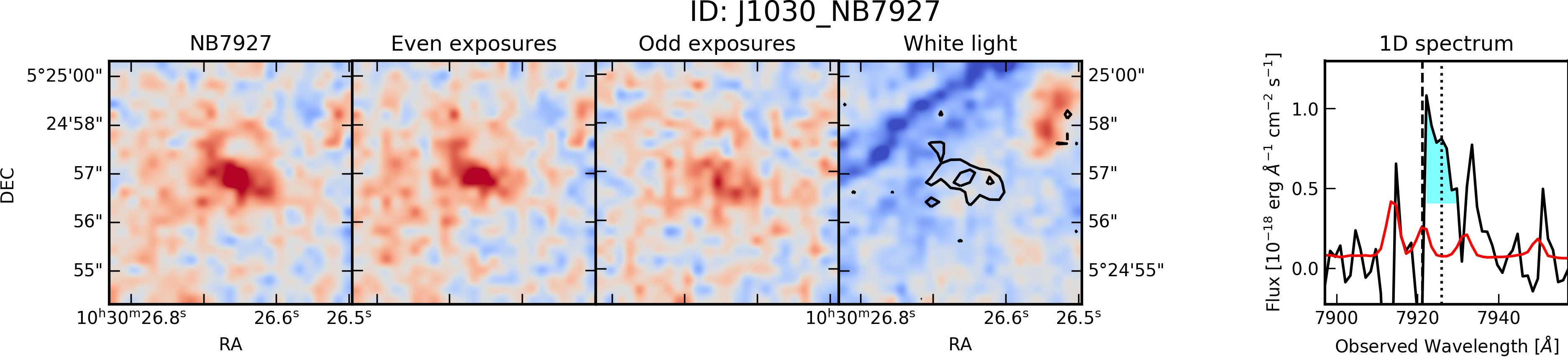} \\
    \includegraphics[width=\textwidth]{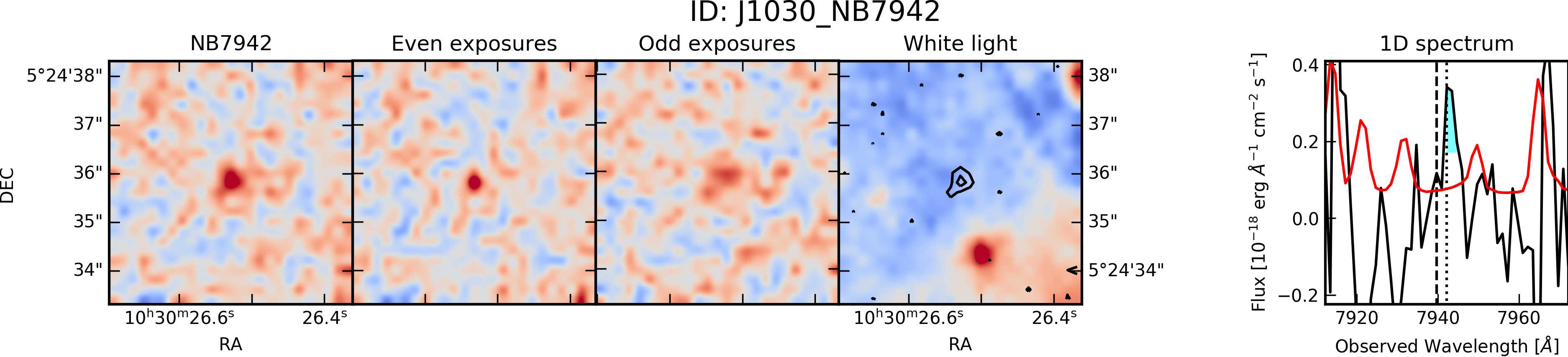} \\
    \includegraphics[width=\textwidth]{Figures/ID_J1030_NB8177.jpeg} \\
    \includegraphics[width=\textwidth]{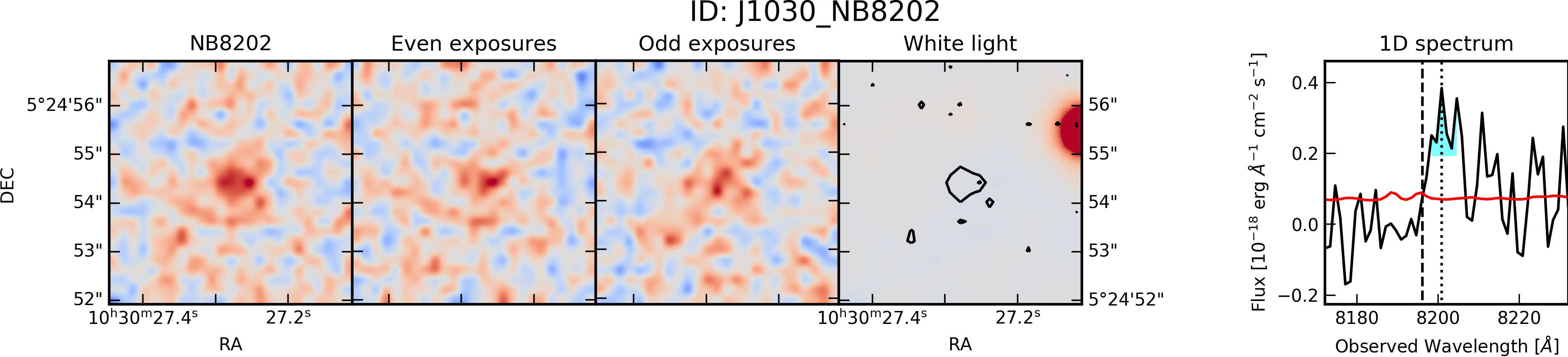} 
    \caption{LAEs detected in the field of J1030 LAEs used in this work}
    \label{fig:J1030_MUSE_detections_LAEs}
\end{figure*}

\begin{figure*}
    \centering
    \includegraphics[width=\textwidth]{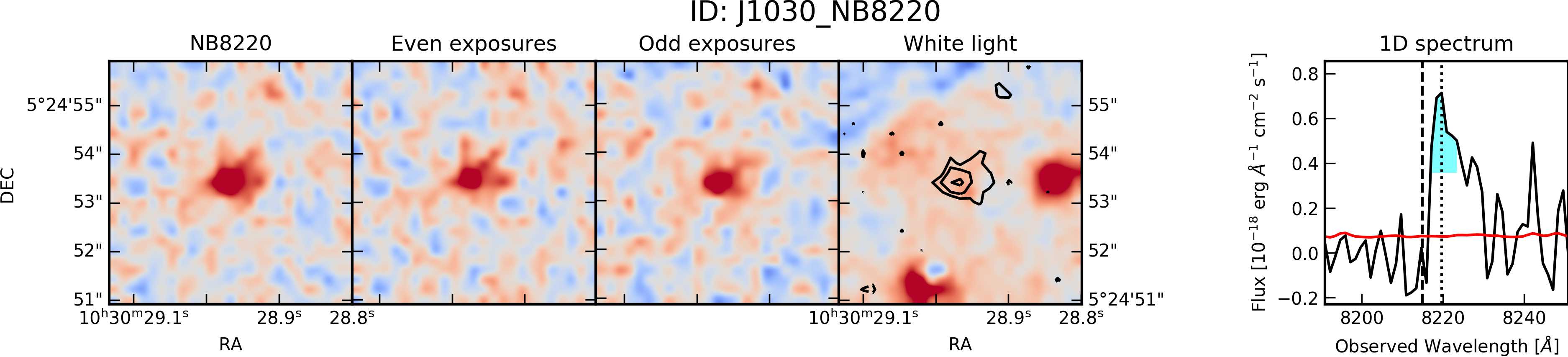} \\ 
    \includegraphics[width=\textwidth]{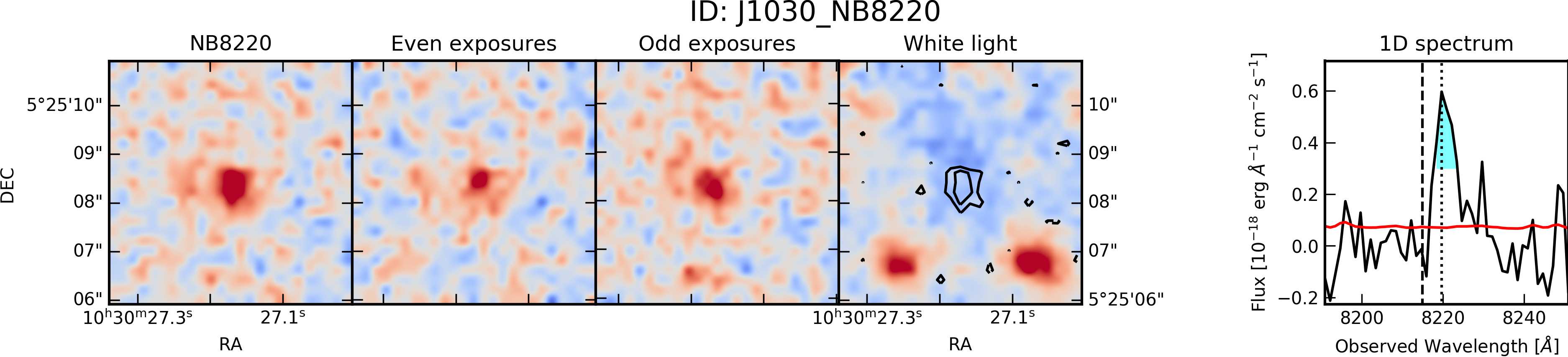}
    \contcaption{LAEs detected in the field of J1030 LAEs used in this work}
\end{figure*}

\begin{figure*}
    \centering
    \includegraphics[width=\textwidth]{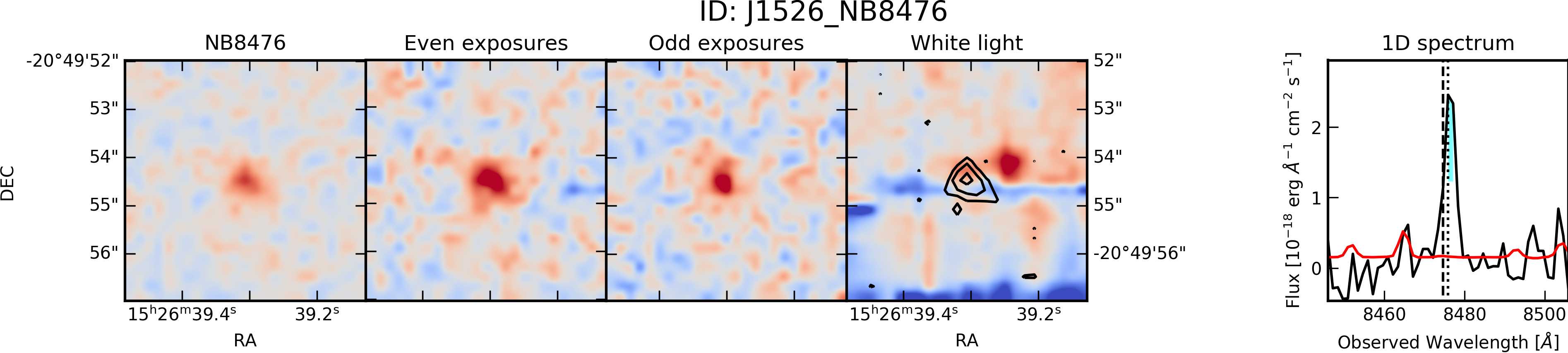}\\ 
    \includegraphics[width=\textwidth]{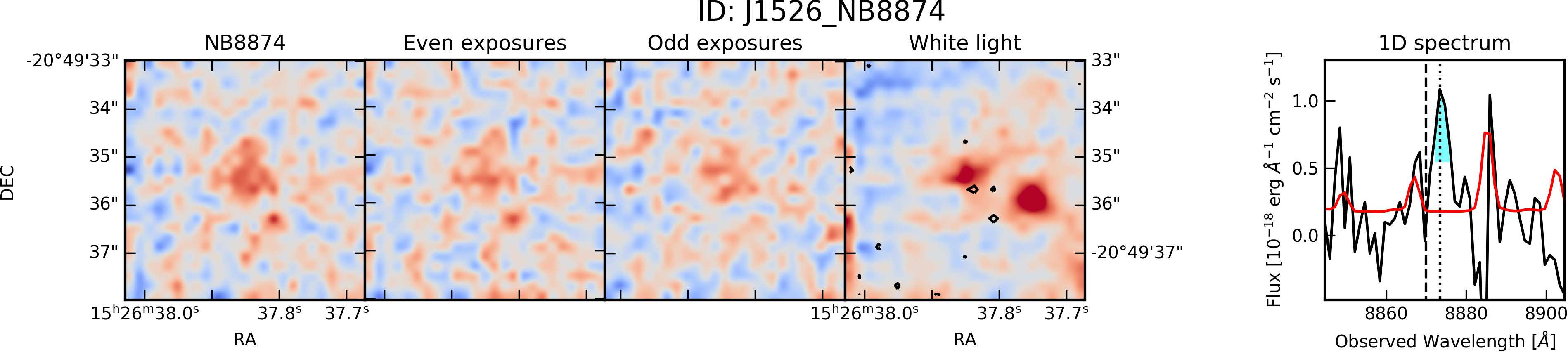} 
    \caption{LAEs detected in the field of J1526 LAEs used in this work}
    \label{fig:J1526_MUSE_detections_LAEs}
\end{figure*}

\begin{figure*}
    \centering
    \includegraphics[width=\textwidth]{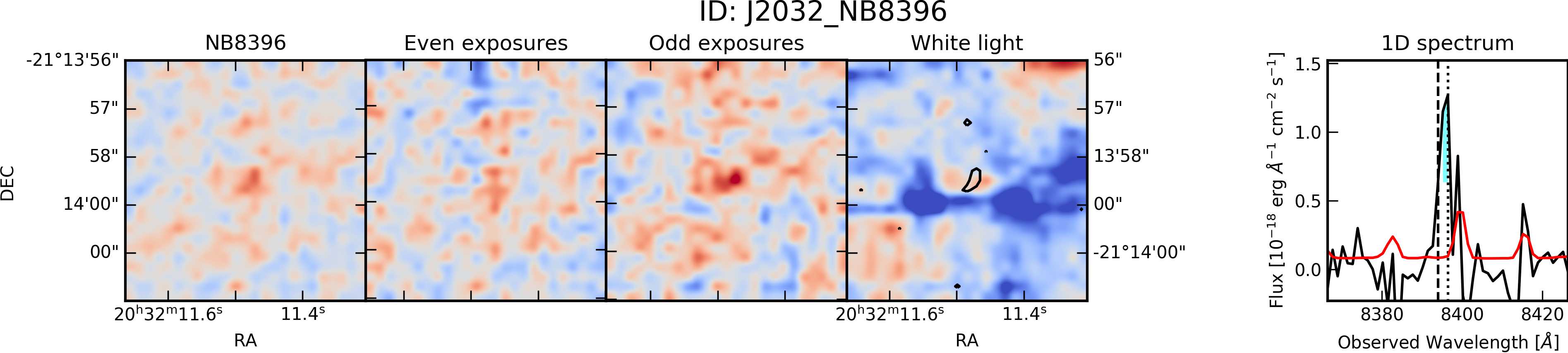}\\ 
    \includegraphics[width=\textwidth]{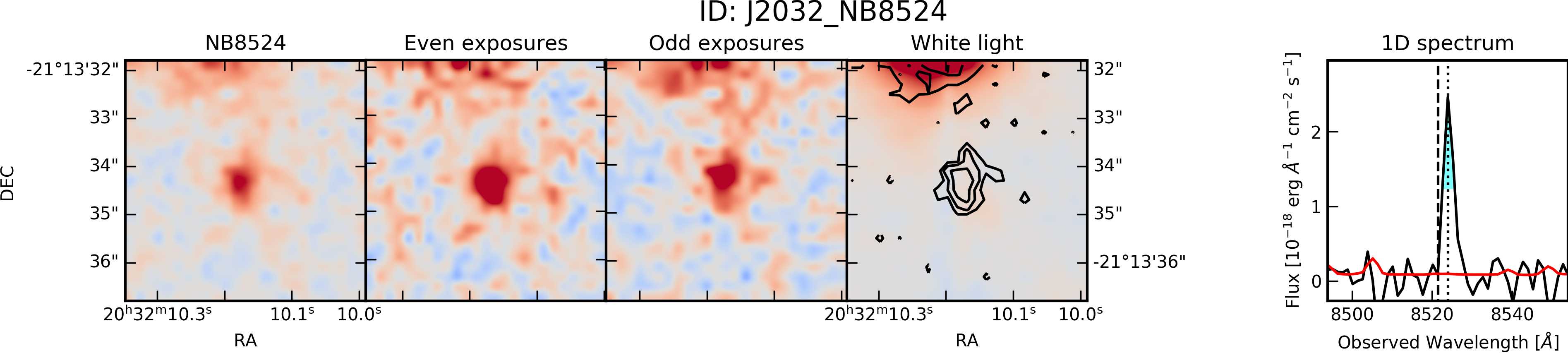}\\ 
    \includegraphics[width=\textwidth]{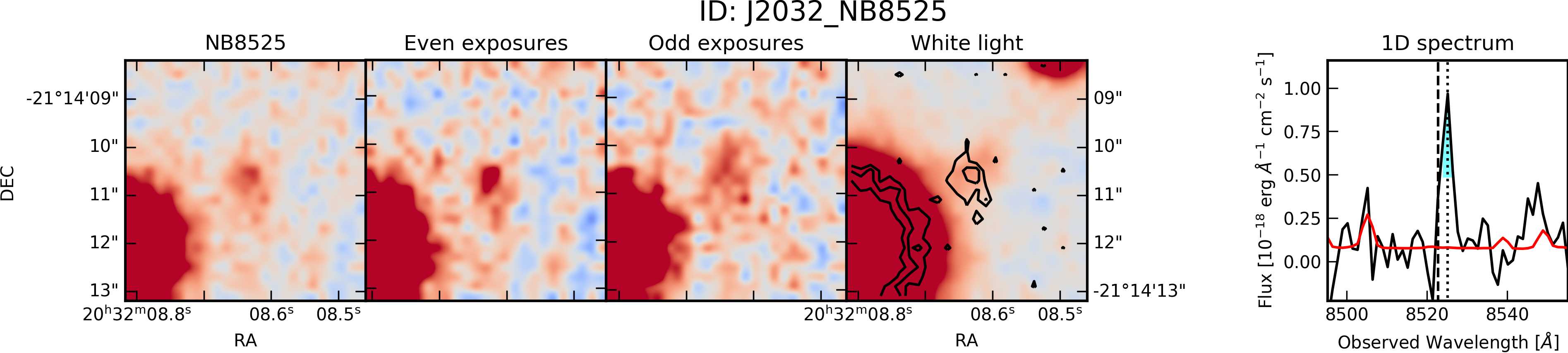}\\ 
    \caption{LAEs detected in the field of J2032 used in this work}
    \label{fig:J2032_MUSE_detections_LAEs}
\end{figure*}

\begin{figure*}
    \centering
    \includegraphics[width=\textwidth]{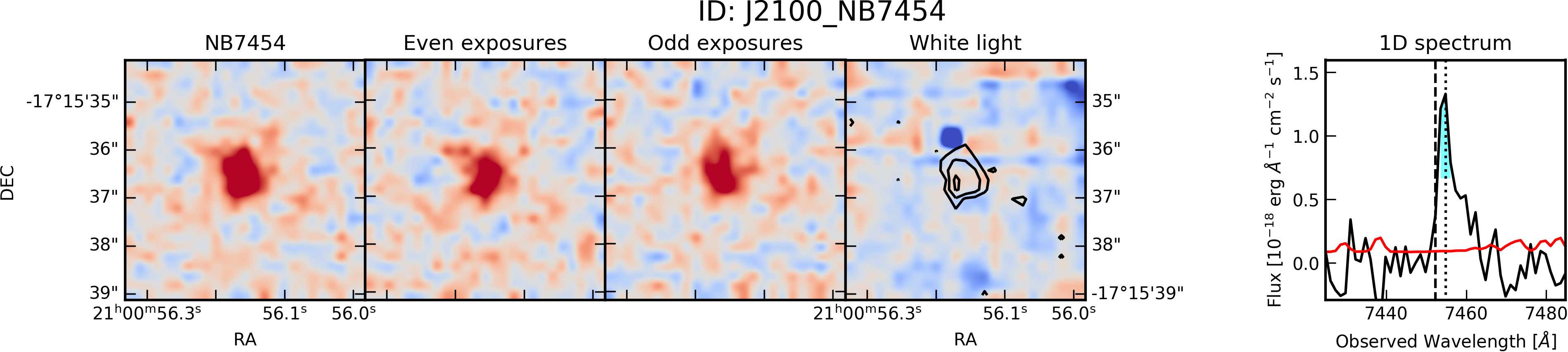}\\ 
    \includegraphics[width=\textwidth]{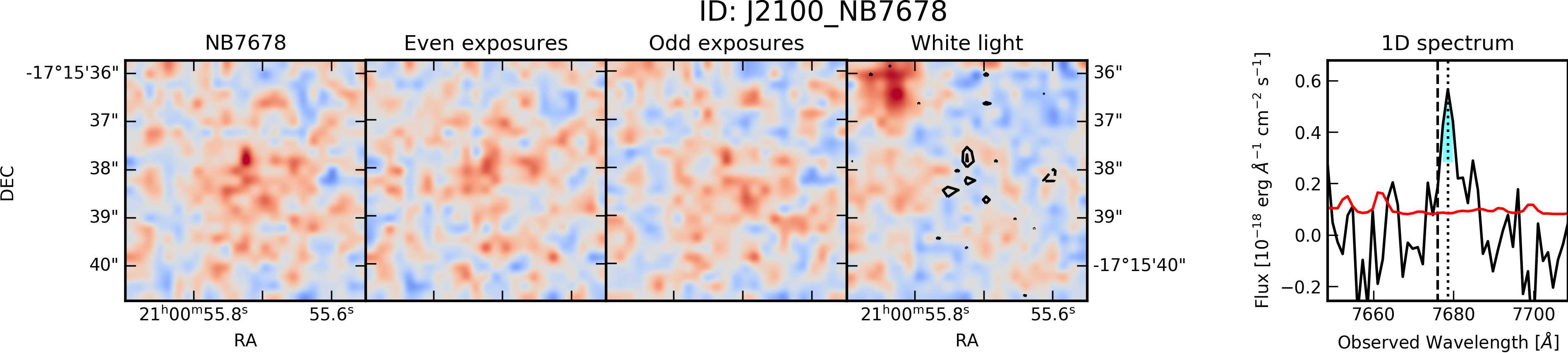}\\ 
    \includegraphics[width=\textwidth]{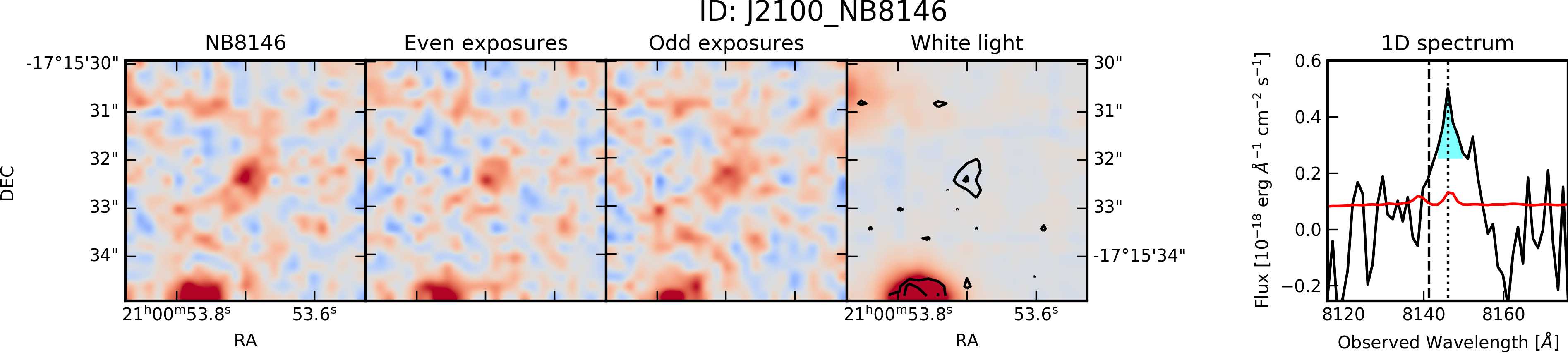}\\ 
    \includegraphics[width=\textwidth]{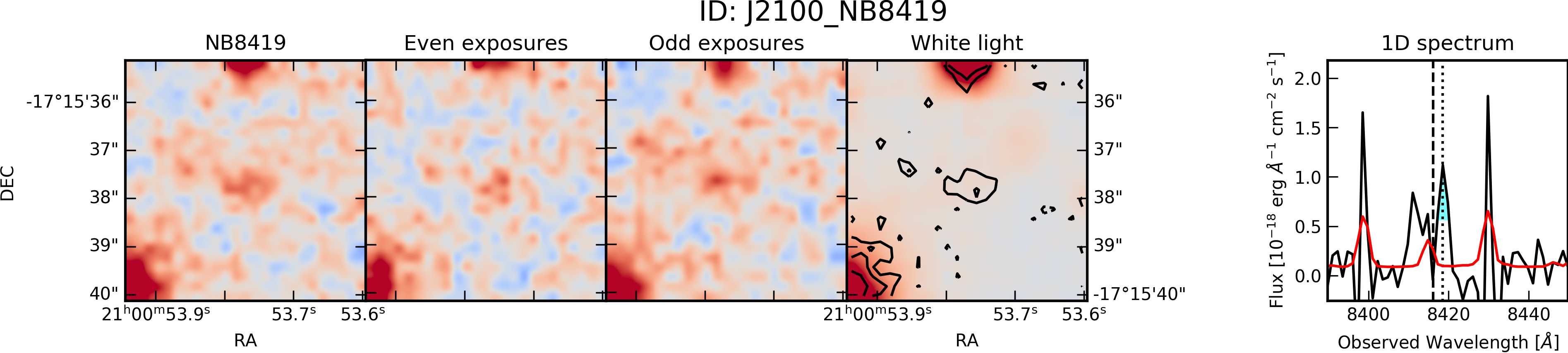}
    \caption{LAEs detected in the field of J2100 used in this work}
    \label{fig:J2100_MUSE_detections_LAEs}
\end{figure*}

\begin{figure*}
    \centering
    \includegraphics[width=\textwidth]{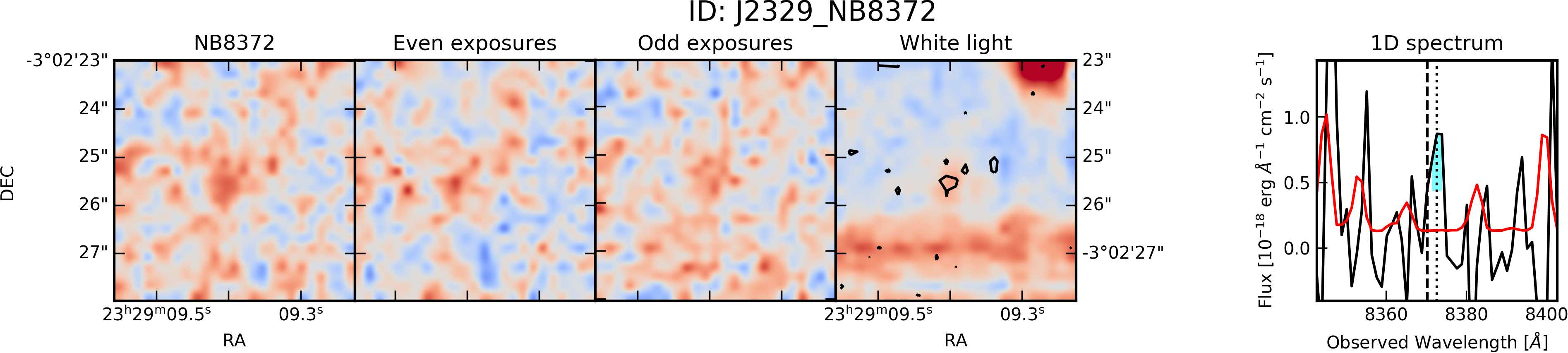} \\
    \includegraphics[width=\textwidth]{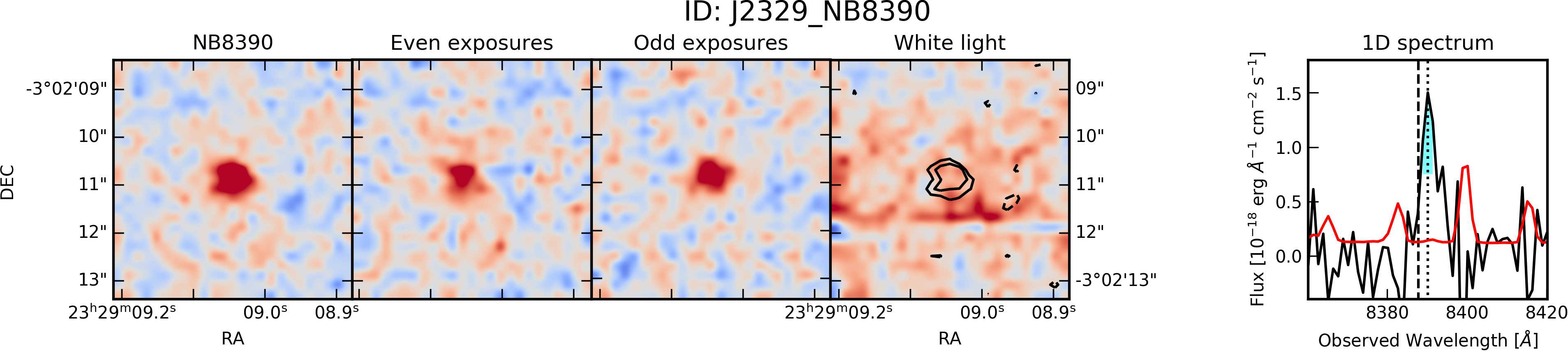} \\
    \caption{LAEs detected in the field of J2329 used in this work}
    \label{fig:J2329_MUSE_detections_LAEs}
\end{figure*}

\begin{figure*}
    \centering
    \includegraphics[width=\textwidth]{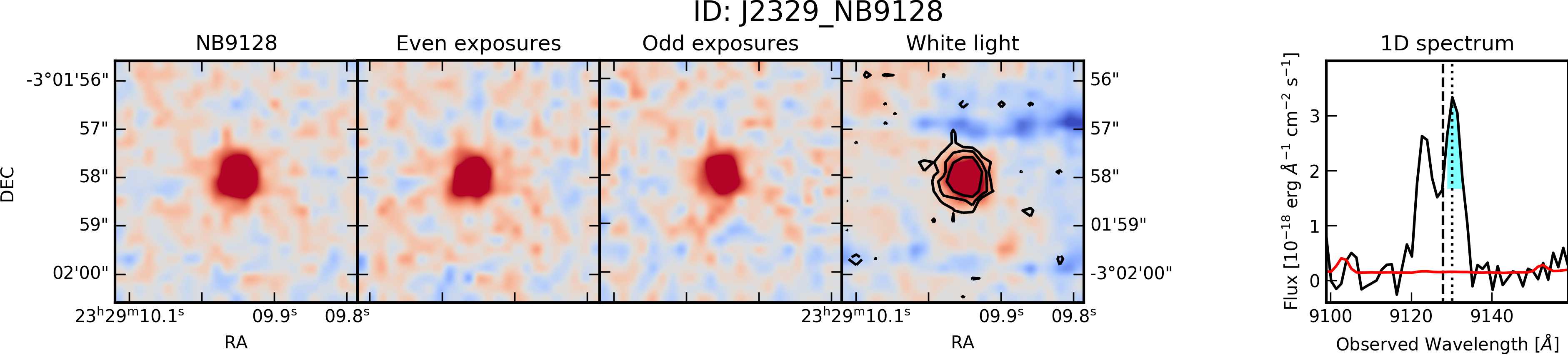} \\
    \includegraphics[width=\textwidth]{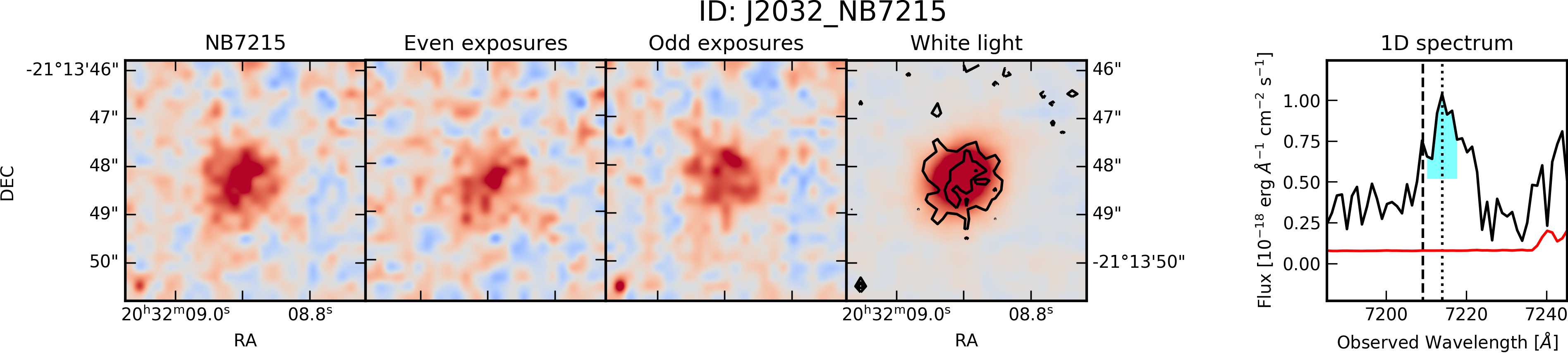} \\ 
    \includegraphics[width=\textwidth]{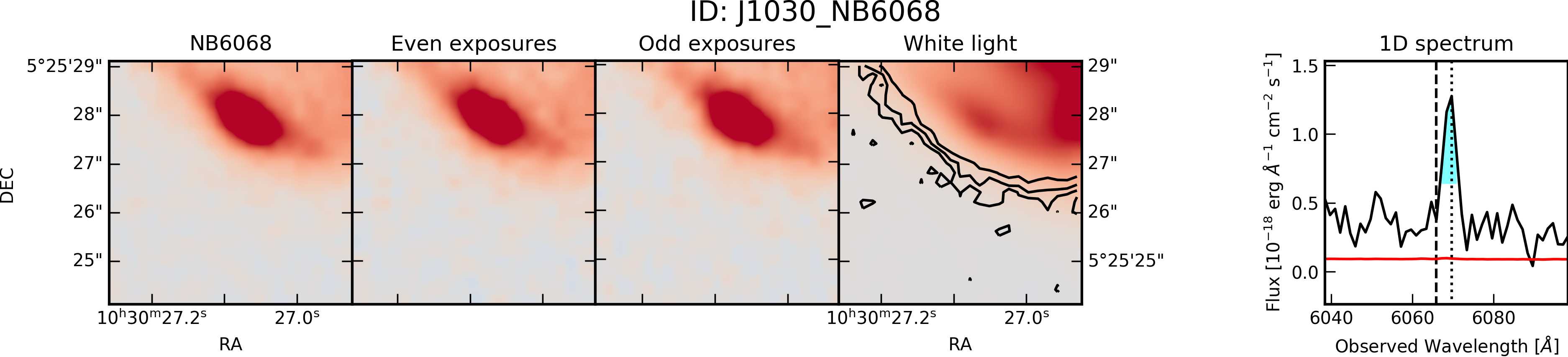} \\ 
    \includegraphics[width=\textwidth]{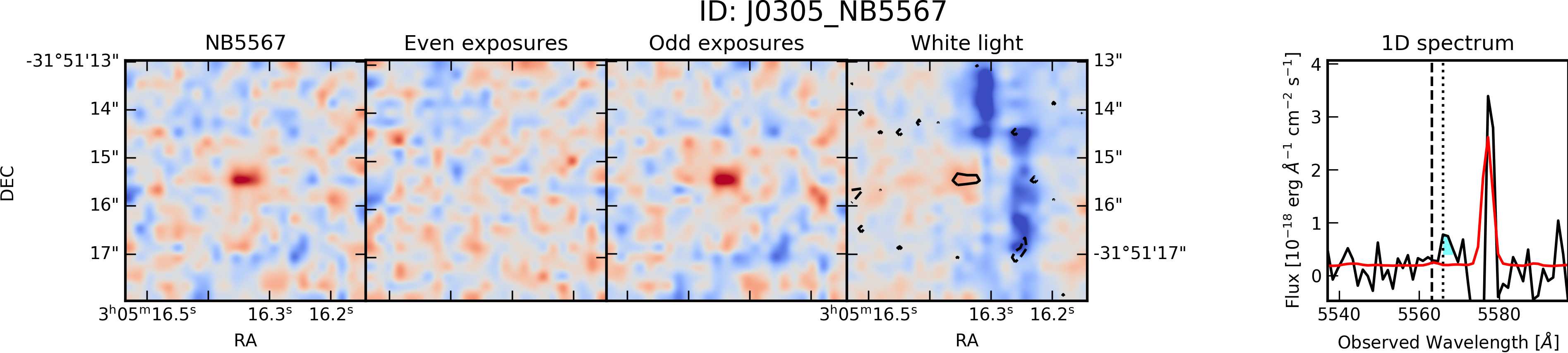} 
    \caption{Examples of typical false \textsc{LSDCat} detections of LAEs in MUSE cubes. \textbf{Top row:} Unsubtracted continuum visible both in the 1D spectra and the white light image. \textbf{2nd row:} [OII]$\lambda\lambda 3727$\,\AA\, emitter with continuum \textbf{3rd  row:} Detection due to a poorly substracted foreground object \textbf{Bottom row:} Faint misdetection due to an artifact or cosmic ray in one pixel in one exposure. The detection is subsequently only seen in the even or odd exposures cubes. }
    \label{fig:MUSE_misdet_examples}
\end{figure*}
\clearpage

\section{The mean transmission in the Lyman-$\alpha$ forest around LBGs and LAEs}\label{appendix:mean_transmission}
\begin{figure*}
    \centering
   \includegraphics[width= \textwidth]{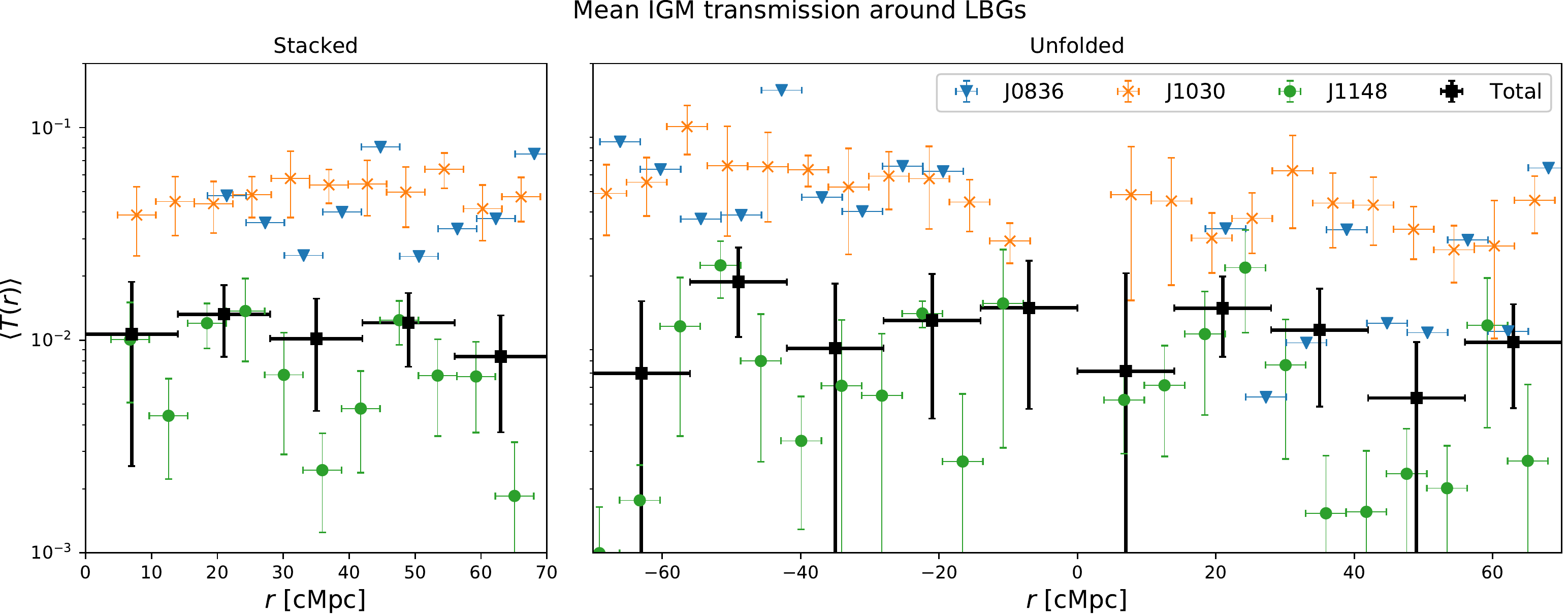}
   \caption{Mean IGM transmission around LBGs in the three different quasar fields (blue triangles: J0836, orange squares: J1030, green circles: J1148) surveyed with DEIMOS to confirm LBGs with Lyman-$\alpha$ emission. The points are slightly offset in the x-axis direction for clarity. Error-bars on the transmission represent the 1$\sigma$ confidence interval from bootstrapping, and thus bins to which only one LBG contributes have no uncertainty. The measurement introduced \citetalias{Kakiichi2018} presents a large scatter between sightlines as well as intrinsic scatter for poorly sampled sigthlines such as J0836.}
   \label{fig:stacked_flux_LBGs}
\end{figure*}

\begin{figure*}
    \centering
    \includegraphics[width= \textwidth]{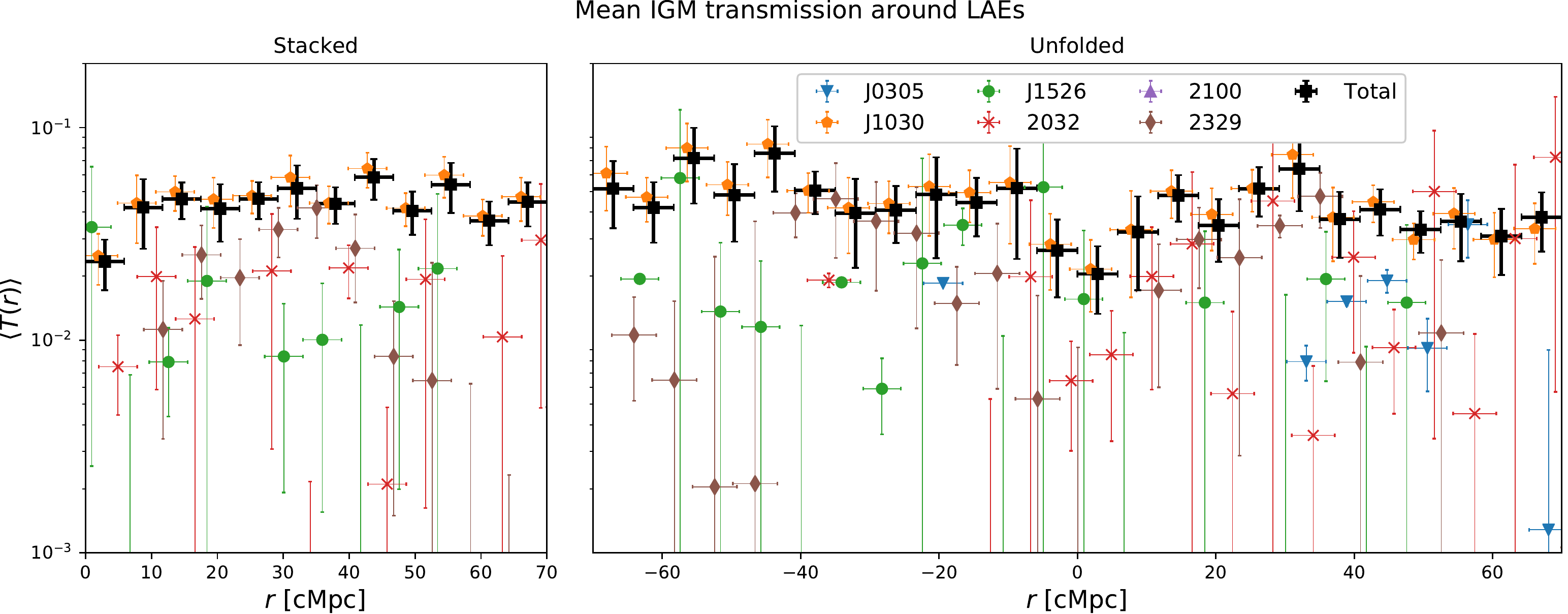}
    \caption{Mean IGM transmission around LAEs averaged in $6$ sightlines. Errorbars on the transmission represent the 1$\sigma$ confidence interval from bootstrap resampling. The signal is dominated by the sightline to J1030, which contains two to three times more LAEs than other fields due to a significantly deeper MUSE observation and a potential overdensity of sources. The stacked flux only shows mild evidence for increased absorption on small scales, and no excess on larger scales of $\sim 10-20$ cMpc.}
    \label{fig:stacked_flux_LAEs}
\end{figure*}

The first measurement of the correlation between galaxies and the reionising IGM was performed in \citetalias{Kakiichi2018} by computing the mean flux at distance $r$ from detected LBGs. We follow the same methodology here, computing the transverse and line-of-sight distance of every pixel in the Lyman-$\alpha$ forest of quasars to the detected LBGs and LAEs. We then bin the observed transmission in segment of a few cMpc, weighting each point by the inverse of the squared error. We present the results in Fig. \ref{fig:stacked_flux_LBGs} and Fig. \ref{fig:stacked_flux_LAEs}. 

We do not find any evidence for increased transmission close to LBGs with our stack of $3$ sightlines and $13$ LAEs, unlike the tantalising signal presented in \citetalias{Kakiichi2018}. Although small number statistics might be biasing the measurement, we argue that we would not expect the observed average transmission to be enhanced near LBGs. The reason for this is twofold and demonstrated by Fig. \ref{fig:stacked_flux_LBGs}. First of all, the signal is dominated by the sightlines with the largest number of LBGs and greatly affected by cosmic variance in these sightlines. The impact of cosmic variance was also demonstrated in theoretical work with simulated Lyman-$\alpha$ skewers which concluded that normalising the transmission in each sightline as in \citetalias{Meyer2019} was necessary to obtain a consistent signal across simulations boxes (or across the sky). Secondly, when the numbers of LBGs grows, the signal is dominated by the evolution of the mean opacity of the IGM with redshift. This is evidenced in the signal of the sightline with the largest number of LBGs (J1030, orange squares points in Fig. \ref{fig:stacked_flux_LBGs}), which increased at negative separations from the LBG (defined as in the direction of the observer, i.e. towards lower redshift IGM), and conversely decreases at positive separations (towards the quasar, i.e. higher redshift IGM). It is therefore no surprising that the mean transmission around our large sample of LAEs does not show any enhancement of the transmission close to LAEs. It is worth noting that the absorption on small scales due to enhanced gas overdensities close to galaxies is still detected. 

\section{Impact of the redshift correction on the 2PCCF signal}
\begin{figure*}
    \centering
    \includegraphics[width=0.8\textwidth]{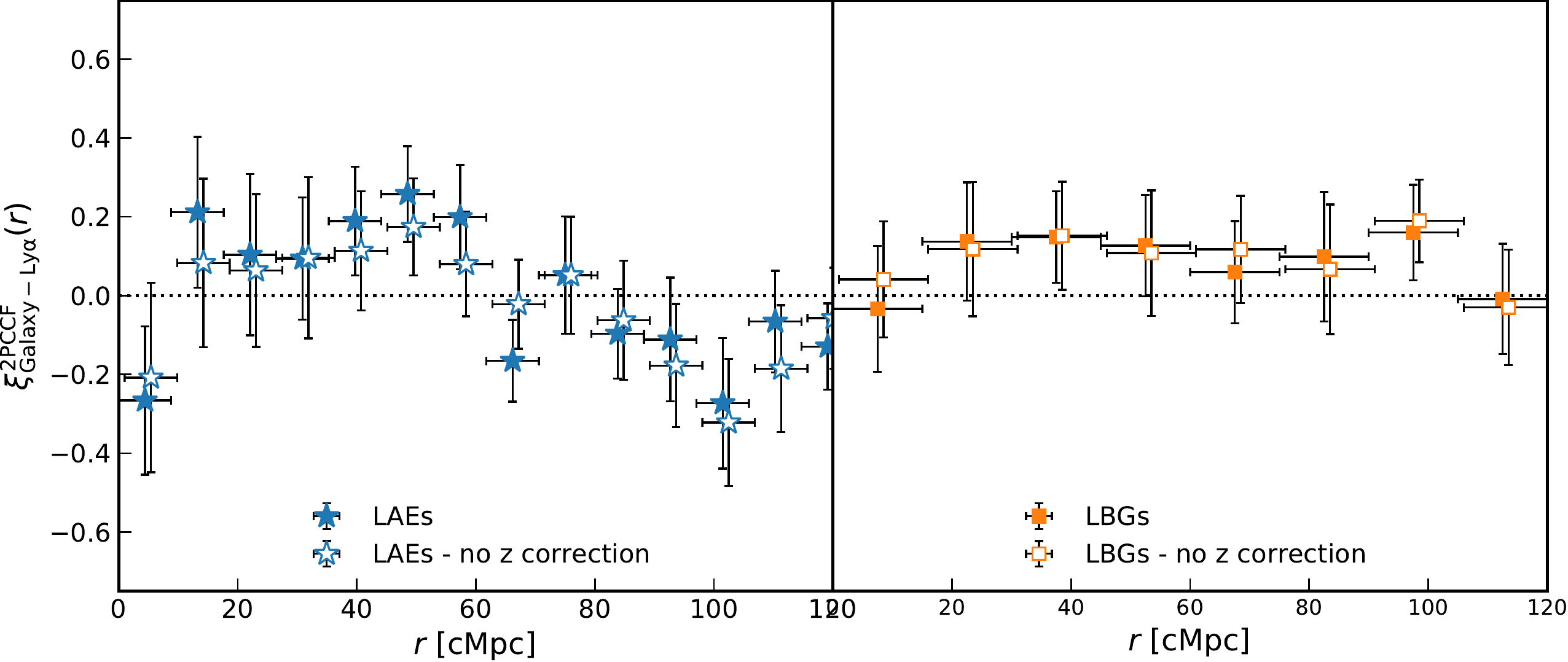}
    \caption{2PCCF of LAE/LBG with the Lyman-$\alpha$ transmission spike with (full symbols) and without (empty symbols) the redshift correction based on the FWHM of the Lyman-$\alpha$ line. Removing the redshift correction flattens significantly the LAE 2PCCF.}
    \label{fig:2PCCF_z_correction}
\end{figure*}

We present in Figure \ref{fig:2PCCF_z_correction} the 2PCCF with and without the redshift correction detailed in Section \ref{sec:z_corr}. Not correcting for the Lyman-$\alpha$ offset to systemic reduces the signal by $-1.5\sigma (-0.2\sigma)$ for the LAE(LBG) 2PCCF. The lesser impact on the LBG 2PCCF can be explained by their larger transverse distance. Indeed, an error on the systemic redshift only affects the line-of-sight distance to transmission spikes. For LAEs, the line-of-sight direction dominates the 3D distance as $r_\perp \lesssim 7 \text{cMpc}$ whereas LBGs are further apart ($r_\perp \approx 10-20 \text{cMpc}$), reducing the impact of an error in the line-of-sight direction.

\section{Illustris gas overdensity PDF}
\label{app:PDF_illustris}
\begin{figure}
\centering
    \includegraphics[width=0.43\textwidth]{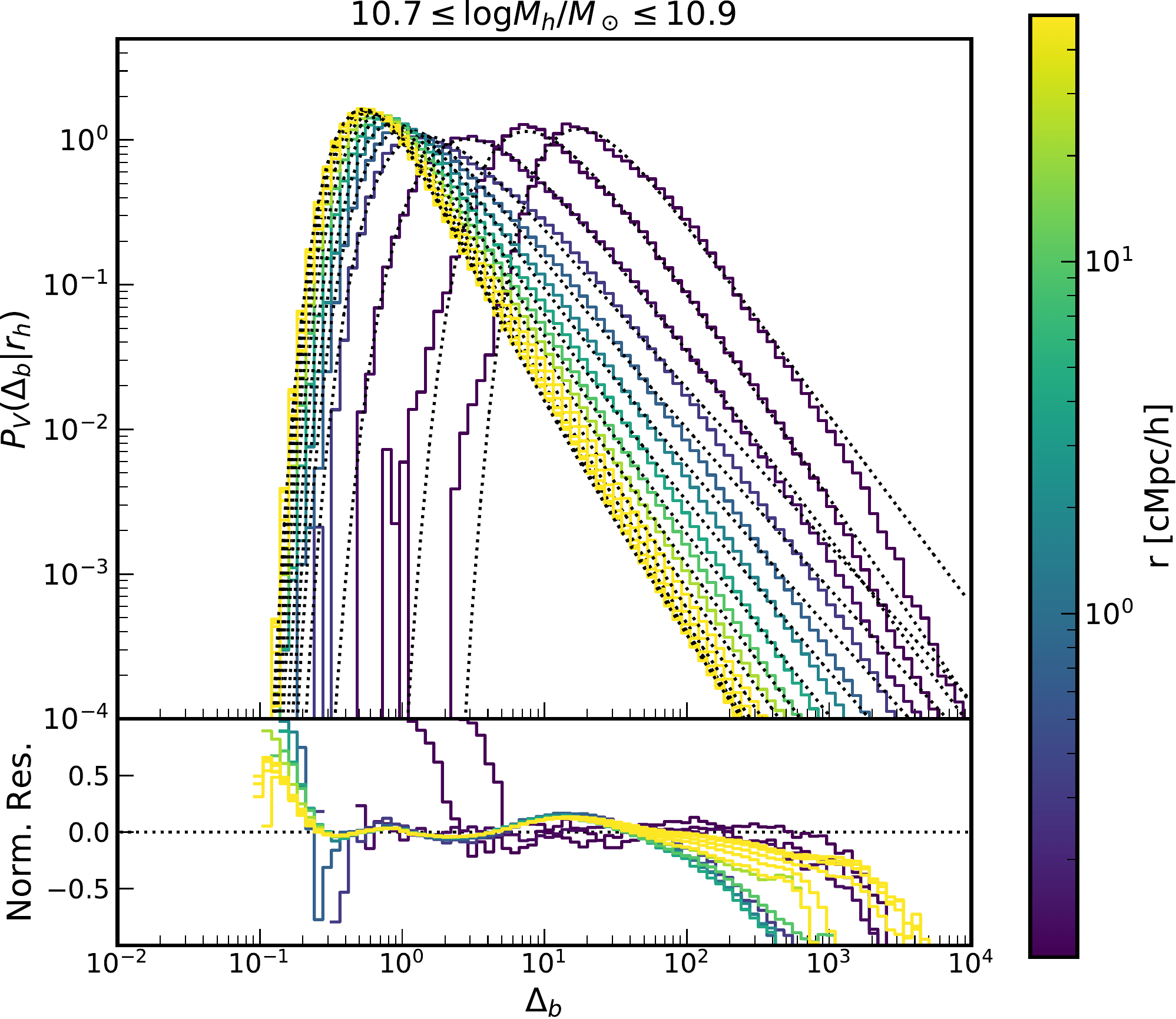}
    \includegraphics[width=0.43\textwidth]{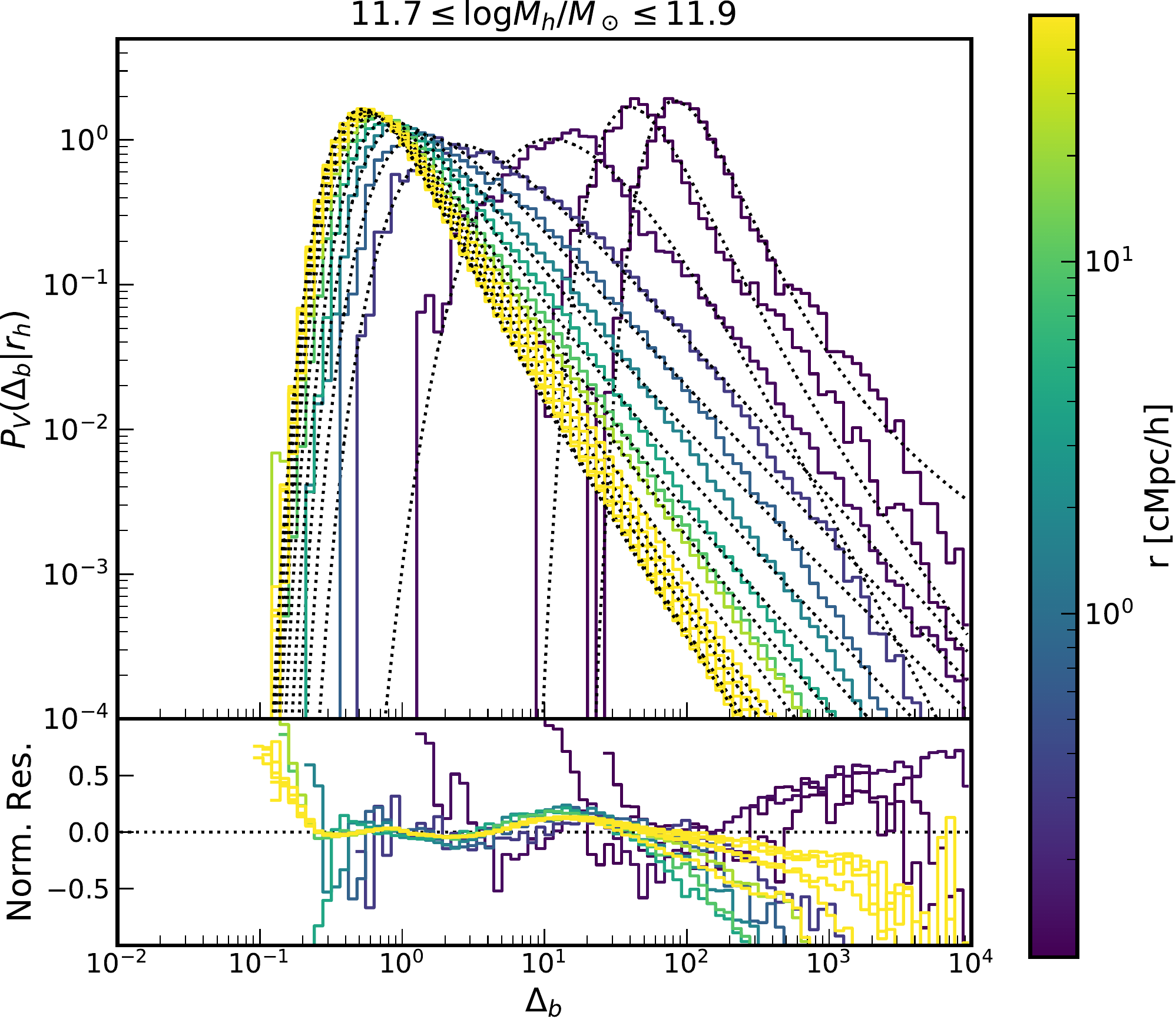}
    \includegraphics[width=0.43\textwidth]{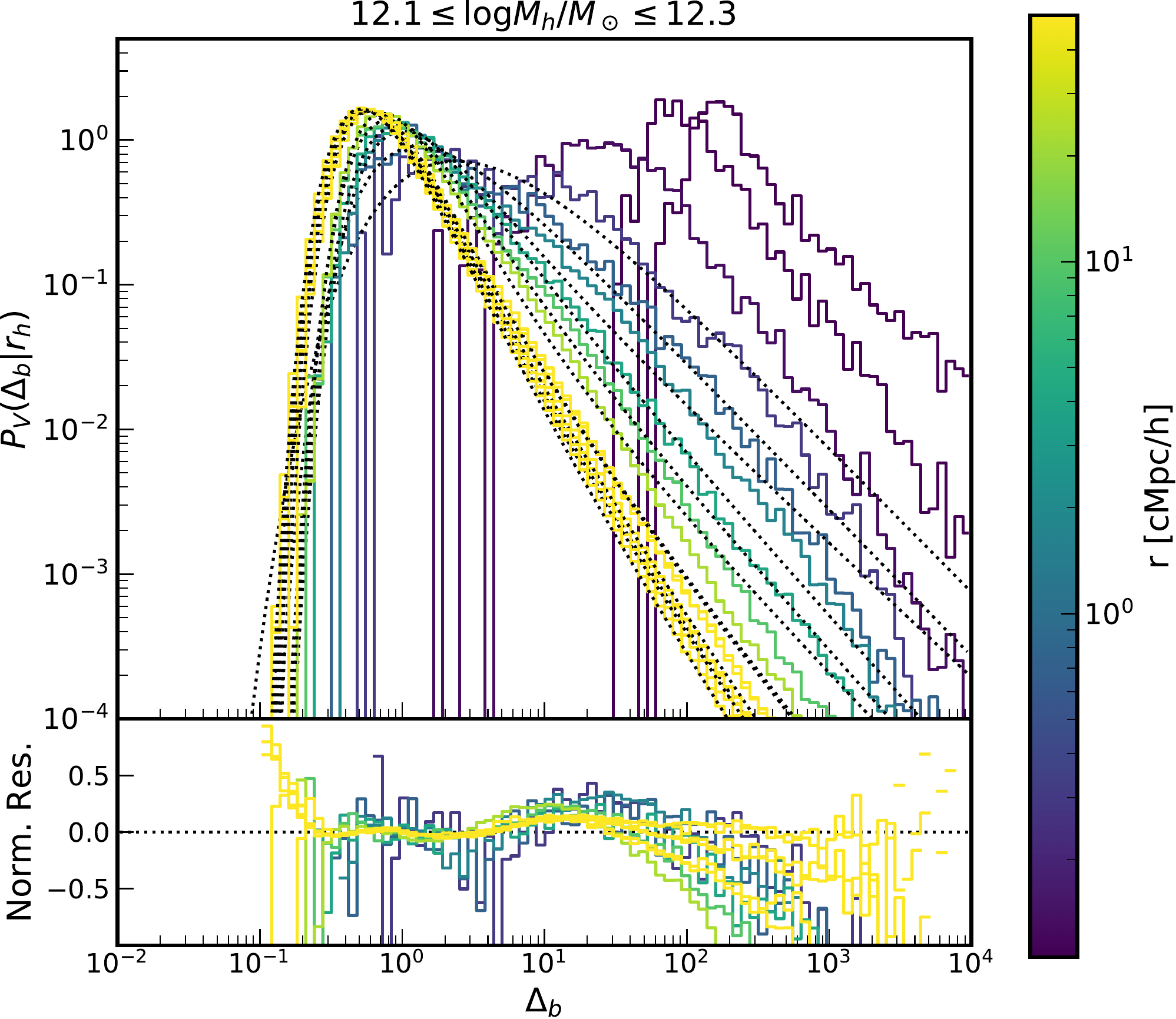}
    \caption{  \textbf{Upper panels:} The fits (dotted black ) of our chosen analytical form (Eq. \ref{eq:analytical_form_PDF}) are overlaid on the extracted PDF from the IllustrisTNG simulation boxes for one mass range and a given redshift, coloured by distance from the centre of the halo.  \textbf{Lower panels:} Residuals of the PDF fit, coloured by distance from the centre of the halo, showing good agreement on the validity limit of the prescribed analytical form between $10^{-1} \leq \Delta_b \leq 10^{2}$. Only a quarter of the raw TNG PDF and associated fits are shown for clarity.}
    \label{fig:appendix_fit_pdf_illustrisTNG}
\end{figure}

\begin{figure}
    \includegraphics[width=0.5\textwidth]{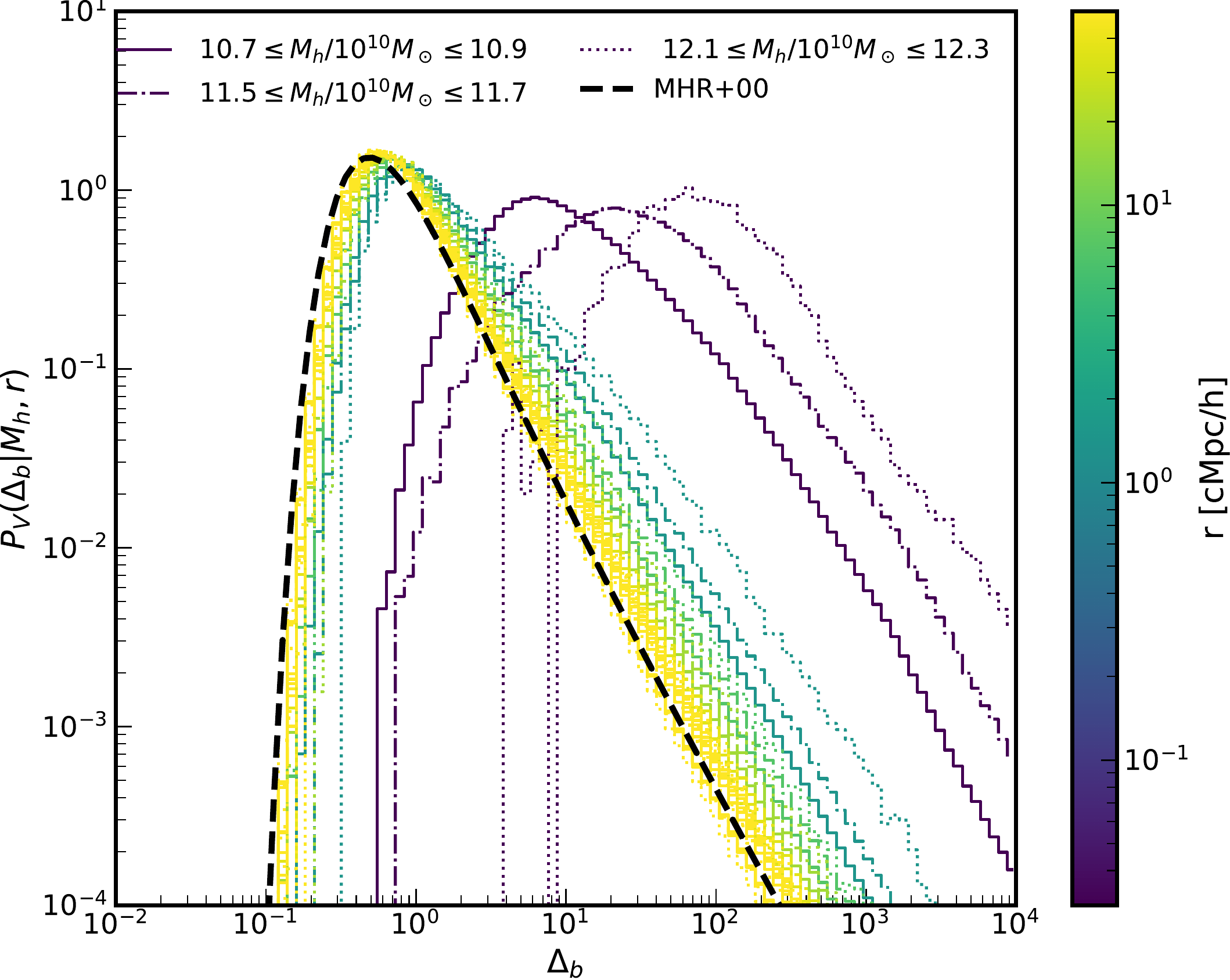}
    \caption{ Baryonic overdensity probability density function $P_V(\Delta_b| r, M_h)$ and its dependence on the mass and the radius from the nearest dark matter halo at $z\sim 5.85$ extracted from the IllustrisTNG100-2 cosmological simulation box. The PDFs are similar except close to the centre of the halo where gas overdensities are preferentially found. On large scales, the PDF matches the analytical fits of \citep[][]{Miralda-Escude2000} and \citep[][dotted black]{Pawlik2009} used in \citetalias{Kakiichi2018} and \citetalias{Meyer2019}.}
    \label{fig:pdf_illustrisTNG}
\end{figure}

We present in Fig.\ref{fig:appendix_fit_pdf_illustrisTNG} fits of the Illustris gas PDF for additional masses. We contrast the effect of the host halo mass of on the PDF in Fig.\ref{fig:pdf_illustrisTNG}. We also list the parameters of the analytical fit (Eq. \ref{eq:analytical_form_PDF}) at all masses and at varied distances from the host halo centre in Table \ref{tab:value_pdf_illustris}. Additional parameterisation and quality plots are available upon request.

\begin{table*}
   \centering
    \begin{tabular}{ccccccccccccc}
      & \multicolumn{4}{c}{$r=1$ cMpc/h} & \multicolumn{4}{c}{$r=15$ cMpc/h} &\multicolumn{4}{c}{$r=50$ cMpc/h}   \\ \hline
       $\log M/M_{\odot}$ & A & C0 & $\delta_0$ & $\beta$ & A & C0 & $\delta_0$ & $\beta$ & A & C0 & $\delta_0$ & $\beta$  \\ \hline
10.6 & 0.6671 & 0.838 & 0.659 & 2.131 & 0.424 & 0.920 & 0.976 & 2.433 & 0.418 & 0.923 & 0.982 & 2.447\\ 
10.8 & 0.6611 & 0.815 & 0.686 & 2.093 & 0.422 & 0.921 & 0.977 & 2.439 & 0.419 & 0.923 & 0.981 & 2.444\\ 
11.0 & 0.6699 & 0.814 & 0.677 & 2.101 & 0.423 & 0.922 & 0.975 & 2.437 & 0.419 & 0.922 & 0.981 & 2.444\\ 
11.2 & 0.6928 & 0.830 & 0.630 & 2.119 & 0.423 & 0.921 & 0.975 & 2.440 & 0.419 & 0.923 & 0.980 & 2.443\\ 
11.4 & 0.7180 & 0.835 & 0.585 & 2.098 & 0.427 & 0.920 & 0.971 & 2.428 & 0.418 & 0.923 & 0.983 & 2.448\\ 
11.6 & 0.7162 & 0.828 & 0.583 & 2.061 & 0.423 & 0.921 & 0.975 & 2.442 & 0.420 & 0.923 & 0.977 & 2.446\\ 
11.8 & 0.6681 & -0.775 & 0.685 & 1.969 & 0.431 & -0.915 & 0.971 & 2.422 & 0.418 & -0.924 & 0.978 & 2.451\\ 
12.0 & 1.1519 & -0.915 & 0.180 & 2.163 & 0.427 & -0.916 & 0.973 & 2.440 & 0.420 & -0.919 & 0.987 & 2.430\\ 
12.2 & 0.6338 & -0.768 & 0.677 & 1.823 & 0.444 & -0.899 & 0.974 & 2.375 & 0.414 & -0.929 & 0.990 & 2.430\\ 
    \end{tabular}
    \caption{Parameters of the analytical fits (Eq. \ref{eq:analytical_form_PDF}) to the volume-weighted gas overdensity PDF extracted from IllustrisTNG100-2 simulation box from a snapshot at $z=5.85$. $\log M$ indicates the mid-point of logarithmic mass bin with $\Delta \log M = 0.2$.}
    \label{tab:value_pdf_illustris}
\end{table*}

\section{Dependence of the cross-correlation on the escape fraction}
\label{app:fesc}
We find that the 2PCCF is most sensitive to the escape fraction. Surprisingly perhaps, decreasing the escape fraction increases the correlation at the redshift and opacity considered here. We now investigate this behaviour by looking back at our cross-correlation model
\begin{align}
\xi^{\rm 2PCCF}_{\rm Gal-Ly\alpha}(r) & = \frac{ P[<\Delta_b^{\text{\tiny max}} (\langle \Gamma_{\text{HI}}^{\text{\tiny CL}}(r) \rangle) | r,M_h] }{ P[<\Delta_b^{\text{\tiny max}}(\bar{\Gamma}_{\rm HI}) | r\rightarrow\infty,M_h ]} - 1     \nonumber  \\ 
 & = \frac{\int_0^{\Delta_b^{\text{\tiny max, CL}}(r)} P_V(\Delta_b | r, M_h) \text{d}\Delta_b}{ \int_0^{\overline{\Delta_b^{\text{\tiny max}}}} P_V(\Delta_b | r, M_h ) \text{d}\Delta_b} - 1
 \text{   ,  }
\end{align}
where we have substituted $\Delta_b^{\text{\tiny max, CL}}(r) = \Delta_b^{\text{\tiny max}} (\langle \Gamma_{\text{HI}}^{\text{\tiny CL}}(r)  \rangle) $ and $\overline{\Delta_b^{\text{\tiny max}}}=\Delta_b^{\text{\tiny max}}(\overline{\Gamma}_{\text{HI}})$. The maximum gas overdensity at which a transmission spike is detected depends on the photoionisation rate $\Delta_b \propto \Gamma^{0.56}$, and the enhanced photoionisation rate is proportional to the mean UVB $\Gamma_{\rm HI}^{\rm CL} (r) \propto \overline{\Gamma_{\rm}}\times (1+\zeta(r))$, where we have subsumed the boosting effect of the clustering faint sources in one function $\zeta(r)$ for convenience.
Therefore the maximum underdensity permitted in order to get a spike around a galaxy is proportional, but slightly higher, than at a random position. The cross-correlation than can be simplified as
\begin{align}
& \xi^{\rm 2PCCF}_{\rm Gal-Ly\alpha}(r) \simeq \nonumber  \\
~~~~~~ & \frac{\int_0^{\overline{\Delta_b^{\text{\tiny max}}}} P_V(\Delta_b| r, M_h) \text{d}\Delta_b + \int_{\overline{\Delta_b^{\text{\tiny max}}}}^{\Delta_b^{\text{\tiny max, CL}}(r)} P_V(\Delta_b| r, M_h) \text{d}\Delta_b }{ \int_0^{\overline{\Delta_b^{\text{\tiny max}}}} P_V(\Delta_b| r, M_h) \text{d}\Delta_b} - 1 \nonumber \\
& = \frac{\int_{\overline{\Delta_b^{\text{\tiny max}}}}^{\Delta_b^{\text{\tiny max, CL}}(r)} P_V(\Delta_b| r, M_h) \text{d}\Delta_b }{ \int_0^{\overline{\Delta_b^{\text{\tiny max}}}} P_V(\Delta_b| r, M_h) \text{d}\Delta_b } \text{   .   } \label{eq:fraction_xcorr} 
\end{align}
When the escape fraction increases, it increases the photoionisation rate, in turn increasing both $\overline{\Delta_b^{\text{\tiny max}}}$ and $\Delta_b^{\text{\tiny max, CL}}(r) \propto \overline{\Delta_b^{\text{\tiny max}}} (1+\zeta(r))^{0.56}$. To understand how the escape fraction impacts the cross-correlation, we consider how the nominator and the denominator of eq. \ref{eq:fraction_xcorr} react to a small change in $\Delta_b$. At first order,

\begin{align}
& \int_{\overline{\Delta_b^{\text{\tiny max}}} + \delta\Delta_b}^{\Delta_b^{\text{\tiny max, CL}}(r) + \delta\Delta_b} P_V(\Delta_b| r, M_h) \text{d}\Delta_b  - \int_{\overline{\Delta_b^{\text{\tiny max}}}}^{\Delta_b^{\text{\tiny max, CL}}(r)} P_V(\Delta_b| r, M_h) \text{d}\Delta_b \nonumber  \\
& \simeq P_V(\Delta_b^{\text{\tiny max, CL}}| r, M_h)\delta\Delta_b  - P_V(\overline{\Delta_b^{\text{\tiny max}}}| r, M_h)\delta\Delta_b
\end{align}

At $\Delta_b \lesssim 1$, which is the regime we probe, $P_V$ is an increasing function, and therefore increasing the escape fraction increases the nominator. However, it is easy to see that the denominator increases by a greater amount

\begin{align}
&\int_0^{\overline{\Delta_b^{\text{\tiny max}}} + \delta\Delta_b} P_V(\Delta_b| r, M_h) \text{d}\Delta_b - \int_0^{\overline{\Delta_b^{\text{\tiny max}}}} P_V(\Delta_b| r, M_h) \text{d}\Delta_b   \nonumber \\
& \simeq P_V(\Delta_b^{\text{\tiny max, CL}}| r, M_h)\delta\Delta_b \text{  ,}
\end{align}
therefore the cross-correlation decreases as the mean UVB increases. In fact, the decrease rate depends on the average maximum gas overdensity to detect a spike, meaning that the cross-correlation is maximally sensitive to changes in the ionisation background when $\overline{\Delta_b^{\rm max}}\sim 0.1-1$. At $z\sim 5.5-6$ we are roughly in that range, although this could be improved with next generation instrument to reach larger opacities. This also implies that as IGM temperatures and the UVB increase at lower redshift, $\overline{\Delta_b^{\rm max}}$ becomes greater than $1$ and the cross-correlation is insensitive to UVB fluctuations.

\bsp
\label{lastpage}
\end{document}